\documentclass[prd,amsmath,amssymb,nofootinbib,superscriptaddress,showkeys,reprint]{revtex4-2}
\usepackage{graphicx}
\usepackage{booktabs}
\usepackage{array}
\usepackage{multirow}
\usepackage{dcolumn}
\usepackage{bm}
\usepackage{cancel}

\usepackage[caption=false]{subfig}
\usepackage[table]{xcolor}
\definecolor{MAGENTA}{named}{magenta}
\usepackage[colorlinks=true, linkcolor=blue, citecolor=blue, urlcolor=blue]{hyperref}
\usepackage{orcidlink}
\usepackage{siunitx}
\newcolumntype{B}{>{\bfseries}c}

\begin{document}

\title{AdS Black Hole Solution with a Dark Matter Halo Surrounded by a Cloud of Strings}

\author{Faizuddin Ahmed\orcidlink{0000-0003-2196-9622}} 
\email{faizuddinahmed15@gmail.com} 
\affiliation{Department of Physics, Royal Global University, Guwahati, 781035, Assam, India}

\author{Abdelmalek Bouzenada\orcidlink{0000-0002-3363-980X}}\email{ abdelmalekbouzenada@gmail.com}
\affiliation{Laboratory of Theoretical and Applied Physics, Echahid Cheikh Larbi Tebessi University 12001, Algeria}

\author{Edilberto O. Silva\orcidlink{0000-0002-0297-5747}}
\email{edilberto.silva@ufma.br (Corresp. author)}
\affiliation{Departamento de F\'{\i}sica, Universidade Federal do Maranh\~{a}o, 65085-580 S\~{a}o Lu\'{\i}s, Maranh\~{a}o, Brazil}

\date{\today}

\begin{abstract}
We derive and analyze a Schwarzschild-like Anti-de Sitter (AdS) BH (BH) obtained as a static, spherically symmetric solution of Einstein’s equations sourced by a cloud of strings (CoS) and a dark matter (DM) halo modeled by a Dehnen-type density profile. We first study the geodesic motion of massless and massive test particles, emphasizing how the CoS parameter $\alpha$ and the DM halo parameters $(\rho_s,r_s)$ influence photon spheres, circular orbits, the BH shadow, and the innermost stable circular orbit (ISCO). We then examine scalar perturbations via the effective potential and the associated quasinormal-mode (QNM) spectra, showing how $\alpha$ and $(\rho_s,r_s)$ deform oscillation frequencies and damping rates, thereby affecting stability diagnostics. Furthermore, we investigate the thermodynamics in the extended phase space, deriving the Hawking temperature, equation of state, Gibbs free energy, and specific heat capacity, and establishing a consistent first law and Smarr relation with natural work terms for $\alpha$ and $(\rho_s,r_s)$. We find that the interplay between the CoS and the DM halo produces quantitative and sometimes qualitative changes in both dynamical and thermodynamical properties, including shifts of the Hawking-Page transition and heat-capacity divergences, thus reshaping the phase structure of Schwarzschild-AdS BHs.
\end{abstract}

\maketitle

\section{Introduction}\label{Sec:I}

In general relativity (GR), BHs arise naturally as exact solutions to Einstein’s field equations. Although these solutions were proposed over a century ago, compelling observational evidence has only recently emerged through revolutionary instruments such as the Event Horizon Telescope (EHT) and the LIGO-VIRGO collaborations \cite{AA1,AA2,AA3}. These groundbreaking observations not only confirm the existence of BHs but also open new avenues for testing fundamental aspects of gravity and probing the unknown properties of BHs. Nonetheless, several open questions remain, particularly concerning the nature of BHs and their interactions with the surrounding environment. 

One of the most persistent mysteries in modern astrophysics and cosmology is the nature and existence of dark matter (DM). As a result, identifying signatures of DM in the vicinity of BHs has become a significant pursuit. In many astrophysical scenarios, especially in galactic centers, supermassive BHs (SMBHs) are thought to be embedded in dense matter distributions, including DM halos. There is substantial observational support for this view; SMBHs are widely believed to power active galactic nuclei (AGNs) \cite{AA5,AA6} and to reside within DM halos \cite{AA7,AA8}.

In addition to the direct association with SMBHs, DM plays a fundamental role in shaping the universe at galactic and cosmological scales. Its influence is evident in the unexpectedly flat rotation curves of spiral and elliptical galaxies \cite{AA9}, the dynamics of cluster mergers such as the Bullet Cluster \cite{AA10}, and the formation of large-scale structures \cite{AA11}. The discovery of flat galactic rotation curves, in particular, was pivotal in postulating the existence of DM. Current astrophysical observations suggest that DM comprises about 90\% of a galaxy's mass, while the remaining 10\% is attributed to luminous baryonic matter \cite{AA12}. The evolution of DM suggests that it was initially concentrated near galactic centers, facilitating star formation and clustering, before redistributing into halo structures through dynamical processes. Observations further indicate that SMBHs, sometimes even in binary systems, are often situated at the center of these DM halos \cite{AA13,AA14}. The necessity of DM to explain various cosmic phenomena is further supported by measurements of the cosmic microwave background, which indicate that DM constitutes approximately 27\% of the universe, with dark energy and baryonic matter comprising 68\% and 5\%, respectively.

Numerous theoretical models have been proposed to explain DM, including several candidates arising from physics beyond the Standard Model, such as weakly interacting massive particles (WIMPs), axions, and sterile neutrinos \cite{AA15,AA16,AA17,AA18}. Given the presumed weak interaction between DM and Standard Model particles, gravitational effects provide a promising way to probe its properties. In particular, DM is expected to be highly concentrated around SMBHs, influencing the dynamics of extreme \cite{AA19} and intermediate mass-ratio inspirals \cite{AA20,AA21}. These interactions could help map the DM distribution near BHs and reveal new physics. Understanding the interplay between DM halos and BHs is thus a key focus area. DM significantly alters galactic rotation curves and is prominently implicated in cluster-scale dynamics, such as the Bullet Cluster event \cite{AA8}. Despite DM’s elusive nature, a growing body of evidence strongly supports its gravitational presence \cite{AA23}.

Research within the Dehnen-type DM halo model has yielded further insights into BH-DM interactions. For example, the influence of DM density slopes on the survival of star clusters post-gas expulsion has been explored \cite{AA31}. Studies have also examined stellar distributions using Plummer and Dehnen profiles, highlighting different central cusp behaviors. Moreover, Dehnen-type DM halo solutions have been applied to study ultra-faint dwarf galaxies \cite{AA33}, and new BH solutions embedded within such halos have recently been proposed \cite{AA32,AA34}. These works analyze various aspects, including thermodynamics, null geodesics \cite{AA32}, and constraints on halo parameters \cite{AA42}. More recently, the influence of DM halos on observational features like quasinormal modes, photon sphere radius, BH shadow \cite{AA43,AA44}, and gravitational waveforms from periodic orbits has been explored within the BH-Dehnen halo framework \cite{AA45}. These studies are crucial for advancing our understanding of how DM environments modify BH spacetimes and the associated observables.

Dark matter (DM) remains a fundamental puzzle in modern theoretical physics, crucial for unifying particle physics and cosmology \cite{b1}. Astrophysical and cosmological evidence, from galaxy dynamics and gravitational lensing to cosmic microwave background observations, strongly supports its existence \cite{b5,b6,b7}. Approximately one-quarter of the universe’s energy density is attributed to DM, essential for explaining anisotropies and large-scale structure formation. While modified gravity theories have been proposed \cite{b8}, many observations contradict these alternatives \cite{b9}. Consequently, DM is widely hypothesized to be non-baryonic, weakly interacting particles, highlighting a particle-physics basis for its gravitational effects \cite{b13}.

The stability of a dark matter (DM) halo is largely determined by the gravitational field of its host once the halo attains sufficient mass, though its exact mass remains uncertain due to complex formation processes \cite{QSDE1}. To address this, halo mass is often treated as a free parameter constrained by observations, with the halo size dependent on particle mass $m$ and the astrophysical context. For example, Earth’s halo radius exceeds the planet’s size if $m \lesssim 10^{-9}$ eV, allowing potential detection via surface or near-orbit experiments; heavier particles yield smaller, less accessible halos \cite{DMH1}. Neutrinos serve as valuable probes, with ultra-light DM inducing time-dependent modifications in neutrino masses and mixing parameters studied in terrestrial, astrophysical, and cosmological settings \cite{DMH2}. These effects are enhanced by local DM overdensities, and their absence constrains DM-DM-neutrino interactions, especially for halos with $m \gtrsim 10^{-10}$ eV where oscillation frequencies exceed current detector sensitivity \cite{DMH5}. Hence, constraints on neutrino-DM couplings primarily arise indirectly, providing robust limits on new physics in local DM environments \cite{DMH10}.

In 1978, Letelier introduced a novel BH solution describing a Schwarzschild BH surrounded by a cloud of strings \cite{PSL}. In this model, the cloud is treated as a closed system, ensuring that the associated stress-energy tensor is conserved. Since then, numerous studies have explored the physical and geometrical implications of string clouds in various contexts, including accretion dynamics, BH thermodynamics, and quasinormal mode analysis. Multiple exact solutions have been proposed within this framework. Notably, in certain cases, the inclusion of a string cloud in a previously regular BH geometry leads to the reappearance of singularities \cite{SS14,SS15}. Glass and Krisch \cite{SS1} explored the effects of a one-dimensional string atmosphere on the geometry of relativistic stellar structures. This work laid the groundwork for later studies investigating similar distributions in the context of BH spacetimes. Ganguly \textit{et al.} \cite{SS2} extended the string cloud model to BH environments by analyzing accretion dynamics onto a Schwarzschild BH surrounded by a cloud of strings. Their results highlighted modifications in the accretion rates due to the string cloud, offering potential astrophysical signatures. Toledo and Bezerra \cite{SS3} studied a BH in Lovelock gravity surrounded simultaneously by quintessence and a cloud of strings. They obtained exact solutions and analyzed the effect of these exotic fields on the BH's thermodynamics, further enriching the phenomenology of such composite systems. In a related work, the same authors investigated the Reissner-Nordström BH encased in a similar environment \cite{SS4}. They focused on the system’s thermodynamics and quasinormal modes, revealing notable deviations from classical BH behavior due to the presence of the string cloud and quintessence field. Morais Graça \textit{et al.} \cite{SS5} explored the quasinormal mode spectrum of a Gauss-Bonnet BH modified by the presence of a string cloud. Their analysis indicated distinctive oscillation modes, which could serve as potential observational signatures in gravitational wave astronomy. Costa \textit{et al.} \cite{SS6} analyzed Letelier’s string cloud solution in the presence of quintessence, studying its implications on BH thermodynamics and Hawking radiation. Their findings illustrate how combined exotic fields influence energy emission processes. Chabab and Iraoui \cite{SS7} extended the thermodynamic analysis to higher-dimensional AdS BHs enveloped by both quintessence and a cloud of strings. They identified Van der Waals-like phase transitions, revealing rich critical behavior in such configurations. Cai and Miao \cite{SS8} investigated the spectral properties and quasinormal modes of a Schwarzschild BH immersed in a string cloud within Rastall gravity. Their work emphasizes the non-trivial influence of alternative gravitational theories on observable BH properties. Ghosh \textit{et al.} \cite{SS9} considered the impact of a string cloud in third-order Lovelock gravity, presenting exact BH solutions and exploring their causal structure and thermodynamic behavior. The study demonstrates the viability of string cloud matter in higher-order gravity. In a follow-up study, Ghosh and Maharaj \cite{SS10} incorporated radiating BHs within Lovelock gravity, analyzing the dynamical evolution in the presence of a string cloud. Their work enriches the understanding of time-dependent geometries in extended gravity theories. Al-Badawi \textit{et al.} \cite{SS11} presented a Schwarzschild BH solution enveloped by both quintessence and a string cloud in modified gravity, revealing modifications to the horizon structure and BH shadow properties relevant for EHT-like observations. In another investigation, Al-Badawi {\it et al.} \cite{SS12} studied the thermodynamics and stability conditions for such BHs, demonstrating how the combined presence of quintessence and string clouds affects heat capacity and phase transitions. Shaymatov \textit{et al.} \cite{SS13} analyzed BH shadows and quasinormal modes in a background including both string cloud and quintessence, highlighting their impact on observational signatures potentially detectable by near-future interferometers. Liu {\it et al.} \cite{Liu} studied the quasinormal mode and greybody factor of Bardeen BHs with a cloud of strings via the WKB approximation. Veer {\it et al.} studied a BH solution for the Einstein gravity in the presence of Ayón-Beato-García nonlinear electrodynamics and a cloud of strings. Al-Badawi {\it et al.} \cite{Al-badawi} studied an effective quantum gravity BH with a cloud of strings surrounded by a quintessence field. Moreover, a cylindrical BH surrounded by a cloud of strings and the quintessence field in \cite{epjc2025} and a Ayón-Beato-García BH coupled with a cloud of strings in \cite{cjphy2025} have  recently been investigated.

Photon rings, or light rings, are closed null geodesics around BHs that play a crucial role in shaping the observable BH shadow. These rings possess distinct topological features that provide insight into the global structure of spacetime. In spherically symmetric BHs, the photon ring reduces to a single circular orbit, corresponding to the photon sphere \cite{chandra,perlick}. However, in more general BH spacetimes, such as axisymmetric or rotating solutions, the topology of photon rings becomes richer: instead of a single orbit, one encounters continuous families of light rings forming a photon region, whose structure depends sensitively on parameters like spin, charge, or external fields \cite{frolov}.

In recent years, topological methods have emerged as powerful tools for analyzing BH solutions. A notable development in this area is Duan’s $\phi$-mapping theory, which reveals a deep connection between topological defects and critical points within BH systems. These defects arise at locations where a vector field vanishes, often indicating significant phase transitions. At such points, a conserved topological current is generated, reflecting key geometric features of the underlying field. This current gives rise to a topological invariant, a global quantity that encapsulates the system’s phase structure. The invariant is calculated using geometric characteristics of the vector field, such as its local twisting and turning in spacetime, thereby offering valuable insights into the dynamics of the system. For detailed studies and applications across various BH configurations, readers are referred to~\cite{BB1}.

The thermodynamics of BHs (BHs) has emerged as a fundamental topic in gravitational physics, providing deep connections between geometry, quantum field theory, and statistical mechanics~\cite{Therm1,Therm2}. In this framework, the surface gravity at the event horizon determines the Hawking temperature ($T_H$), while the entropy is proportional to the horizon area. A key semiclassical insight is that Hawking radiation leads to thermal instability: as the BH evaporates, its temperature rises, accelerating the process. This motivates the study of thermal stability, which assesses the system's response to small perturbations in thermodynamic variables~\cite{Therm8}. A widely used criterion involves the specific heat at constant pressure, given by $C_p = T\left(\partial S / \partial T\right)_p$, where $C_p > 0$ indicates local stability~\cite{Therm9,Therm10}. Additionally, the behavior of $C_p$ can signal phase transitions, either through discontinuities or divergences that characterize critical points. Such analyses have been extensively applied to BH systems, offering insights into critical phenomena and phase structure~\cite{Therm13,Therm14}.

In this work, we investigate a modified Schwarzschild-AdS BH spacetime influenced by two physically motivated external structures: a cloud of cosmic strings and a surrounding dark matter (DM) halo. These modifications are not arbitrary; both string clouds and DM halos arise naturally in various astrophysical and cosmological scenarios. Clouds of strings represent topological defects that may form in the early universe, while DM halos are essential for explaining galactic dynamics and large-scale structure formation. Embedding such structures into a BH geometry allows for a more realistic modeling of astrophysical BH environments. We begin by constructing the modified spacetime metric incorporating the effects of both the string cloud and the DM halo, and we analyze how these ingredients alter the underlying geometry. Next, we examine the geodesic structure of the spacetime by studying both null (photon) and timelike (massive particle) geodesics. This analysis shows how observational features such as light bending, photon sphere, BH shadow, orbital stability, and particle motion are influenced by the surrounding environment. Building on this, we investigate the topology and properties of light rings, which are crucial in determining the shadow of a BH and have become directly relevant through observations by the Event Horizon Telescope. We further probe the response of the BH spacetime to external disturbances by studying scalar field perturbations. From this, we derive the associated effective potential and compute the quasinormal modes (QNMs), which offer insight into the stability and ringdown behavior of the system, key aspects in gravitational wave phenomenology. The analysis is then extended to the thermodynamic properties of the BH. We extract thermodynamic quantities from the horizon data, develop the extended phase space framework incorporating the cosmological constant as pressure, and formulate the corresponding first law. The resulting equation of state is analyzed for signatures of critical behavior, such as phase transitions analogous to those in van der Waals fluids. We further study heat capacity, Gibbs free energy, and the Hawking-Page transition, identifying stability regimes and thermodynamic phases. 

The structure of the paper is organized as follows. In Section~\ref{Sec:I}, we present the introduction and motivation for the study. Section~\ref{Sec:II} is dedicated to the construction of the background geometry, where we incorporate the effects of both a cloud of strings and a surrounding dark matter halo into the Schwarzschild-AdS BH spacetime. In Section~\ref{Sec:III}, we analyze the geodesic motion of test particles and photons in the modified geometry. Section~\ref{Sec:IV} focuses on the topological features of light rings. In Section~\ref{Sec:V}, we investigate scalar field perturbations and compute the associated quasinormal modes (QNMs). Section~\ref{Sec:VI} deals with the thermodynamic properties of the BH, starting with fundamental thermodynamic quantities derived from horizon data. Finally, Section~\ref{Sec:VII} summarizes the main findings and provides concluding remarks, with a discussion on the influence of model parameters and possible future directions.

\section{Background Geometry: Schwarzschild-AdS BH spacetime with a CoS and a Dark-Matter Halo}\label{Sec:II}

Throughout this section, we set $G=c=1$ and $8\pi=1$, so Einstein’s equations read $G_{\mu\nu}+\Lambda g_{\mu\nu}=T_{\mu\nu}$. 
Here, we consider a Schwarzschild-like BH spacetime surrounded by a DM halo characterized by a Dehnen-type density profile. Moreover, the BH solution is coupled with a cloud of strings. In Ref.~\cite{UU}, the authors presented the BH spacetime involving the DM distribution. The resulting spacetime metric describing the BH-DM solution \begin{align}
	ds^2&=-\left(1-\frac{2M}{r}-\rho_s\,r_s^2\,\ln\!\left(1+\frac{r_s}{r}\right)\right)\,dt^2\notag\\&+\frac{dr^2}{\left(1-\frac{2M}{r}-\rho_s\,r_s^2\,\ln\!\left(1+\frac{r_s}{r}\right)\right)}\notag\\&+r^{2}\,\left(d\theta ^{2}+\sin ^{2}\theta d\phi ^{2}\right).\label{aa1}
\end{align}
Here, $\rho_s$ and $r_s$ are the characteristic density and characteristic scale of the DM halo, respectively.

Moreover, the solution of Einstein’s equations describing a BH with a cloud of strings was obtained by Letelier~\cite{PSL}. The energy-momentum tensor is given by
\begin{align}
T^t_{\ t} (\mbox{CoS})= T^r_{\ r} (\mbox{CoS}) = -\frac{\alpha}{r^2},\; T^\theta_{\ \theta} (\mbox{CoS})=0= T^\phi_{\ \phi} (\mbox{CoS}), \label{aa2}
\end{align}
where $\alpha$ is an integration constant which is related to the presence of the cloud of strings. The resulting spacetime metric reads as
\begin{align}
ds^2&=-\left(1-\alpha-\frac{2M}{r}\right)\,dt^2+\frac{dr^2}{\left(1-\alpha-\frac{2M}{r}\right)}\notag\\&+r^{2}\,\left(d\theta ^{2}+\sin ^{2}\theta d\phi ^{2}\right).\label{aa3}
\end{align}
With $8\pi=1$, the effective CoS energy density is $\rho_{\rm CoS}=\alpha/r^2$ (since $\rho=-T^t{}_t$). One can verify consistency directly: for $f(r)=1-\alpha-2M/r$, the identity $G^t{}_t=(rf'+f-1)/r^2$ gives $G^t{}_t=-\alpha/r^2$; because $G^t{}_t= T^t{}_t-\Lambda$ and here $\Lambda=0$, it follows $T^t{}_t=-\alpha/r^2$ as in \eqref{aa2}.

We work in geometrized units $G=c=1$, and denote the AdS curvature radius by $\ell_p$ (used here purely as the AdS length, not as the Planck length), so that $\Lambda=-3/\ell_p^{2}$.

Inspired by these works and assuming no direct coupling between the cloud of strings and the DM halo, we consider a static and spherically symmetric geometry of the form
\begin{eqnarray}
	ds^2=-f(r)\,dt^2+\frac{dr^2}{f(r)}+r^{2}\left(d\theta ^{2}+\sin ^{2}\theta d\phi ^{2}\right),\label{metric}
\end{eqnarray}
with metric function
\begin{equation}
	f(r)=1-\alpha-\frac{2M}{r}-\rho_s\,r_s^2\,\ln\!\left(1+\frac{r_s}{r}\right)+\frac{r^2}{\ell^2_p}.\label{function}
\end{equation}

We are not merely postulating the form of $f(r)$; we now show that it solves Einstein’s equations for a composite source consisting of a cloud of strings and a dark-matter halo in AdS. 
Write
\begin{equation}
f(r)=1-\alpha-\frac{2m(r)}{r}+\frac{r^{2}}{\ell_p^{2}},   
\end{equation}
so that, the $tt$-equation $G^{t}{}_{t}+\Lambda=T^{t}{}_{t}$ implies $m'(r)=\tfrac{1}{2}\,r^{2}\,\rho_{\rm tot}(r)$ with $\rho_{\rm tot}=\rho_{\rm CoS}+\rho_{\rm DM}$ (the $\Lambda$ contribution is already contained in $+r^2/\ell_p^2$).
The string cloud contributes $T^{t}{}_{t}=T^{r}{}_{r}=-\alpha/r^{2}$ (Letelier), while the halo (parametrized by $(\rho_s,r_s)$) generates the logarithmic correction in $f(r)$.
Comparing with Eq.~\eqref{function}, we obtain
\begin{align}
m(r)=M+\frac{r}{2}\,\Phi(r),\qquad \Phi(r)\equiv \rho_s r_s^2\ln\!\Big(1+\frac{r_s}{r}\Big).   
\end{align}
Therefore,
\begin{equation}
m'(r)=\frac{1}{2}\,\Phi(r)+\frac{r}{2}\,\Phi'(r),    
\end{equation}
with
\begin{equation}
\Phi'(r)=-\frac{\rho_s r_s^3}{r(r+r_s)}, 
\end{equation}
so that
\begin{align} 
m'(r)=\frac{\rho_s r_s^2}{2}\ln\!\Big(1+\frac{r_s}{r}\Big)-\frac{\rho_s r_s^3}{2(r+r_s)}.  
\end{align}
Using $m'(r)=\tfrac{1}{2}r^{2}\rho_{\rm tot}(r)$ (excluding $\Lambda$ and subtracting the CoS piece), we isolate the halo density as
\[
\rho_{\rm DM}(r)=\frac{1}{r^{2}}
	\left[\rho_s r_s^2\ln\!\Big(1+\frac{r_s}{r}\Big)-\frac{\rho_s r_s^3}{r+r_s}\right]\ge 0,
\]
which is precisely the source of the logarithmic term in Eq.~\eqref{function}. 
Hence, the $f(r)$ used here is not just an ansatz; it is a solution consistent with Einstein’s equations for the composite source $T^\mu{}_\nu(\text{CoS})+T^\mu{}_\nu(\text{DM})-\Lambda g^\mu{}_\nu$.
In what follows, we consistently use $8\pi=1$; in particular, the extended-thermodynamics identification is $P\equiv-\Lambda=3/\ell_p^2$.

\paragraph*{Additional remarks:}
	\begin{itemize}
		\item Mass function and halo mass. From $m(r)=M+\frac{r}{2}\Phi(r)$ and $\Phi(r)\sim \rho_s r_s^3/r$ as $r\to\infty$, we have
        \begin{equation}
		m(r)=M+\frac{\rho_s r_s^3}{2}+O(r^{-1}),   
        \end{equation}
		so the halo produces a finite asymptotic offset $M_{\rm halo}=\rho_s r_s^3/2$.
		\item Asymptotics of $\rho_{\rm DM}$. For $r\gg r_s$, we have
        \begin{equation}
    \ln\!\Big(1+\frac{r_s}{r}\Big)=\frac{r_s}{r}-\frac{r_s^2}{2r^2}+O(r^{-3})\ 
    \end{equation}
    so that
    \begin{equation}
    \rho_{\rm DM}(r)=\frac{\rho_s r_s^4}{2}\,\frac{1}{r^4}+O(r^{-5}),   
    \end{equation}
	i.e., the density decays as $r^{-4}$ and the total halo mass is finite.
	\item Central behaviour. For $r\ll r_s$, we have
    \begin{equation}
	\rho_{\rm DM}(r)\simeq \frac{\rho_s r_s^2}{r^2}\Big[\ln(r_s/r)-1\Big],   
    \end{equation}
	while the enclosed halo mass $m_{\rm DM}(r)=\tfrac{r}{2}\Phi(r)\sim \tfrac{\rho_s r_s^2}{2}\,r\ln(r_s/r)\to0$ as $r\to0$ (no central mass divergence).
	\item Energy conditions for the CoS. $\rho_{\rm CoS}=\alpha/r^2>0$, $p_r=-\rho_{\rm CoS}$, $p_t=0$, which satisfies the null energy condition along radial null directions ($\rho+p_r=0$).
	\end{itemize}

\begin{figure*}[ht!]
\centering	\includegraphics[width=0.45\linewidth]{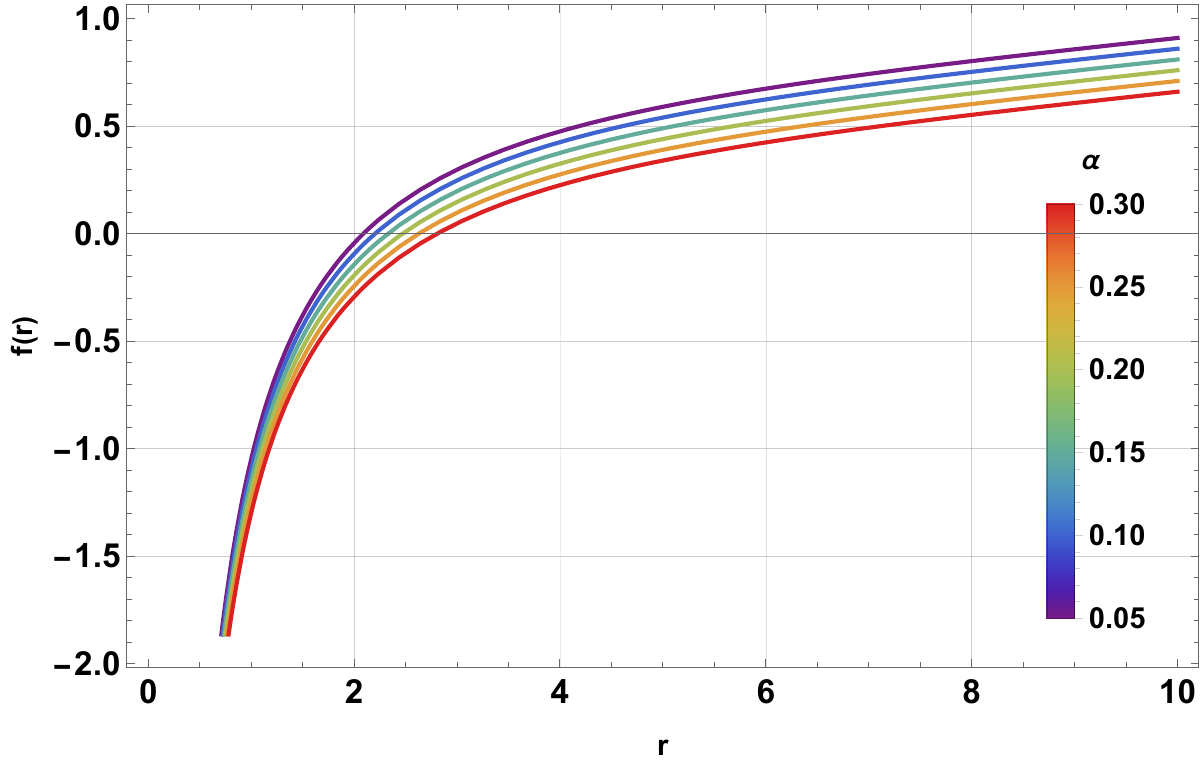}\qquad
\includegraphics[width=0.45\linewidth]{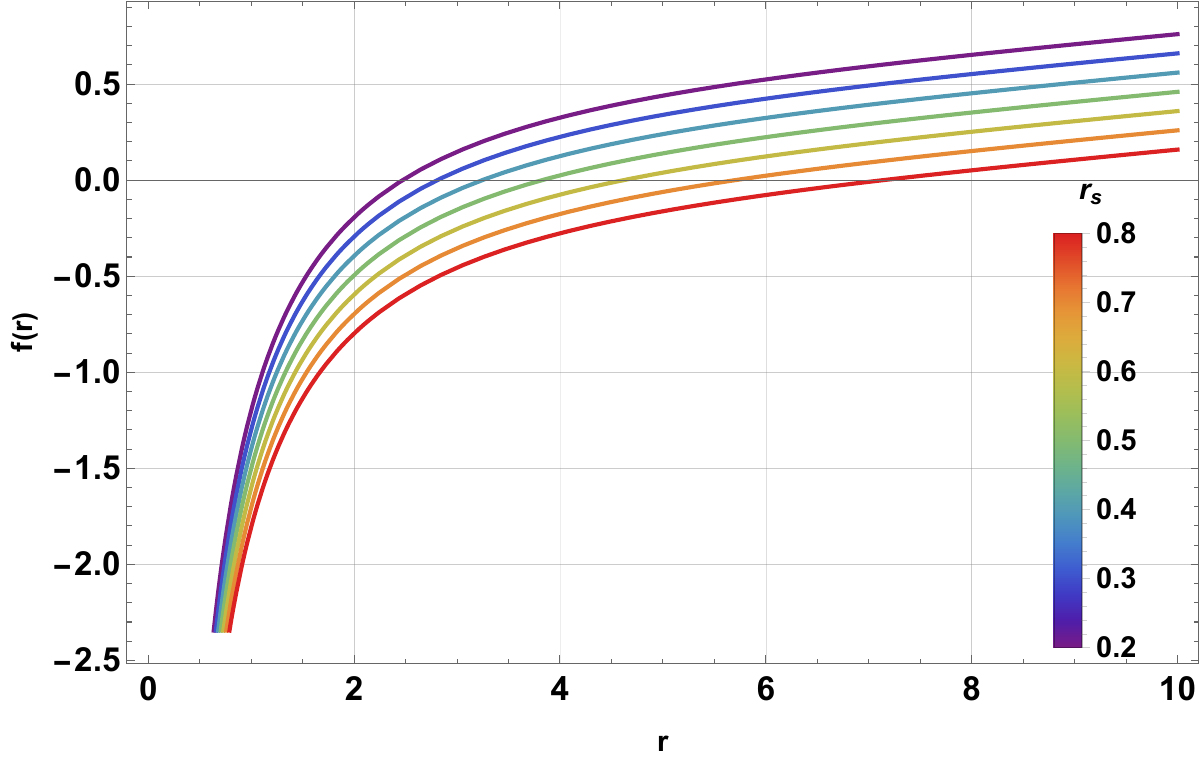}\\
(a) $r_s=0.2,\,\rho_s=0.02$ \hspace{5cm} (b) $\alpha=0.1,\,\rho_s=0.01$
\caption{\footnotesize The metric function as a function of $r$ for different values of CoS parameter $\alpha$. Here, $M=1,\,\ell_p=25$.}
\label{fig:function}
\end{figure*}

In the limit $r_s \to 0$ or $\rho_s \to 0$, one recovers the Letelier-AdS BH, which further reduces to the Schwarzschild-AdS solution for $\alpha=0$. Moreover, in the limit $\alpha=0$ and $\ell_p \to \infty$, we recover the BH-DM halo solution reported in Ref. \cite{UU}.

Figure \ref{fig:function} illustrates the behavior of the metric function $f(r)$ by varying the CoS parameter $\alpha$ and the halo radius $r_s$.

\section{Geodesic Motion Around a BH}\label{Sec:III}

Geodesic motion provides the cleanest description of how photons and test particles move in the strong-gravity regime of BHs, showing the geometry of spacetime through observable signatures. In static, spherically symmetric backgrounds, spacetime symmetries yield conserved energy and angular momentum, which organize the dynamics into bound, plunging, and scattering trajectories for massive particles, and into capture versus escape for photons \cite{wald}. For timelike geodesics, stable circular motion can exist down to an innermost stable circular orbit; below this threshold, small perturbations trigger an inevitable plunge. Periapsis precession and strong-field light deflection arise naturally in this picture and underpin phenomena such as relativistic precession of stellar orbits, hot-spot motion in accretion flows, and gravitational lensing. For null geodesics, an unstable photon region (often called the photon sphere in spherical cases) acts as a separatrix between rays that fall in and those that return to infinity; its critical impact parameters delineate the boundary of the BH shadow and govern strong-lensing observables \cite{chandra}. Altogether, geodesic analysis links fundamental geometry to measurements, providing a common language for interpreting lensing, shadows, spectral lines, and variability near compact objects.

Geodesics describe the trajectories of freely falling particles and photons in curved spacetime, determined by the geodesic equation
\begin{equation}
\frac{d^2 x^\mu}{d \tau^2} + \Gamma^\mu_{\;\nu\rho}\,\frac{dx^\nu}{d\tau}\,\frac{dx^\rho}{d\tau} = 0,\label{bb1}
\end{equation}
where $\Gamma^\mu_{\;\nu\rho}$ are the Christoffel symbols and $\tau$ is the affine parameter (proper time for massive particles and an affine parameter for photons).

We use the Lagrangian approach to analyze the geodesic motion of massless and massive test particles. The Lagrangian density is
\begin{equation}
    \mathcal{L}=\frac{1}{2}\,g_{\mu\nu}\,\dot{x}^{\mu}\,\dot{x}^{\nu},\label{bb2}
\end{equation}
where the dot denotes the derivative with respect to the affine parameter and $g_{\mu\nu}$ is the metric tensor.

For the metric \eqref{metric}, the Lagrangian becomes
\begin{equation}
    \mathcal{L}=\frac{1}{2}\,\left[-f(r)\,\dot{t}^2+\frac{\dot{r}^2}{f(r)}+r^{2}\left(\dot{\theta}^{2}+\sin ^{2}\theta\, \dot{\phi} ^{2}\right)\right].\label{bb3}
\end{equation}

In a spherically symmetric and static spacetime, the timelike $\xi_{(t)} \equiv \partial_t$ and rotational Killing vectors $\xi_{(\phi)} \equiv \partial_{\phi}$ lead to two conserved quantities. Defining the four-velocity (or tangent) $u^\mu=dx^\mu/d\tau$ and imposing $g_{\mu\nu}u^\mu u^\nu=\varepsilon$ with $\varepsilon=-1$ (timelike) and $\varepsilon=0$ (null), the conserved energy $\mathrm{E}$ and angular momentum $\mathrm{L}$ are
\begin{align}
\mathrm{E} &= -g_{tt}\,\dot{t}, \label{bb4}\\
\mathrm{L} &= g_{\phi\phi}\,\dot{\phi},\label{bb5}
\end{align}
where $g_{tt}$ and $g_{\phi\phi}$ are metric components, and $\tau$ is the proper time for massive particles or an affine parameter for photons \cite{wald}. Physically, $\mathrm{E}$ is the total conserved energy as measured at infinity, while $\mathrm{L}$ is the conserved angular momentum about the symmetry axis.

Without loss of generality, by spherical symmetry, we restrict motion to the equatorial plane $\theta=\pi/2$. The equations for $t$, $r$ and $\phi$ read
\begin{align}
\dot{t}&=\frac{\mathrm{E}}{f(r)},\label{bb6}\\
\dot{\phi}&=\frac{\mathrm{L}}{r^2},\label{bb7}\\
\dot{r}^2&=\mathrm{E}^2-V_\text{eff}(r),\label{bb8}
\end{align}
where $V_\text{eff}(r)$ is the effective potential of the system and is given by
\begin{align}
&V_\text{eff}(r)=\left(-\varepsilon+\frac{\mathrm{L}^2}{r^2}\right)\,f(r)
=\left(-\varepsilon+\frac{\mathrm{L}^2}{r^2}\right)\notag\\&\times \left(1-\alpha-\frac{2M}{r}-\rho_s\,r_s^2\,\ln\!\left(1+\frac{r_s}{r}\right)+\frac{r^2}{\ell^2_p}\right).\label{bb9}
\end{align}
One can see that geometric and physical parameters such as the BH mass $M$, the curvature radius $\ell_p$, the string-cloud parameter $\alpha$, and the DM-halo profile $(r_s,\rho_s)$ significantly influence the curvature of spacetime. Consequently, these parameters shape the effective potential, which in turn governs the dynamics of photons and massive test particles.

\subsection{\large {\bf Photon Dynamics}}\label{Sec:III-1}

Photon dynamics around BHs provides crucial insights into the structure of spacetime and forms the theoretical foundation for many astrophysical observations. Since photons follow null geodesics, their trajectories are strongly influenced by the gravitational field of the BH. In spherically symmetric spacetimes, photons experience an effective potential that determines whether the event horizon captures them, escapes to infinity, or remains in unstable circular orbits at the so-called photon sphere \cite{chandra,wald}.

The photon sphere represents an unstable set of trajectories: any small perturbation causes photons to either spiral into the BH or escape outward. This unstable nature manifests observationally through phenomena such as strong gravitational lensing, where light rays can loop multiple times around the BH before reaching a distant observer \cite{perlick}. The critical impact parameter associated with the photon sphere defines the angular size of the BH shadow, an observable quantity now directly imaged by the Event Horizon Telescope for supermassive BHs such as M87* and Sgr A* \cite{perlick}. 

\begin{figure*}[ht!]
\centering
\includegraphics[width=0.45\linewidth]{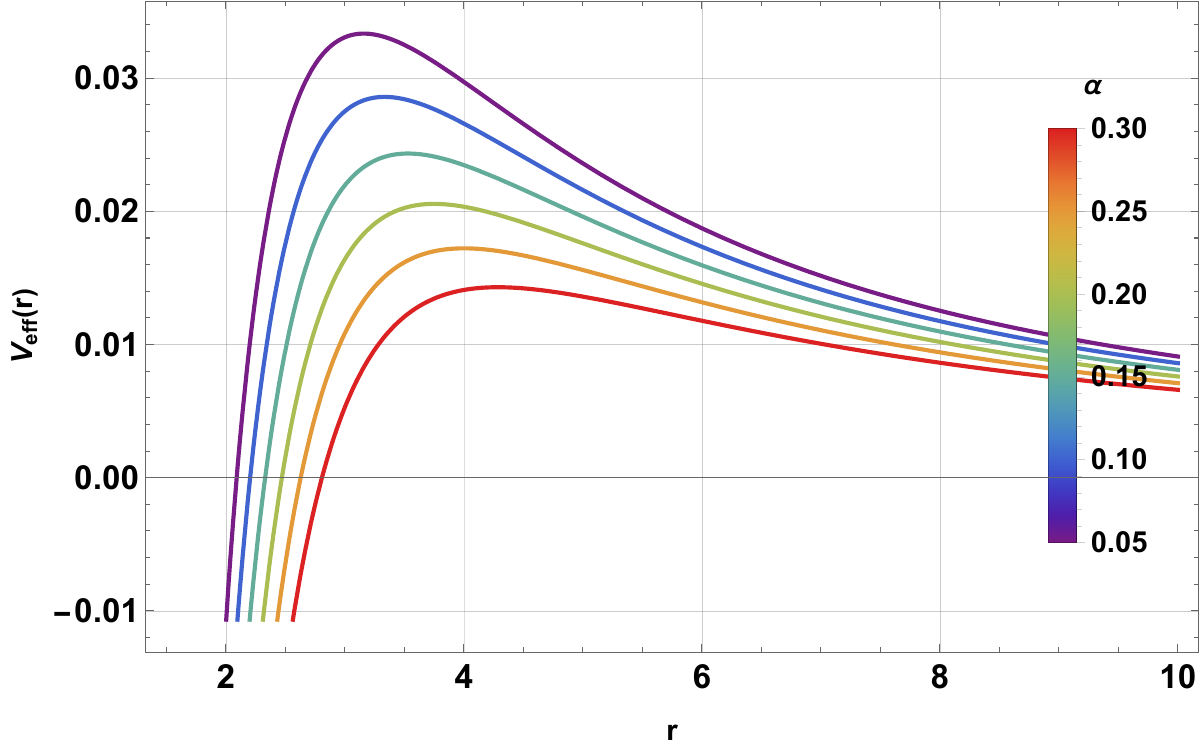}\qquad
\includegraphics[width=0.45\linewidth]{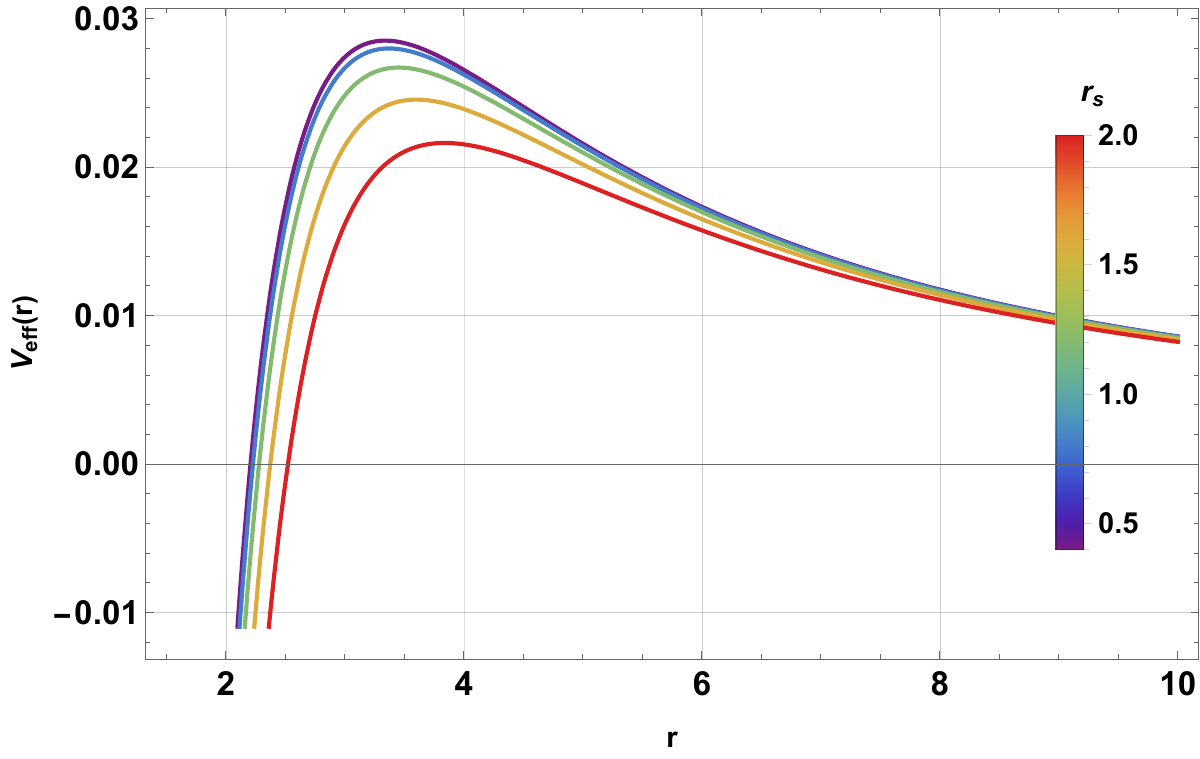}\\
(a) $r_s=0.2,\,\rho_s=0.02$ \hspace{5cm} (b) $\alpha=0.1,\,\rho_s=0.05$
\caption{\footnotesize The behavior of the effective potential for null geodesics as a function of $r$ for different values of CoS parameter $\alpha$ and $r_s$. Here, $M=1,\,\ell_p=25$.}
\label{fig:potential-null}
\end{figure*}

For null geodesic motion, $\varepsilon=0$, the effective potential form Eq. (\ref{bb9}) reduces as
\begin{equation}
V_\text{eff}(r)=\frac{\mathrm{L}^2}{r^2}\,\left(1-\alpha-\frac{2M}{r}-\rho_s\,r_s^2\,\mbox{ln} {\left(1+\frac{r_s}{r}\right)}+\frac{r^2}{\ell^2_p}\right).\label{bb10}
\end{equation}

Equation (\ref{bb10}) governs the dynamics of massless particles, such as photons. With the help of this potential, we will discuss photon trajectories, the effective radial force experienced by photon particles, the photon sphere, and the shadow cast by the BH, and analyze the outcomes. 

In Fig. \ref{fig:potential-null}, we illustrate the behavior of the effective potential that governs the dynamics of photon particles as a function of $r$ by varying the CoS parameter $\alpha$ and the halo radius $r_s$. We observe in both panels that the potential reduces with increasing $\alpha$ and $r_s$, indicating less binding by the gravitational field. 

\begin{center}
\large{\bf Photon Trajectories}
\end{center}

Photon trajectories describe the paths followed by massless particles, such as photons, through spacetime, particularly under the influence of gravity. Photons move along null geodesics, which are curves for which the spacetime interval is zero. These paths are profoundly influenced by the curvature of spacetime caused by massive objects or compact objects.

The equation of the orbit using Eqs. (\ref{bb7}) and (\ref{bb8}) and finally employing (\ref{bb10}) is given by
\begin{equation}
\left(\frac{1}{r^2}\,\frac{dr}{d\phi}\right)^2+\frac{1-\alpha}{r^2}=\frac{1}{\beta^2}-\frac{1}{\ell^2_p}+\frac{2M}{r^3}+\frac{\rho_s\,r^2_s}{r^2}\,\mbox{ln}\left(1+\frac{r_s}{r}\right).\label{bb11}
\end{equation}
Performing a transformation to a new variable via $r=1/u$ into the above equation results
\begin{align}
\left(\frac{du}{d\phi}\right)^2+(1-\alpha)\,u^2&=\frac{1}{\beta^2}-\frac{1}{\ell^2_p}+2\,M\, u^3\notag\\&+\rho_s\,r^2_s\,u^2\,\mbox{ln}(1+r_s\,u).\label{bb12}
\end{align}
Differentiating both sides w. r. to $\phi$ and after simplification results
\begin{align}
\frac{d^2u}{d\phi^2}+(1-\alpha)\,u&=3\,M\, u^2+\rho_s\,r^2_s\,u\,\mbox{ln}(1+r_s\,u)\notag\\&+\frac{1}{2}\,\frac{\rho_s\,r^3_s\,u^2}{1+r_s\,u}.\label{bb13}
\end{align}
Equation (\ref{bb13}) represents a nonlinear second-order differential equation governing photon trajectories in the given gravitational field. It is evident that the dark matter profile, characterized by $(r_s,\rho_s)$, together with the CoS parameter $\alpha$, significantly influences the photon trajectories and consequently modifies the geodesic paths of light propagating around the BH.

\begin{figure*}[tbhp]
  \centering
  \includegraphics[width=\textwidth]{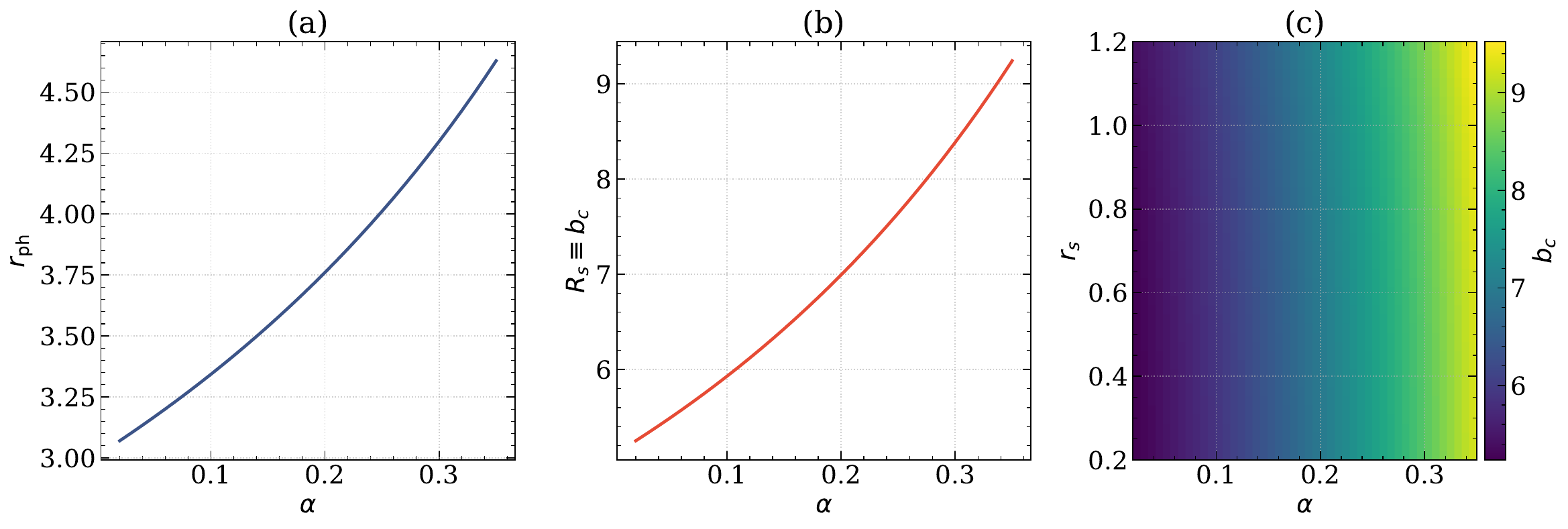}
  \caption{Photon-sphere and shadow diagnostics. Panels: (a) photon-sphere radius $r_{\rm ph}(\alpha)$; (b) shadow radius $R_s(\alpha)\equiv b_c(\alpha)$; (c) heatmap of $b_c(\alpha,r_s)$.
  Unless varied explicitly, the parameters are $M=1$, $\rho_s=0.05$, $r_s=0.5$, and $\ell_p=25$.
  In (a)-(b) we vary $\alpha\in[0.02,0.35]$; in (c) we scan $\alpha\in[0.02,0.35]$ and $r_s\in[0.2,1.2]$, keeping $M=1$, $\rho_s=0.05$, $\ell_p=25$.}
  \label{fig:criticals}
\end{figure*}

\begin{figure}[tbhp]
\centering
\includegraphics[scale=0.45]{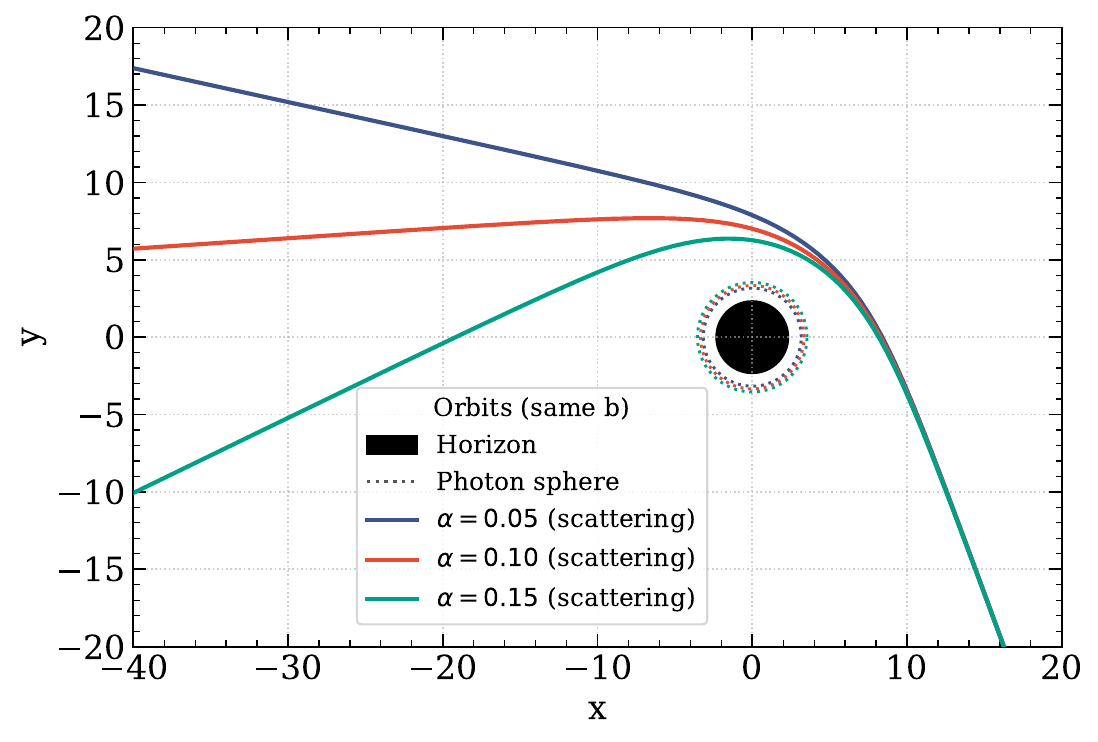}
\caption{Null geodesics with a common impact parameter. Single-panel plot showing photon trajectories for $\alpha=\{0.05,\,0.10,\,0.15\}$, all integrated with the same impact parameter
$b=b_{\rm fixed}=1.35\,b_c(\alpha{=}0.10)\approx 8.005$.
The horizon is shown as a filled disk, and the photon sphere as a dotted circle (both per model). Fixed parameters: $M=1$, $\rho_s=0.05$, $r_s=0.5$, $\ell_p=25$.}
\label{fig:orbits-b-fixed}
\end{figure}
\begin{figure*}[tbhp]
\centering
\includegraphics[scale=0.45]{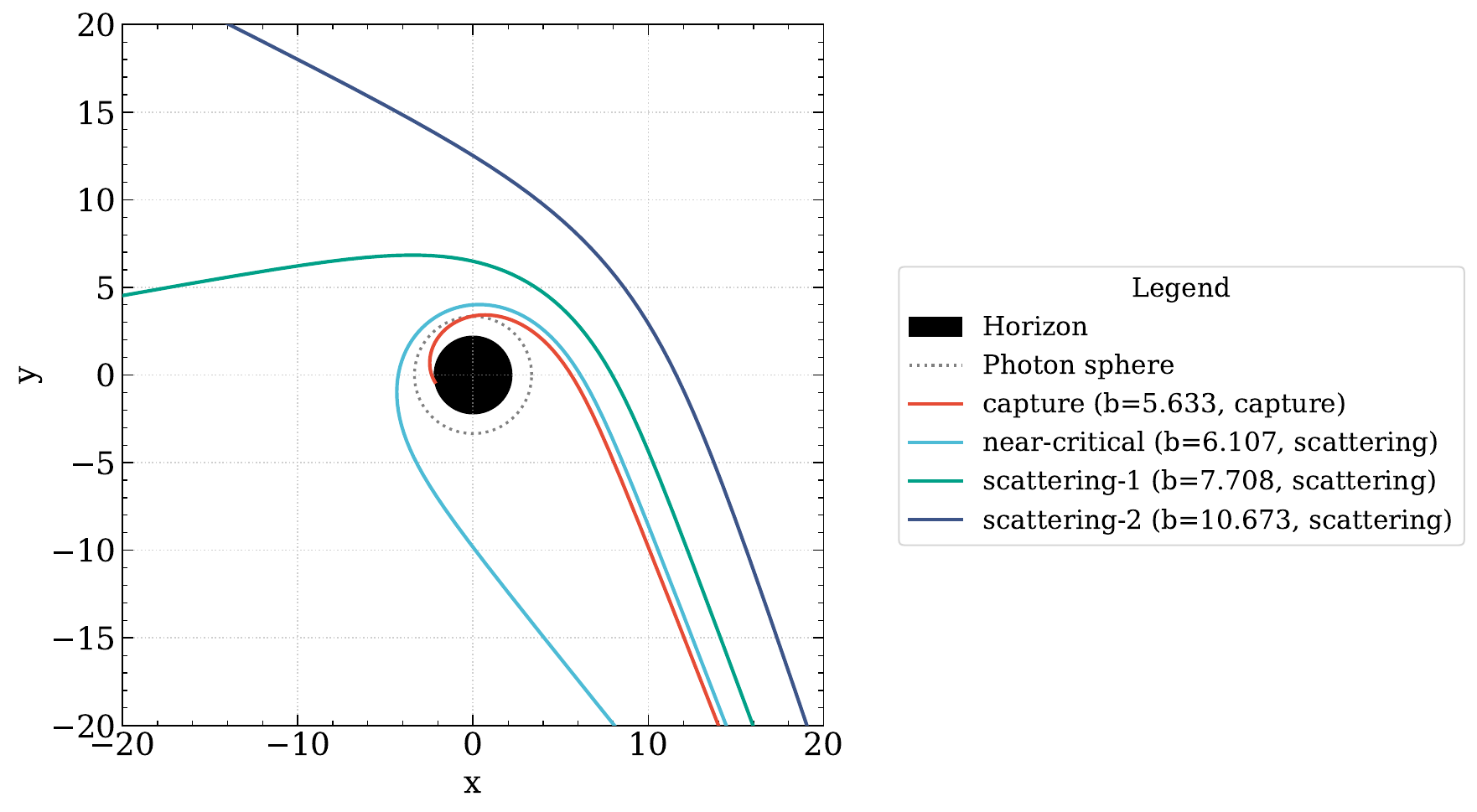}
\caption{Null geodesics for fixed $\alpha$ and varying $b/b_c$. Photon trajectories for $\alpha=0.10$ with $b=\{0.95,\,1.03,\,1.30,\,1.80\}\times b_c(\alpha)$ (capture, near-critical, and two scattering cases). For $\alpha=0.10$ the model gives $b_c\simeq 5.929$, i.e., $b\simeq\{5.633,\,6.107,\,7.708,\,10.673\}$. The horizon (filled disk) and photon sphere (dotted circle) are shown. Fixed parameters: $M=1$, $\rho_s=0.05$, $r_s=0.5$, $\ell_p=25$;
parameter values identical to Fig.~\ref{fig:orbits-b-fixed}.}
\label{fig:orbits-vary-b}
\end{figure*}

In Fig.~\ref{fig:criticals} we show three diagnostic plots that organize the key optical scales of the spacetime for the parameter set $M=1$, $\rho_s=0.05$, $r_s=0.5$, and $\ell_p=25$. Panel (a) displays the photon-sphere radius $r_{\rm ph}(\alpha)$ obtained from the circular-null condition $2\mathcal{F}(r)-r\,\mathcal{F}'(r)=0$. Panel (b) shows the corresponding shadow (critical) radius $R_s(\alpha)\equiv b_c(\alpha)=r_{\rm ph}/\sqrt{f(r_{\rm ph})}$. Panel (c) presents a heatmap of $b_c(\alpha,r_s)$ across a broad range of $\alpha$ and $r_s$, highlighting how the Dehnen-type halo scale competes with the string-cloud parameter in setting the size of the capture cross section and the BH shadow. Together, these panels map where the photon ring exists and how the optical size of the BH responds to controlled deformations of the geometry; the heatmap, in particular, is useful to read off regimes where small changes in either $\alpha$ or $r_s$ produce visible shifts in $b_c$.

In Fig.~\ref{fig:orbits-b-fixed} we show null geodesics for three values of the string-cloud parameter, $\alpha=0.05,\,0.10,\,0.15$, all propagated with the \emph{same} impact parameter $b$ chosen from the baseline $\alpha=0.10$ case as $b\simeq 1.35\,b_c(\alpha{=}0.10)$. The horizon is drawn as a filled disk and the photon sphere as a dotted circle for each $\alpha$, so the eye can compare the geometric scales to the trajectory. Because $b$ is fixed while the geometry changes, the bending angle varies slightly across $\alpha$; in practice, for the baseline values $M=1$, $\rho_s=0.05$, $r_s=0.5$, and $\ell_p=25$, the three curves remain in the scattering regime but display modest shifts in closest approach and final scattering direction. This panel makes explicit that even when orbits are qualitatively the same (all scatter), quantitative changes in the background, either through $\alpha$ or the halo, induce logarithmic terms, which translate into measurable differences in trajectory shape.

\begin{figure}[tbhp]
  \centering
  \includegraphics[scale=0.45]{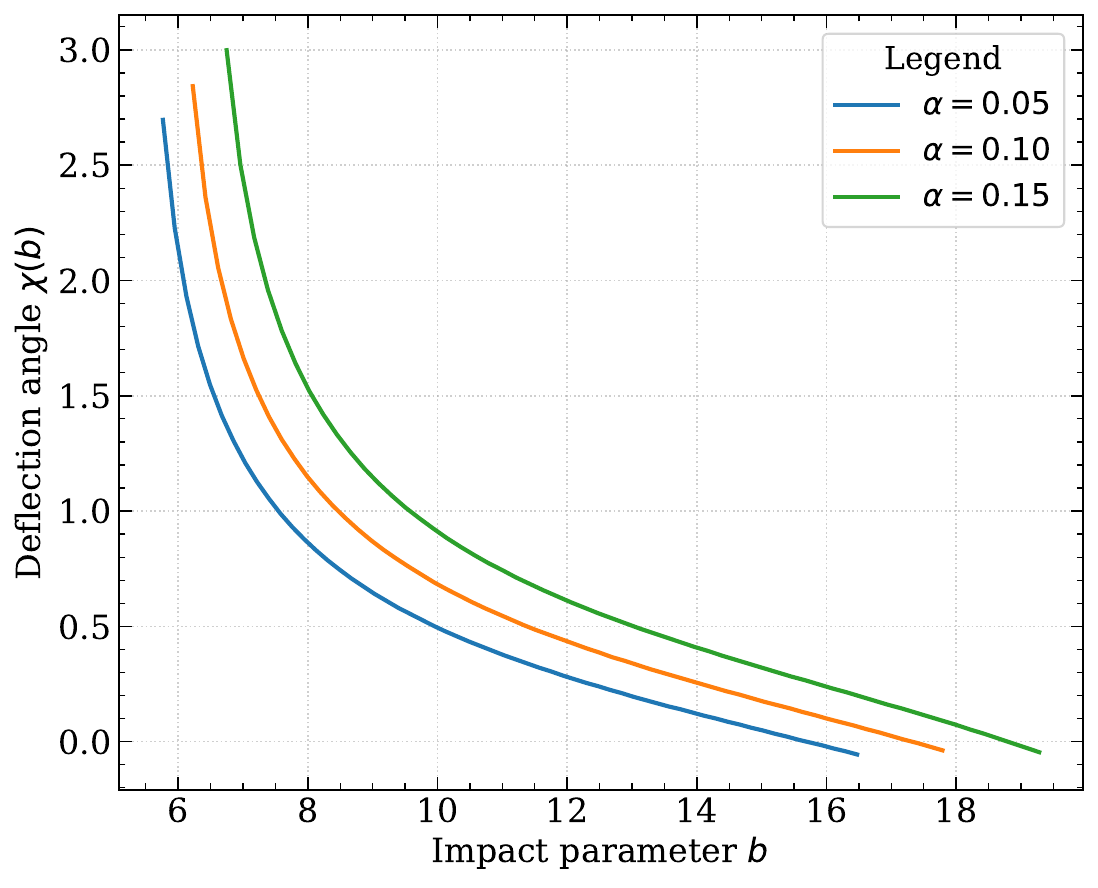}
  \caption{Deflection angle $\chi(b)$ vs impact parameter. Scattering deflection $\chi(b)=(\phi_{\rm out}-\phi_{\rm in})-\pi$ for $\alpha=\{0.05,\,0.10,\,0.15\}$.
  For each $\alpha$ we sample $b\in[1.05\,b_c(\alpha),\,3.0\,b_c(\alpha)]$.
  Fixed parameters: $M=1$, $\rho_s=0.05$, $r_s=0.5$, $\ell_p=25$;
  geodesics integrated over $\phi\in[-20,20]$ with $R_{\infty}=120$.
  Curves terminate at $b\to 1.05\,b_c(\alpha)$ to avoid capture.}
  \label{fig:deflection}
\end{figure}
\begin{figure}[tbhp]
  \centering
  \includegraphics[scale=0.45]{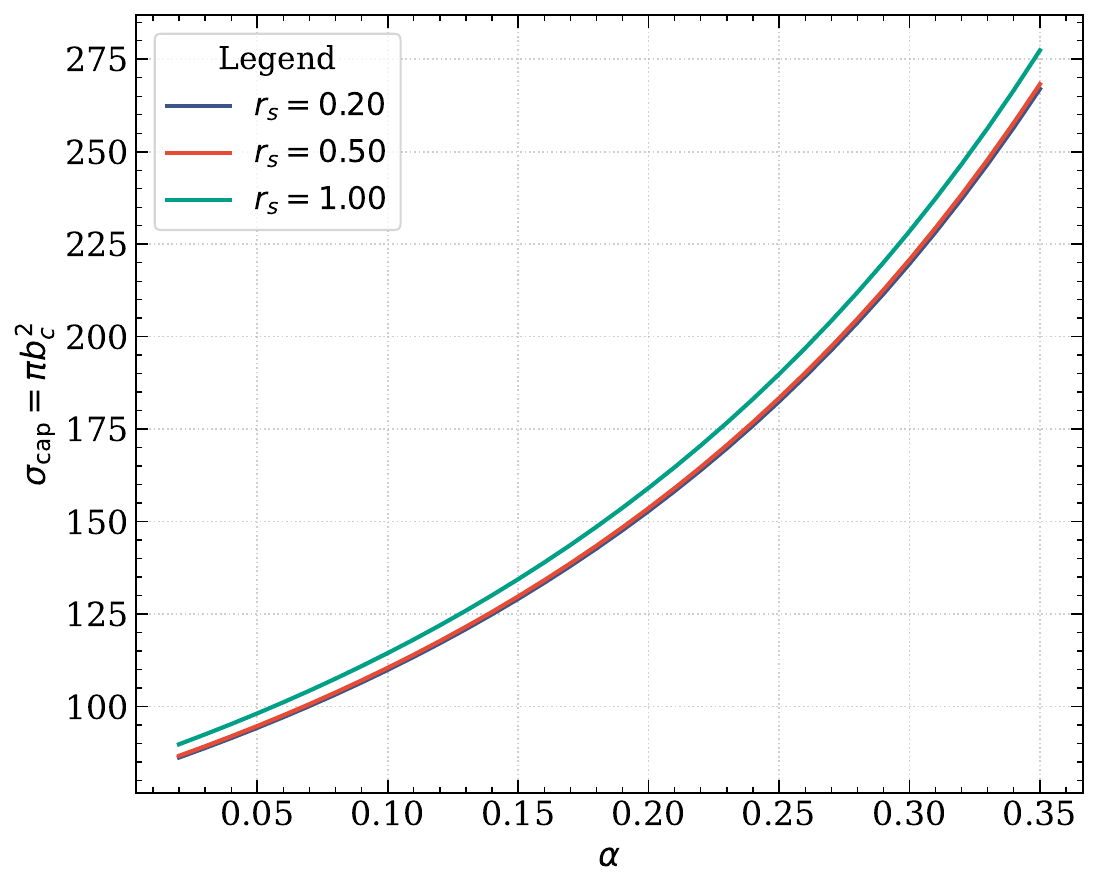}
  \caption{Capture cross-section vs string-cloud parameter.
  $\sigma_{\rm cap}(\alpha)=\pi\, b_c(\alpha)^2$ for three halo scales $r_s\in\{0.2,\,0.5,\,1.0\}$.
  Fixed parameters: $M=1$, $\rho_s=0.05$, $\ell_p=25$.
  We scan $\alpha\in[0.02,0.35]$; for each $(\alpha,r_s)$, $b_c$ is obtained from $r_{\rm ph}(\alpha,r_s)$ via $b_c=r_{\rm ph}/\sqrt{f(r_{\rm ph})}$.}
  \label{fig:capture-xsec}
\end{figure}
In Fig.~\ref{fig:orbits-vary-b} we fix $\alpha=0.10$ and vary the impact parameter around the critical value, plotting representative cases $b/b_c=\{0.95,\,1.03,\,1.30,\,1.80\}$. The $b<b_c$ ray is captured and crosses the horizon; the near-critical ray ($b\approx b_c$) executes multiple whirls near the photon sphere before escaping; and the two $b>b_c$ rays scatter with decreasing deflection as $b$ grows. The horizon is shown as a solid-filled disk, and the photon sphere as a dotted circle, which helps visualize how the unstable circular orbit controls the transition between capture and scattering. This figure is the clearest demonstration of how the single scale $b_c$ stratifies the null geodesics into capture, quasi-bound (whirl), and ordinary scattering, for the same background $(M=1,\rho_s=0.05,r_s=0.5,\ell_p=25)$.

In Fig.~\ref{fig:deflection} we plot the deflection angle $\chi(b)$ versus the impact parameter for $\alpha=0.05,\,0.10,\,0.15$ at fixed $M=1$, $\rho_s=0.05$, $r_s=0.5$, and $\ell_p=25$. Each curve starts just above the corresponding $b_c(\alpha)$ and extends to larger $b$. As expected, $\chi(b)$ grows rapidly as $b\to b_c^+$, reflecting the strong-lensing regime dominated by the unstable photon orbit; for larger $b$, $\chi$ decreases monotonically toward the weak-lensing regime. Comparing the curves across $\alpha$ reveals how the string-cloud term shifts the effective optical size and, hence, the location and sharpness of the strong-lensing rise: changes in $\alpha$ translate into small but systematic horizontal displacements of the $\chi(b)$ profile through $b_c(\alpha)$.

In Fig.~\ref{fig:capture-xsec} we present the capture cross section $\sigma_{\rm cap}(\alpha)=\pi\,b_c(\alpha)^2$ as a function of $\alpha$ for three halo radii $r_s=\{0.2,\,0.5,\,1.0\}$ with $M=1$, $\rho_s=0.05$, and $\ell_p=25$. For each $r_s$, the curve is obtained directly from the critical impact parameter computed in the full background. Comparing the three series makes clear how the DM scale controls the geometric reach of the BH: increasing $r_s$ modifies the logarithmic contribution in the lapse and reshapes the photon-sphere condition, which in turn adjusts $b_c$ and the total capture area. This plot summarizes, in a single observable, the combined imprint of the string cloud and halo scale on high-energy geodesics; it also provides a convenient handle to propagate these effects into shadow diameter and high-frequency absorption estimates.

\begin{center}
\large{\bf Effective Radial Force}
\end{center}

Photons move along null geodesics, but their trajectories can be conveniently understood in terms of an effective radial potential and the corresponding radial force. The effective radial force is not a Newtonian force in the usual sense, but rather a geometrical manifestation of spacetime curvature acting on the photon trajectory. In a static, spherically symmetric spacetime, the conserved energy and angular momentum allow the photon’s radial motion to be described by an effective potential. The gradient of this potential behaves like a force that governs whether photons fall into the BH, escape to infinity, or remain in circular orbits. The condition where this effective force vanishes corresponds to circular photon orbits, forming the so-called photon sphere \cite{chandra,wald}. 

\begin{figure*}[ht!]
\centering
\includegraphics[width=0.45\linewidth]{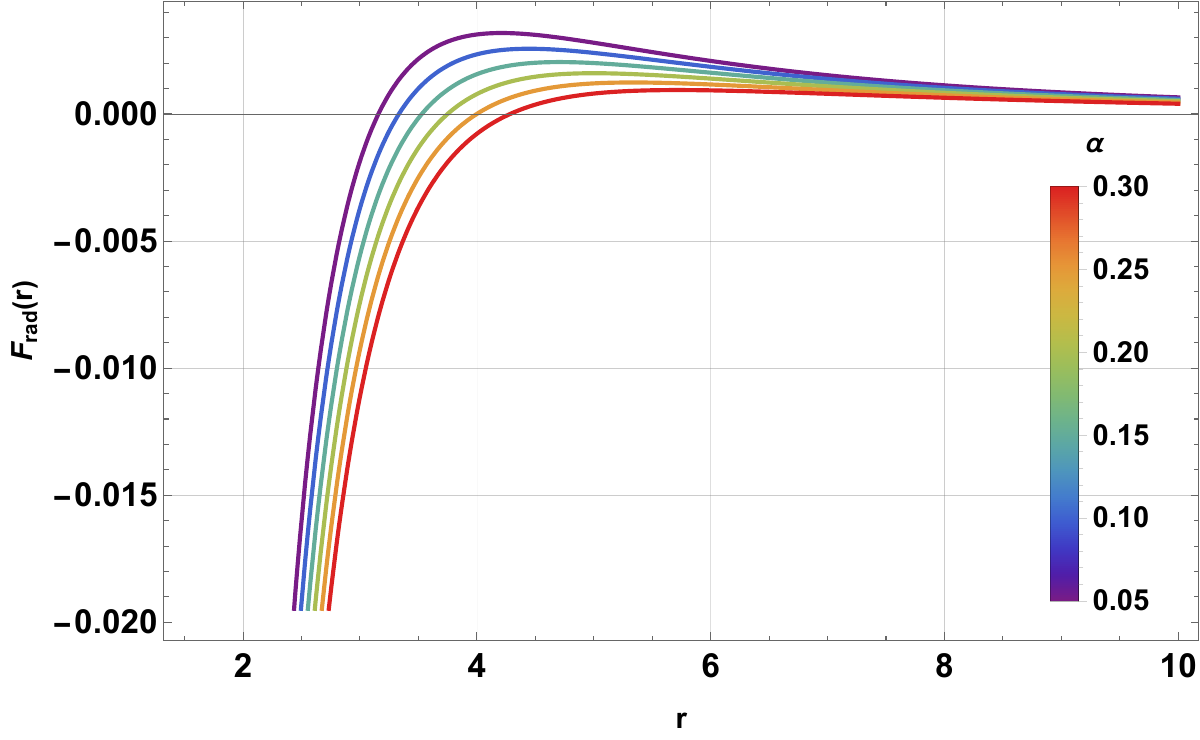}\qquad
\includegraphics[width=0.45\linewidth]{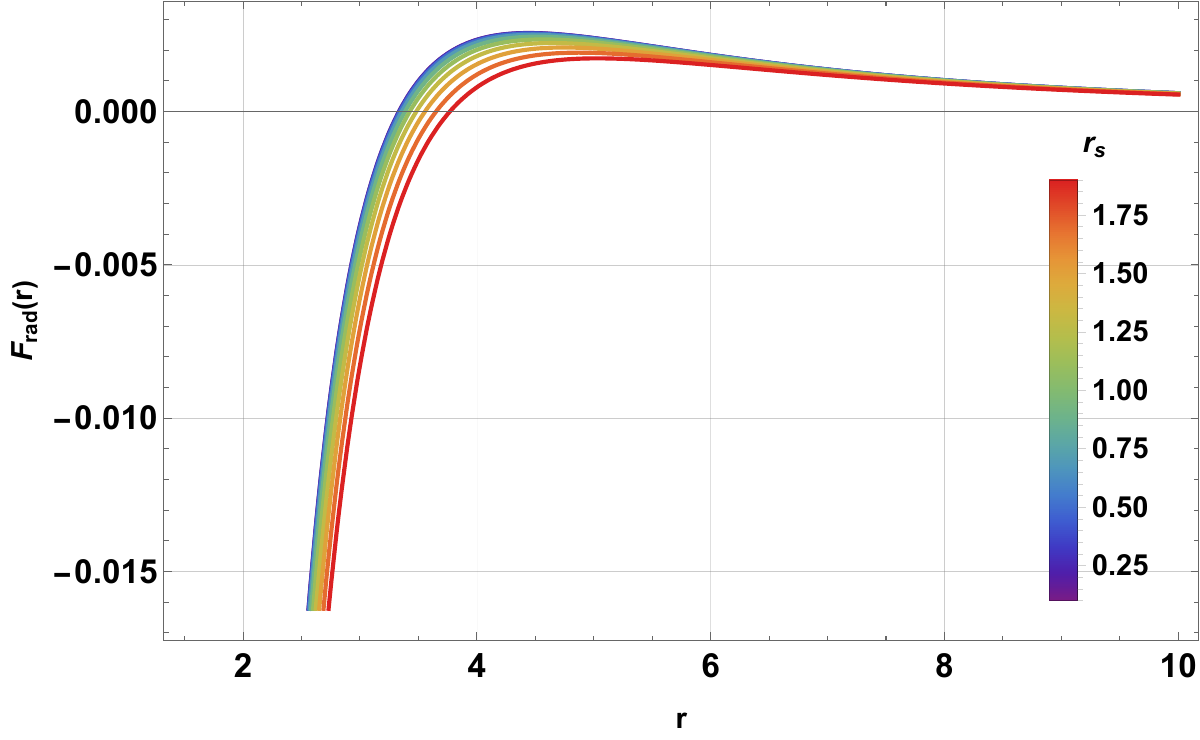}\\
(a) $r_s=0.2,\,\rho_s=0.02$ \hspace{5cm} (b) $\alpha=0.1,\,\rho_s=0.05$
\caption{\footnotesize The behavior of the effective radial force experienced by photons as a function of $r$ for different values of the CoS parameter $\alpha$ and $r_s$. Here, $M=1,\,\ell_p=25$.}
\label{fig:force}
\end{figure*}

One may define an effective radial force associated with the effective potential as
\begin{equation}
    F_\text{rad}=-\frac{1}{2}\,\frac{dV_{\rm eff}}{dr}.\label{bb14}
\end{equation}
Substituting the effective potential for null geodesics given in Eq. (\ref{bb10}), we find
\begin{align}
    &F_\text{rad}=\frac{\mathrm{L}^2}{r^3}\notag\\ & \times \left[1-\alpha-\frac{3M}{r}-\rho_s\,r_s^2\,\mbox{ln} {\left(1+\frac{r_s}{r}\right)}-\frac{\rho_s\,r^3_s}{2\,(r+r_s)}\right].\label{bb15}
\end{align}
One can see that geometric and physical parameters such as the BH mass $M$, the curvature radius $\ell_p$, the string cloud parameter $\alpha$, and the DM halo profile characterized by $(r_s,\rho_s)$ significantly influence the effective radial force experienced by the photon particles.

Thus, the effective radial force formalism provides an intuitive bridge between the abstract geodesic equations and physically observable phenomena such as gravitational lensing, BH shadows, and the propagation of light near compact objects. It highlights the role of spacetime geometry in shaping photon dynamics and connects directly to astrophysical signatures that test general relativity in the strong-field regime \cite{perlick,frolov}.

In Fig. \ref{fig:force}, we generate graphs showing the behavior of the effective radial force experienced by photon particles as a function of $r$ by varying the CoS parameter $\alpha$ and the halo radius $r_s$. We observe in both panels that this radial force decreases with increasing $\alpha$ and $r_s$, indicating the photon particles are less bound by the gravitational field. 

\begin{center}
\large{\bf Photon Sphere and BH shadows}
\end{center}

The photon sphere acts as a boundary between capture and escape: photons with smaller impact parameters are pulled into the BH, while those with larger impact parameters escape to infinity. At the photon sphere itself, the balance between inward and outward tendencies is extremely delicate, making these orbits unstable; any small perturbation sends the photon either spiraling inward or outward. This unstable equilibrium explains why the photon sphere determines the apparent size of the BH shadow, now observed directly by the Event Horizon Telescope \cite{perlick}.
 
For circular null orbits of radius $r$=const., the conditions $\dot{r}=0$ and $\ddot{r}=0$ must be satisfied. The first condition simplifies to $\mathrm{E}^2=V_\text{eff}(r)$, which gives us the critical impact parameter for photon particles. This is using Eqs. (\ref{bb10}) given by
\begin{equation}
b_c=\frac{\mathrm{L}_\text{ph}}{\mathrm{E}_\text{ph}}=\frac{r}{\sqrt{1-\alpha-\frac{2M}{r}-\rho_s\,r_s^2\,\mbox{ln} {\left(1+\frac{r_s}{r}\right)}+\frac{r^2}{\ell^2_p}}}\Bigg{|}_{r=r_c}.\label{bb16}
\end{equation} 

Noted that if $b\,(=\mathrm{L}/\mathrm{E}) < b_{c}$, the photon is captured by the BH and inevitably crosses the event horizon. If $b > b_{c}$, the photon is scattered back to infinity, experiencing gravitational deflection. If $b = b_{c}$, the photon asymptotically approaches the photon sphere, orbiting in an unstable circular trajectory. Thus, the critical impact parameter $b_{c}$ acts as the dividing line between capture and escape, making it a fundamental quantity in defining the apparent BH shadow as seen by distant observers. In practice, $b_{c}$ corresponds to the shadow radius, while $\beta$ labels individual photon trajectories relative to this boundary \cite{perlick}.

Now, we focus on an important feature of the BH called the apparent shadow size cast by the BH. The radius of the BH shadow is equal to the critical impact parameter for a photon particle when it traverses in unstable circular orbits. This is defined by
\begin{equation}
R_s=b_c=\frac{r_\text{ph}}{\sqrt{1-\alpha-\frac{2M}{r_\text{ph}}-\rho_s\,r_s^2\,\mbox{ln} {\left(1+\frac{r_s}{r_\text{ph}}\right)}+\frac{r_\text{ph}^2}{\ell^2_p}}}.\label{bb17}
\end{equation}

\begin{figure*}[ht!]
\centering
\includegraphics[width=0.45\linewidth]{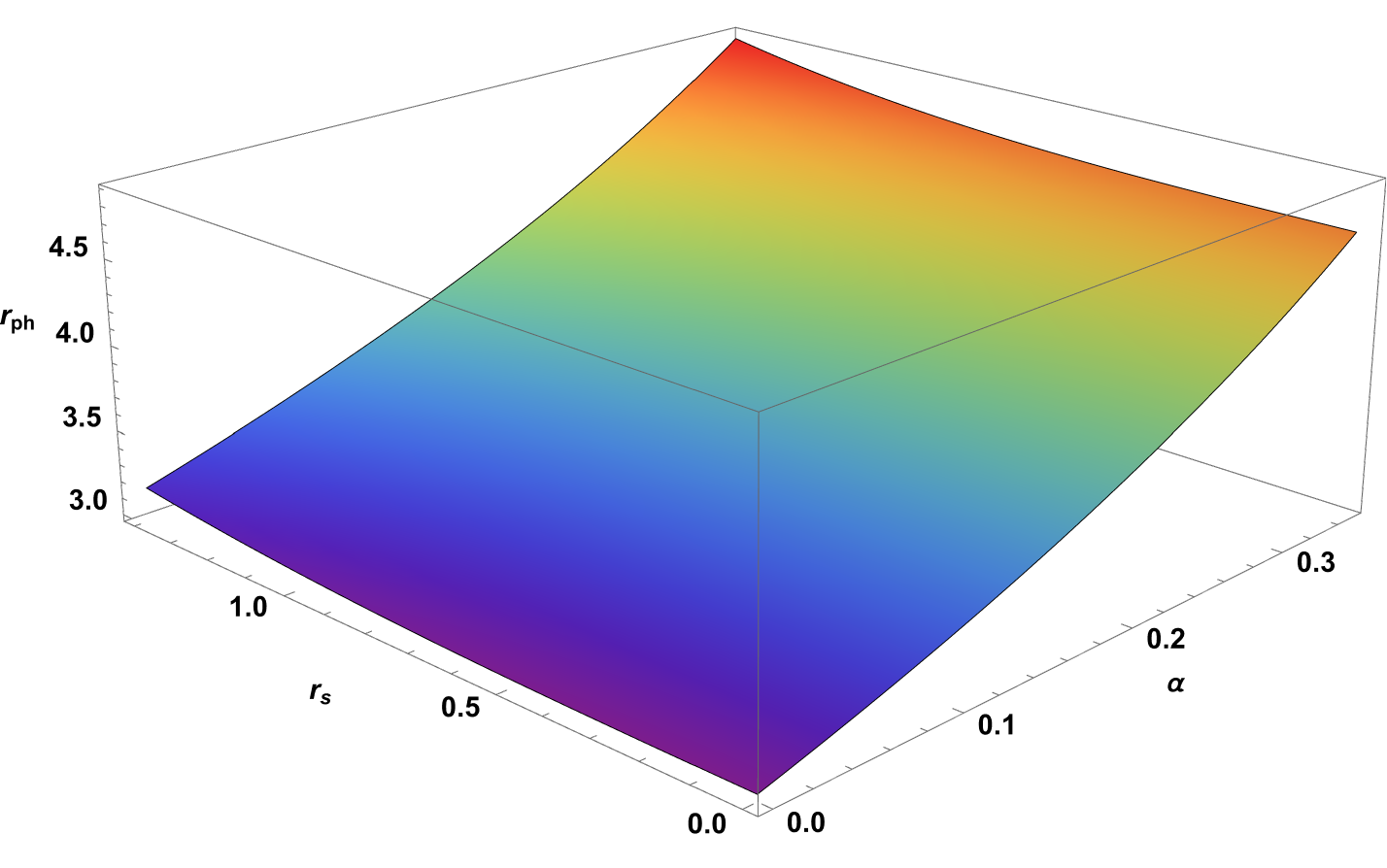}\qquad
\includegraphics[width=0.45\linewidth]{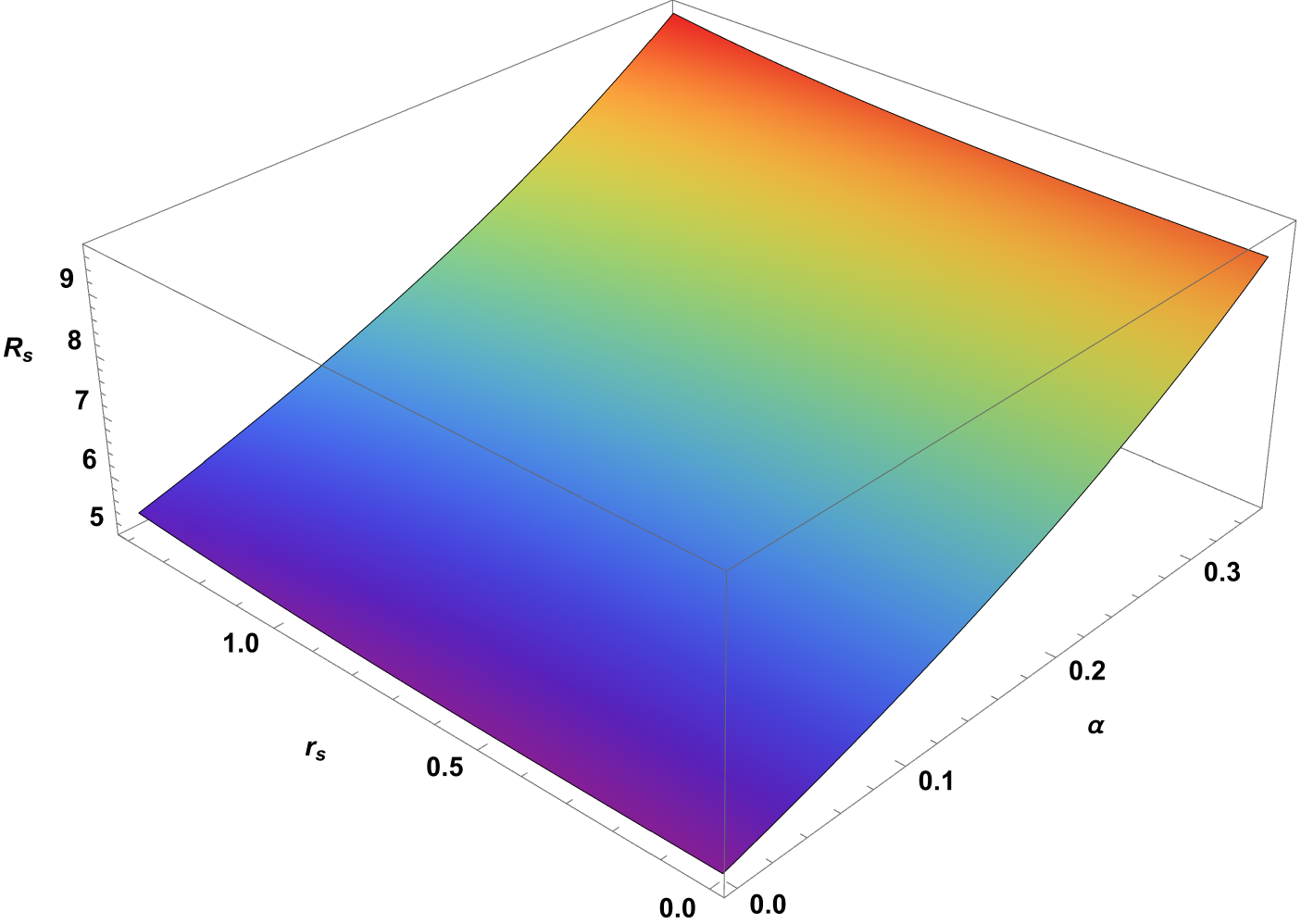}\\
(a) $\rho_s=0.05$ \hspace{5cm} (b) $\rho_s=0.05,\,\ell_p=25$
\caption{\footnotesize 3D plot of the photon sphere radius $r_\text{ph}$ and shadow size $R_s$ as a function of $\alpha$ and $r_s$. Here, $M=1$.}
\label{fig:3d-plot}
\end{figure*}

One can see that geometric and physical parameters such as the BH mass $M$, the curvature radius $\ell_p$, the string cloud parameter $\alpha$, and the DM halo profile characterized by $(r_s,\rho_s)$ modify the critical impact parameter for photon particles. Consequently, the shadow size cast by the BH is also influenced by these factors.

Now, we determine the photon sphere radius $r=r_\text{ph}$ using the second condition stated earlier. This condition $\ddot{r}=0$ implies that 
\begin{equation}
    \frac{dV_\text{eff}(r)}{dr}=0.\label{bb18}
\end{equation}
Substituting potential (\ref{bb10}) into the above relation results 
\begin{equation}
1-\alpha-\frac{3M}{r}-\rho_s\,r_s^2\,\mbox{ln} {\left(1+\frac{r_s}{r}\right)}-\frac{1}{2}\,\frac{\rho_s\,r^3_s}{r+r_s}=0.\label{bb19}
\end{equation}

Equation (\ref{bb19}) represents an infinite polynomial equation in the radial coordinate $r$, for which obtaining an exact analytical solution is highly challenging. Nevertheless, the photon sphere radius $r = r_{\text{ph}}$ can be determined numerically by assigning suitable values to the parameters appearing in the polynomial equation.

\begin{table*}[ht!]
\centering
\begin{tabular}{|c|c|c|c|c|c|c|c|}
\hline
$\alpha (\downarrow) \backslash r_s (\rightarrow)$ & $0.2$ & $0.4$ & $0.6$ & $0.8$ & $1.0$ & $1.2$ & $1.4$ \\
\hline
0.05 & 3.1585 & 3.16256 & 3.17309 & 3.19272 & 3.22382 & 3.26862 & 3.32928 \\
\hline
0.10 & 3.33397 & 3.33828 & 3.34946 & 3.37035 & 3.40351 & 3.45134 & 3.51617 \\
\hline
0.15 & 3.53009 & 3.53467 & 3.54658 & 3.56889 & 3.60435 & 3.65558 & 3.72510 \\
\hline
0.20 & 3.75072 & 3.75561 & 3.76835 & 3.79225 & 3.83031 & 3.88537 & 3.96020 \\
\hline
0.25 & 4.00077 & 4.00601 & 4.01969 & 4.04540 & 4.08642 & 4.14585 & 4.22672 \\
\hline
0.30 & 4.28655 & 4.29218 & 4.30693 & 4.33471 & 4.37912 & 4.44357 & 4.53138 \\
\hline
0.35 & 4.61628 & 4.62237 & 4.63837 & 4.66855 & 4.71688 & 4.78714 & 4.88300 \\
\hline
\end{tabular}
\caption{Numerical results of the photon sphere radius $r_\text{ph}$ for different values of $\alpha$ and $r_s$, with $M=1$, $\rho_s=0.05$.}
\label{tab:1}
\end{table*}

\begin{table*}[ht!]
\centering
\begin{tabular}{|c|c|c|c|c|c|c|c|}
\hline
$\alpha (\downarrow) \backslash r_s (\rightarrow)$ & $0.1$ & $0.3$ & $0.5$ & $0.7$ & $0.9$ & $1.1$ & $1.3$ \\
\hline
0.05 & 5.47561 & 5.47884 & 5.49060 & 5.51586 & 5.55912 & 5.62454 & 5.71606 \\
\hline
0.10 & 5.91326 & 5.91673 & 5.92938 & 5.95658 & 6.00322 & 6.07383 & 6.17267 \\
\hline
0.15 & 6.40917 & 6.41290 & 6.42652 & 6.45586 & 6.50623 & 6.58256 & 6.68946 \\
\hline
0.20 & 6.97376 & 6.97778 & 6.99246 & 7.02415 & 7.07862 & 7.16122 & 7.27699 \\
\hline
0.25 & 7.61956 & 7.62389 & 7.63974 & 7.67398 & 7.73292 & 7.82237 & 7.94778 \\
\hline
0.30 & 8.36152 & 8.36618 & 8.38328 & 8.42027 & 8.48402 & 8.58085 & 8.71662 \\
\hline
0.35 & 9.21716 & 9.22217 & 9.24058 & 9.28048 & 9.34929 & 9.45388 & 9.60052 \\
\hline
\end{tabular}
\caption{Numerical values of the shadow radius $R_s$ for different values of $\alpha$ and $r_s$ with $M=1$, $\rho_s=0.05$, $\ell_p=25$.}
\label{tab:2}
\end{table*}

In Fig. \ref{fig:3d-plot}, we draw a 3D plot of the photon sphere radius $r_\text{ph}$ and shadow size $R_s$ as a function of $(\alpha, r_s)$. 

In Table \ref{tab:1}, we present numerical results for the photon sphere radius $r_\text{ph}$ by varying the CoS parameter $\alpha$ and halo radius $r_s$, while keeping other parameters fixed. Similarly, Table \ref{tab:2} presents the numerical values of the shadow size $R_s$ cast by the BH for different values of $\alpha$ and $r_s$.  

\subsection{\large {\bf Particle Dynamics}}\label{Sec:III-2}

The study of the dynamics of test particles around BHs in the presence of external fields is both important and significant, since it provides crucial insights into the structure of spacetime and the behavior of matter under strong gravity. In particular, the location of the innermost stable circular orbit (ISCO) plays a fundamental role in determining the efficiency of accretion processes, the emission of electromagnetic radiation, and the dynamics of compact objects in binary systems. The ISCO radius is directly related to the background geometry and can serve as a diagnostic tool to distinguish between different BH solutions or to probe deviations from general relativity. Furthermore, ISCO properties are closely connected to astrophysical observations, such as X-ray spectra from accretion disks, quasi-periodic oscillations (QPOs), and gravitational wave signatures from extreme mass-ratio inspirals (EMRIs). The presence of external fields, such as dark matter halo distributions, can significantly alter the location and stability of circular orbits, thereby affecting observable astrophysical phenomena. A detailed analysis of ISCOs thus provides an important window into both theoretical and observational aspects of BH physics \cite{Barack2019}.

For time-like geodesics, $\varepsilon=1$, the effective potential form Eq. (\ref{bb9}) reduces to
\begin{align}
V_\text{eff}(r)&=\left(1+\frac{\mathrm{L}^2}{r^2}\right)\notag\\ &\times \left(1-\alpha-\frac{2M}{r}-\rho_s\,r_s^2\,\mbox{ln} {\left(1+\frac{r_s}{r}\right)}+\frac{r^2}{\ell^2_p}\right).\label{dd1}
\end{align}

\begin{figure*}[ht!]
\centering
\includegraphics[width=0.45\linewidth]{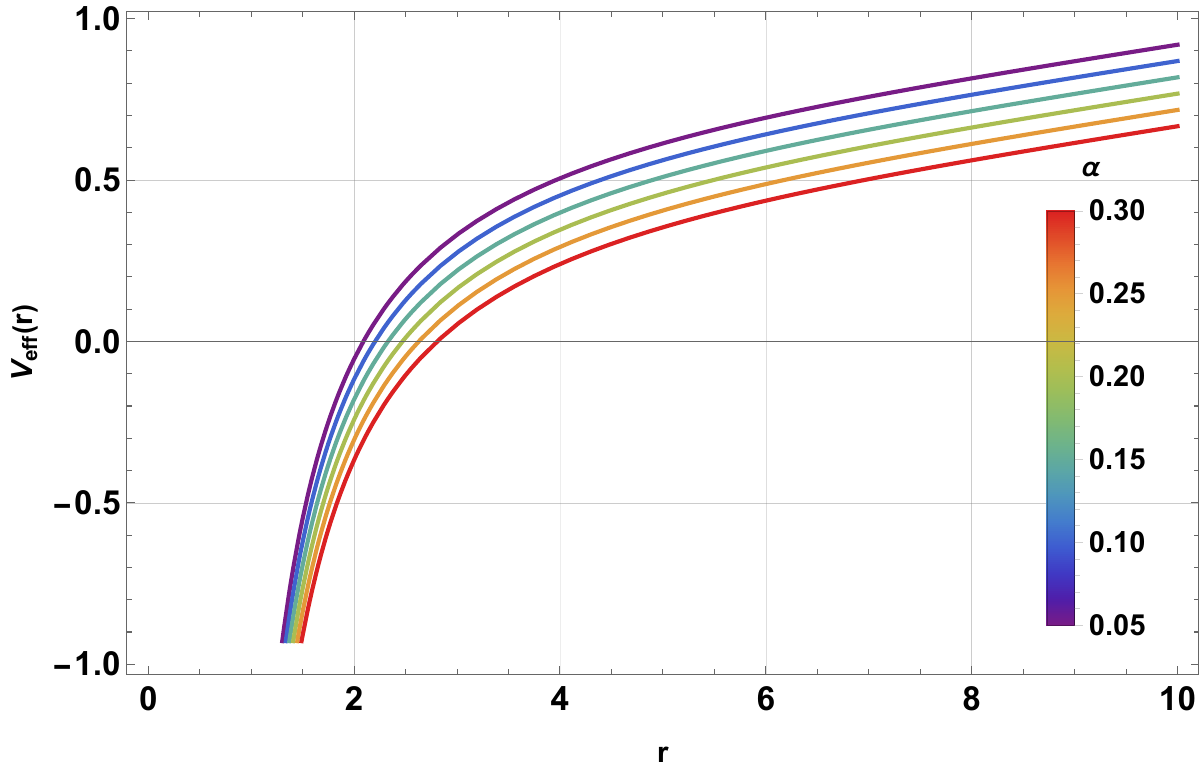}\qquad
\includegraphics[width=0.45\linewidth]{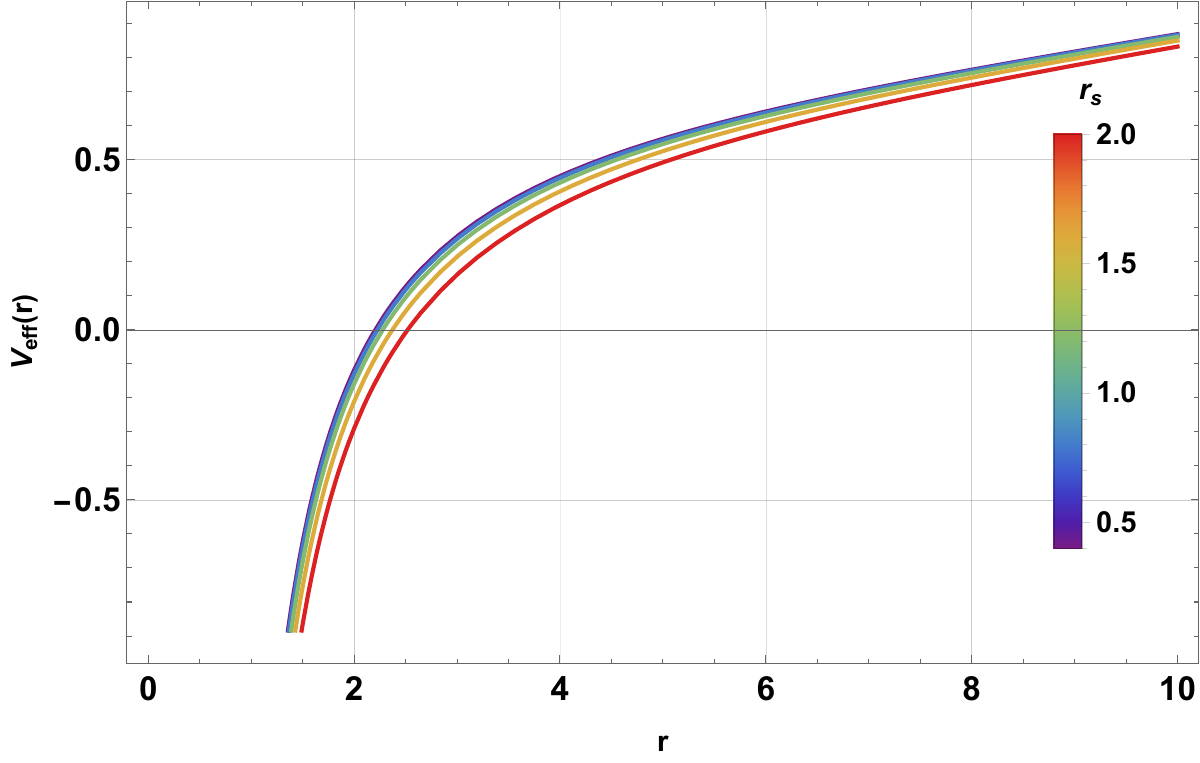}\\
(a) $r_s=0.2,\,\rho_s=0.02$ \hspace{5cm} (b) $\alpha=0.1,\,\rho_s=0.05$
\caption{\footnotesize The behavior of the effective potential for time-like geodesics as a function of $r$ for different values of CoS parameter $\alpha$ and $r_s$. Here, $M=1,\,\ell_p=25$.}
\label{fig:potential-timelike}
\end{figure*}

In Fig. \ref{fig:potential-timelike}, we show the behavior of the effective potential of time-like geodesics as a function of $r$ by varying the CoS parameter $\alpha$ and the halo radius $r_s$. In both panels, we observe that this potential also decreases with increasing $\alpha$ and $r_s$, indicating that the time-like particles are less bound by the gravitational field. 

For circular orbits of radius $r=r_0$, the conditions $\dot{r}=0$ and $\ddot{r}=0$ must be satisfied. These conditions using Eq. (\ref{bb8}) simplify as
\begin{align}
\mathrm{E}^2&=V_\text{eff}(r)=\left(1+\frac{\mathrm{L}^2}{r^2}\right)\notag\\&\times\left(1-\alpha-\frac{2M}{r}-\rho_s\,r_s^2\,\mbox{ln} {\left(1+\frac{r_s}{r}\right)}+\frac{r^2}{\ell^2_p}\right),\label{dd2}
\end{align}
and
\begin{equation}
V'_\text{eff}(r)=0,\label{dd3}
\end{equation}
where prime denotes derivative w. r. to $r$ and $V_\text{eff}$ is given in Eq. (\ref{dd1}).

Simplification of the relation (\ref{dd3}) using Eq. (\ref{dd1}) results
\begin{equation}
    \mathcal{L}_\text{specific}=r\,\sqrt{\frac{\frac{M}{r}+\frac{\rho_s\,r_s^3}{2\,(r+r_s)}+\frac{r^2}{\ell^2_p}}{1-\alpha-\frac{3M}{r}-\rho_s\,r_s^2\,\mbox{ln} {\left(1+\frac{r_s}{r}\right)}-\frac{1}{2}\,\frac{\rho_s\,r^3_s}{r+r_s}}}.\label{dd4}
\end{equation}
Substituting Eq. (\ref{dd4}) into the Eq. (\ref{dd2}) gives us another physical quantity as,
\begin{equation}
    \mathcal{E}_\text{specific}=\pm\,\frac{\left(1-\alpha-\frac{2M}{r}-\rho_s\,r_s^2\,\mbox{ln} {\left(1+\frac{r_s}{r}\right)}+\frac{r^2}{\ell^2_p}\right)}{\sqrt{1-\alpha-\frac{3M}{r}-\rho_s\,r_s^2\,\mbox{ln} {\left(1+\frac{r_s}{r}\right)}-\frac{1}{2}\,\frac{\rho_s\,r^3_s}{r+r_s}}}.\label{dd5}
\end{equation}

Here $\mathcal{L}_\text{specific}$ and $\mathcal{E}_\text{specific}$, respectively, represent the specific angular momentum and specific energy of test particles orbiting around the selected BH. One can see that the geometric and physical parameters, such as the BH mass $M$, the curvature radius $\ell_p$, the string cloud parameter $\alpha$, and the DM halo profile characterized by $(r_s,\rho_s)$, modify these physical quantities.

\begin{figure*}[ht!]
\centering
\includegraphics[width=0.45\linewidth]{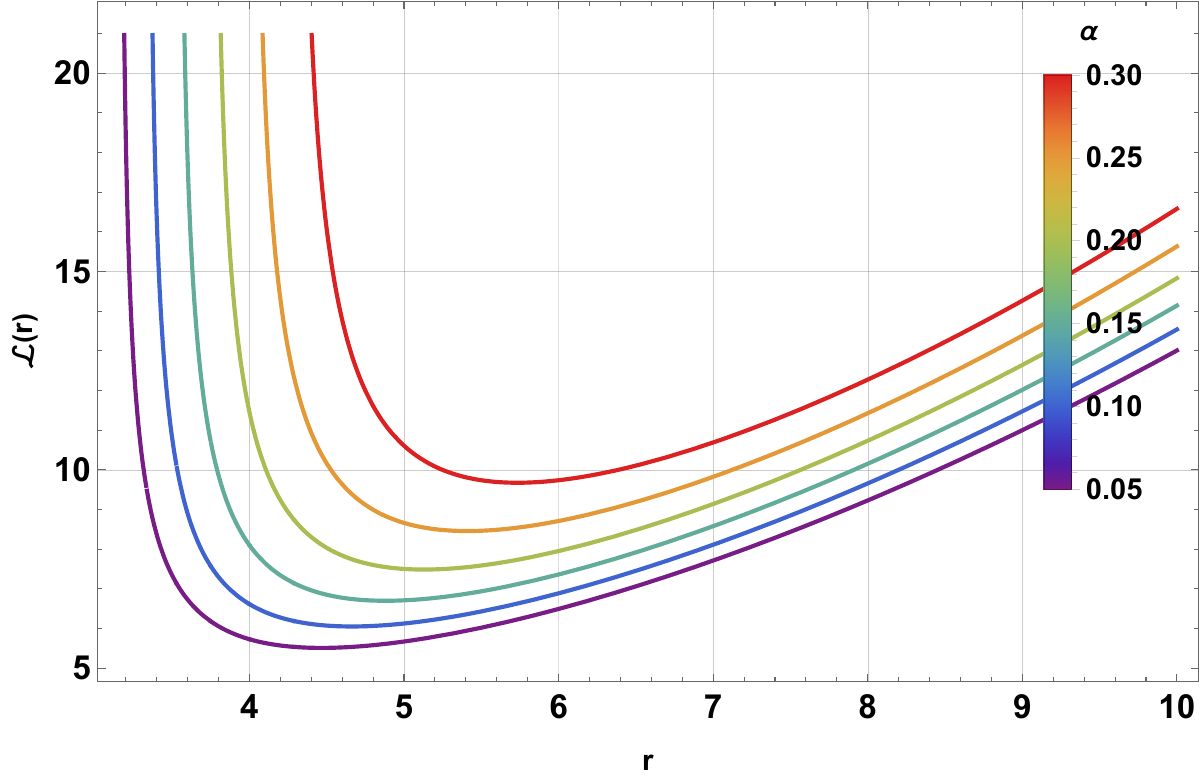}\qquad
\includegraphics[width=0.45\linewidth]{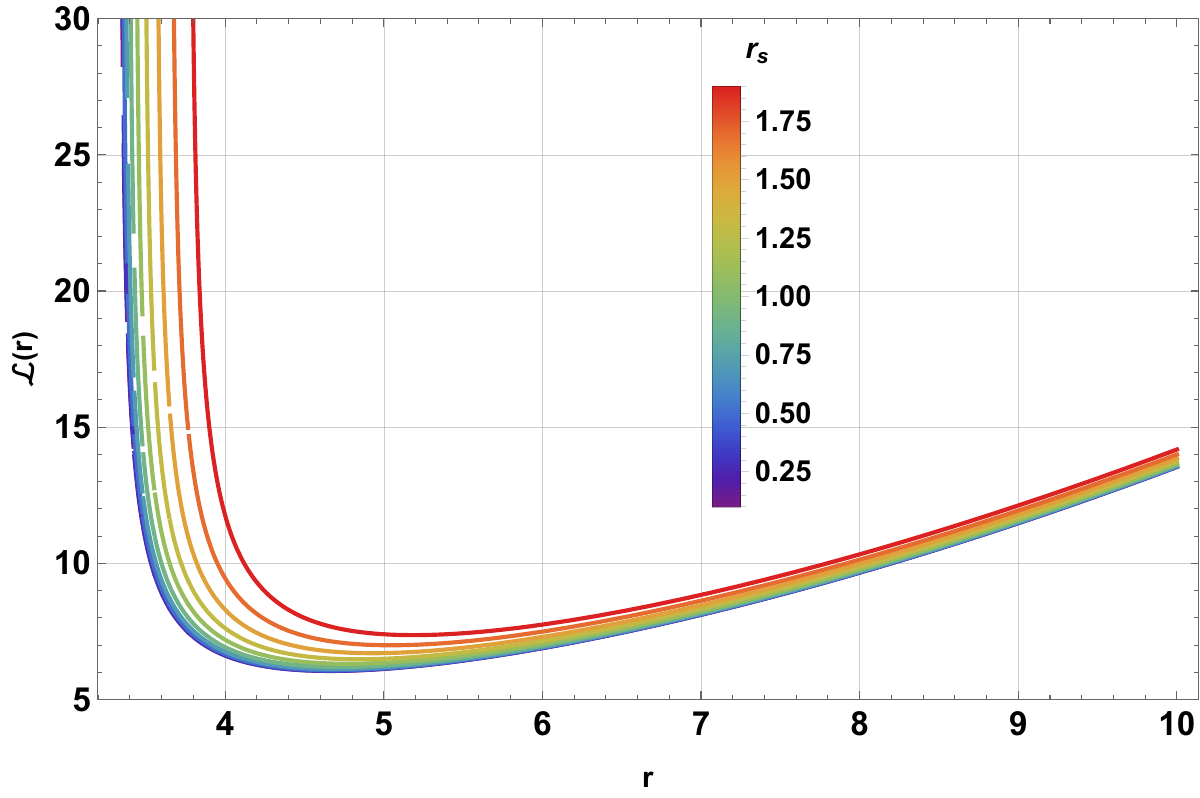}\\
(a) $r_s=0.2,\,\rho_s=0.02$ \hspace{5cm} (b) $\alpha=0.1,\,\rho_s=0.05$
\caption{\footnotesize Behavior of the specific angular momentum $\mathcal{L}$ of test particles as a function of $r$ for different values of CoS parameter $\alpha$ and halo radius $r_s$. Here, $M=1,\,\ell_p=10$.}
\label{fig:momentum}
\hfill\\
\includegraphics[width=0.45\linewidth]{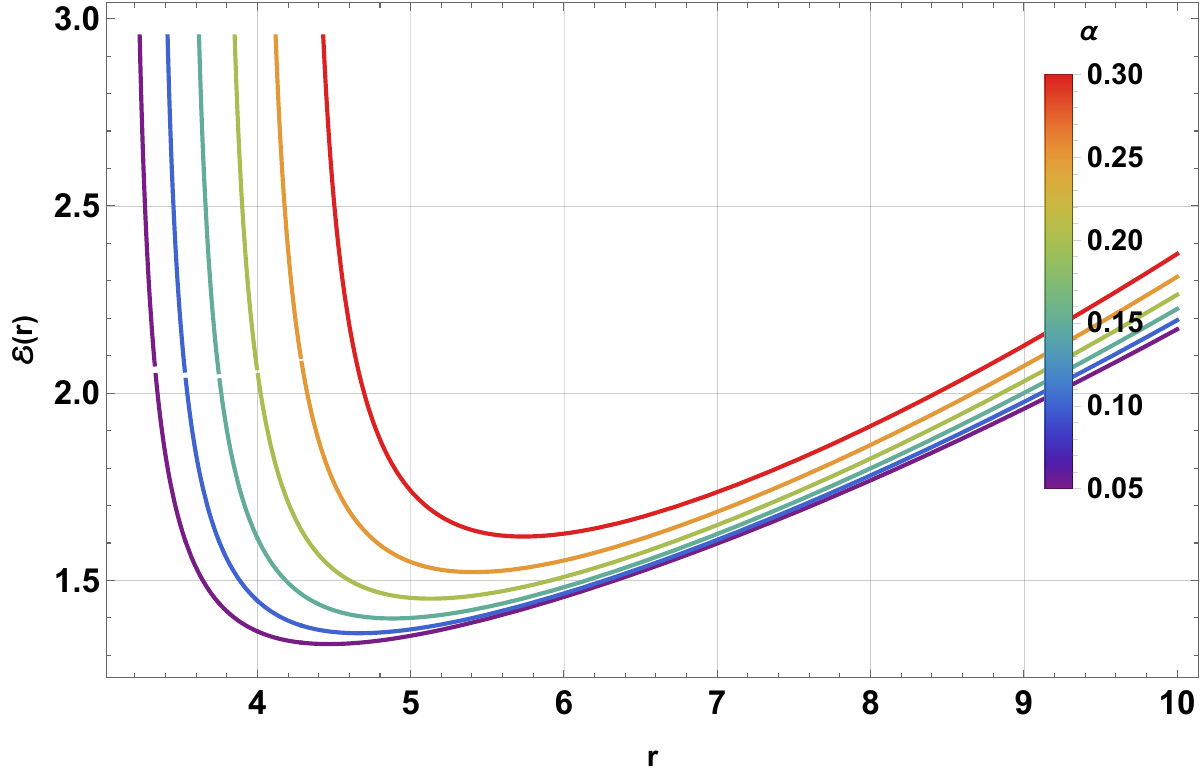}\qquad
\includegraphics[width=0.45\linewidth]{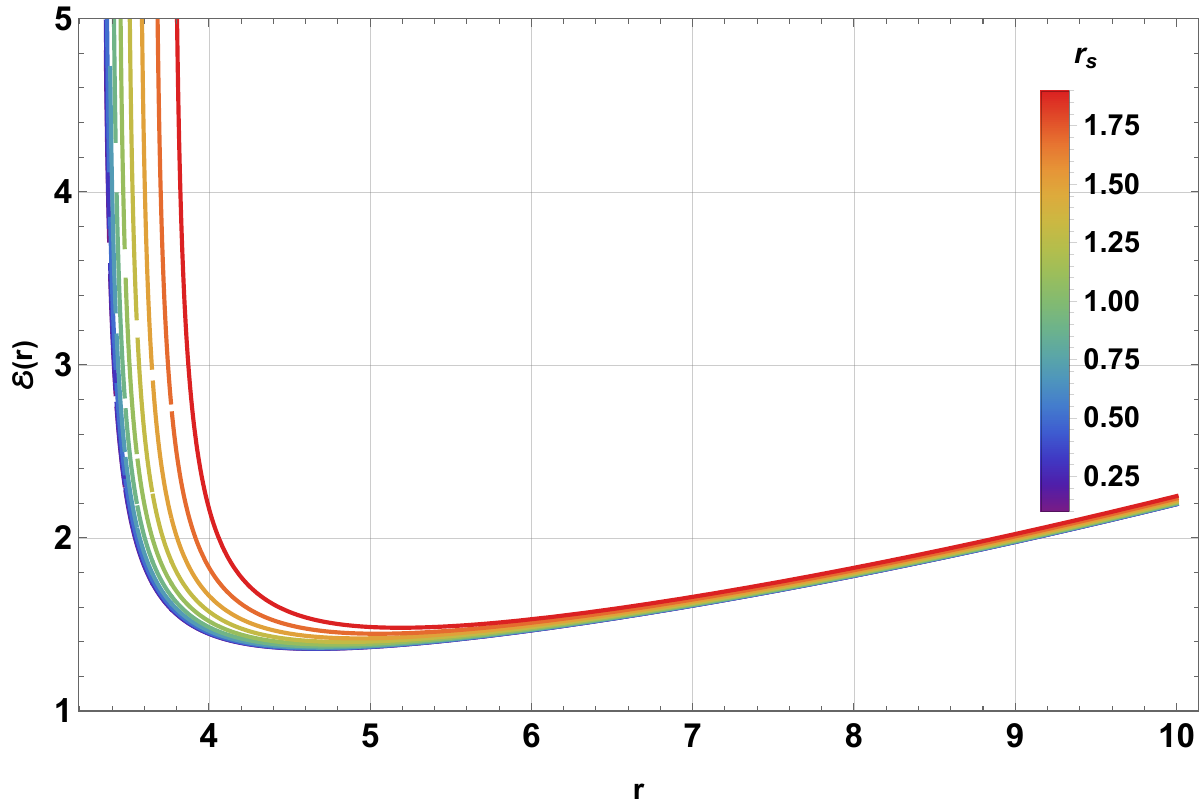}\\
(a) $r_s=0.2,\,\rho_s=0.02$ \hspace{5cm} (b) $\alpha=0.1,\,\rho_s=0.05$
\caption{\footnotesize Behavior of the specific energy $\mathcal{E}$ of test particles as a function of $r$ for different values of CoS parameter $\alpha$ and halo radius $r_s$. Here, $M=1,\,\ell_p=10$.}
\label{fig:energy}
\end{figure*}

In Fig. \ref{fig:momentum}, the behavior of the specific angular momentum of test particles as a function of $r$ is shown by varying the CoS parameter $\alpha$ and the halo radius $r_s$. In both panels, we observe that the angular momentum becomes larger with increasing values of $\alpha$ and $r_s$, indicating that the massive test particles require more angular momentum in traversing around the BH in ISCO.

In Fig. \ref{fig:energy}, the behavior of the specific energy of test particles as a function of $r$ is shown by varying the CoS parameter $\alpha$ and the halo radius $r_s$. In both panels, we observe that energy increases with increasing values of $\alpha$ and $r_s$, indicating that the massive test particles require more energy in traversing around the BH in ISCO.

The next important feature of massive test particles traversing around the BH is the innermost stable circular orbits. The ISCO corresponds to the smallest radius at which a test particle can stably orbit a BH. Inside the ISCO, circular orbits become unstable, and particles either plunge into the BH or move outward. The ISCO thus marks the transition between stable circular motion and dynamical instability.

For a circular orbit at radius $r_{c}$, the following conditions must hold:
\begin{itemize}
\item Existence of circular orbit: 
\[\frac{dV_{\text{eff}}}{dr}\Big|_{r=r_{c}} = 0.\]
  
\item Stability of circular orbit:
\[
\frac{d^{2}V_{\text{eff}}}{dr^{2}}\Big|_{r=r_{c}} > 0.\]
  
\item ISCO condition: The ISCO corresponds to the marginally stable orbit, where stability is lost, i.e.,
\[\frac{d^{2}V_{\text{eff}}}{dr^{2}}\Big|_{r=r_{\text{ISCO}}} = 0.\]
\end{itemize}

Substituting the effective potential given in Eq. (\ref{dd1}) into the ISCO condition results in the following polynomial equation:
\begin{align}
&\left(1-\alpha-\frac{2M}{r}-\rho_s\,r_s^2\,\mbox{ln} {\left(1+\frac{r_s}{r}\right)}+\frac{r^2}{\ell^2_p}\right)\notag\\ & \times\left[\frac{2M}{r^3} + \rho_s r_s^3 \,\frac{3(r^2+r r_s) - (2r+r_s)}{(r^2+r r_s)^2} + \frac{8}{\ell_p^2}\right] \notag \\ &-2\,\left(\frac{2M}{r^2} + \frac{\rho_s r_s^3}{r(r+r_s)} + \frac{2r}{\ell_p^2}\right)^2=0.\label{dd6}
\end{align}

\begin{figure*}[ht!]
\centering
\includegraphics[width=0.55\linewidth]{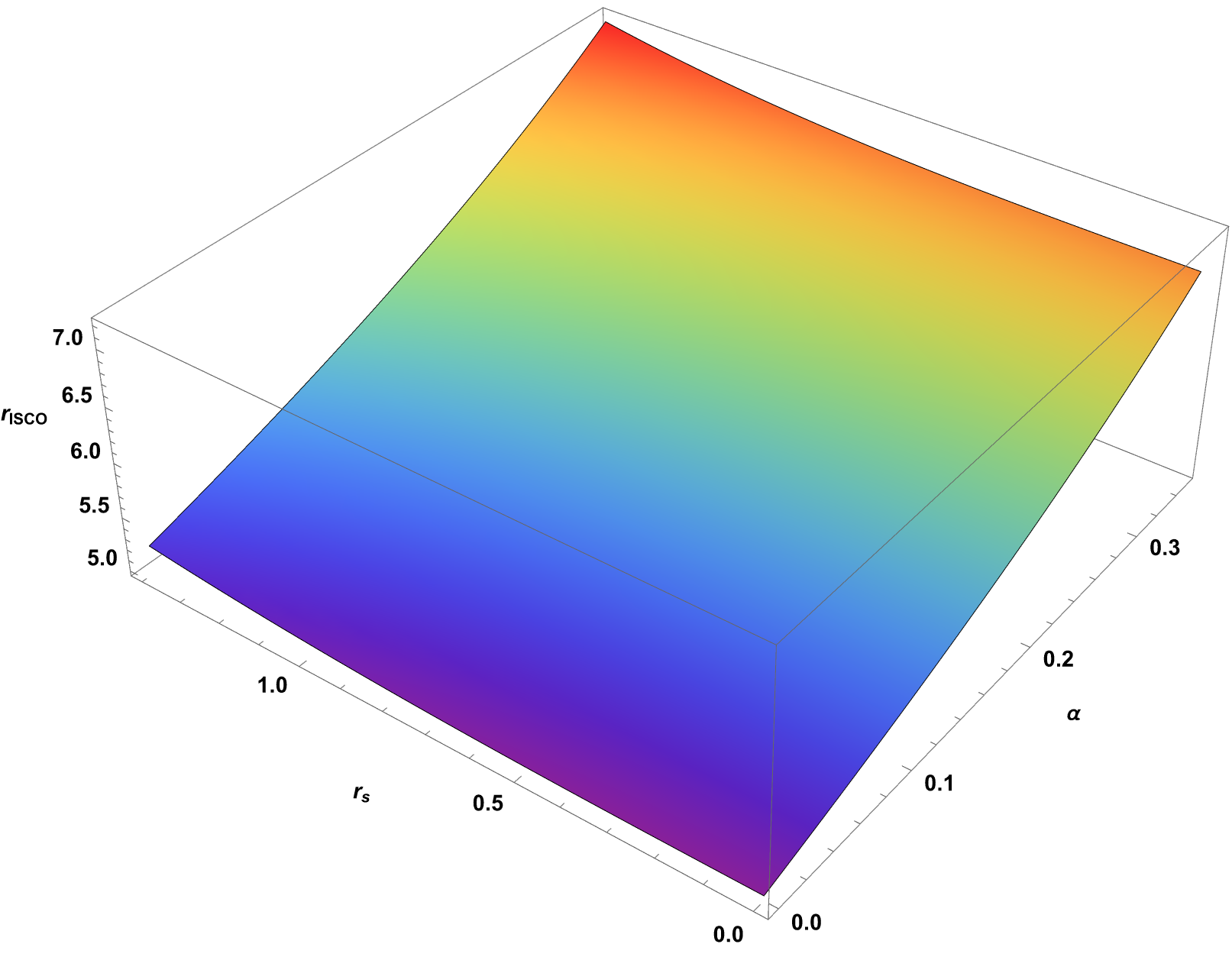}
\caption{\footnotesize 3D plot of the ISCO radius. Here $M=1,\,\rho_s=0.05,\,\ell_p=25$.}
\label{fig:ISCO}
\end{figure*}

Equation~(\ref{dd6}) represents a polynomial of infinite degree in $r$, for which an exact analytical solution is generally unattainable. However, the ISCO radius $ r = r_\text{ISCO} $ can still be determined numerically by assigning suitable values to the various geometric and physical parameters involved in the equation. In Fig. \ref{fig:ISCO}, we generate a 3D plot of the ISCO radius as a function of CoS parameter $\alpha$ and the halo radius $r_s$, while keeping other parameters fixed.

In Table \ref{tab:ISCO-radius}, we present numerical results for ISCO radius $r_\text{ISCO}$ by varying the CoS parameter $\alpha$ and halo radius $r_s$, while keeping other parameters fixed.

\begin{table*}[ht!]
\centering
\begin{tabular}{|c|c|c|c|c|c|c|c|}
\hline
$\alpha (\downarrow) \backslash r_s (\rightarrow)$ & $0.2$ & $0.4$ & $0.6$ & $0.8$ & $1.0$ & $1.2$ & $1.4$ \\ \hline
0.05 & 5.19317 & 5.17762 & 5.13858 & 5.07084 & 4.97603 & 4.86261 & 4.74332 \\ \hline
0.10 & 5.40638 & 5.39040 & 5.35018 & 5.28017 & 5.18182 & 5.06370 & 4.93908 \\ \hline
0.15 & 5.64049 & 5.62407 & 5.58266 & 5.51034 & 5.40834 & 5.28532 & 5.15508 \\ \hline
0.20 & 5.89948 & 5.88262 & 5.83999 & 5.76529 & 5.65948 & 5.53130 & 5.39506 \\ \hline
0.25 & 6.18849 & 6.17117 & 6.12727 & 6.05006 & 5.94022 & 5.80651 & 5.66379 \\ \hline
0.30 & 6.51430 & 6.49648 & 6.45120 & 6.37129 & 6.25708 & 6.11734 & 5.96746 \\ \hline
\end{tabular}
\caption{\footnotesize Computed values of ISCO radius $r_\text{ISCO}$ for different $\alpha$ and $r_s$ with $M=1$, $\rho_s=0.05$, and $\ell_p=25$.}
\label{tab:ISCO-radius}
\end{table*}

\begin{figure*}[ht!]
\centering
\includegraphics[width=0.45\linewidth]{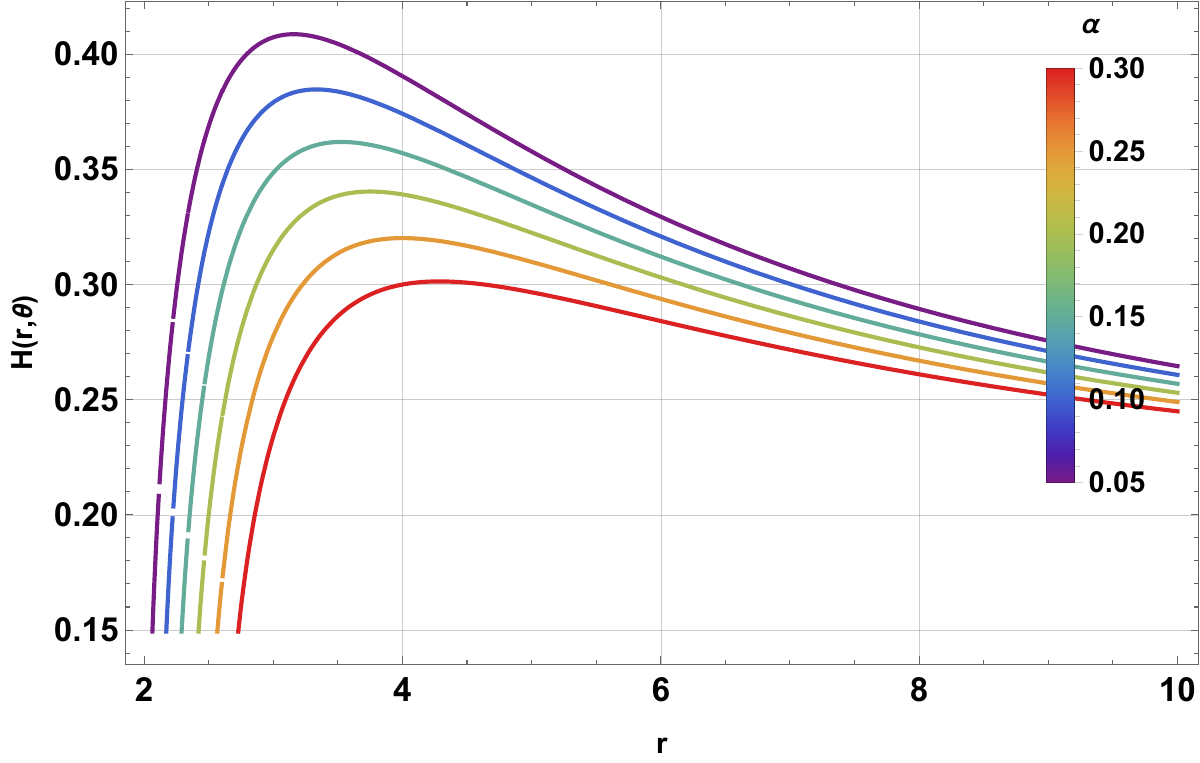}\qquad
\includegraphics[width=0.45\linewidth]{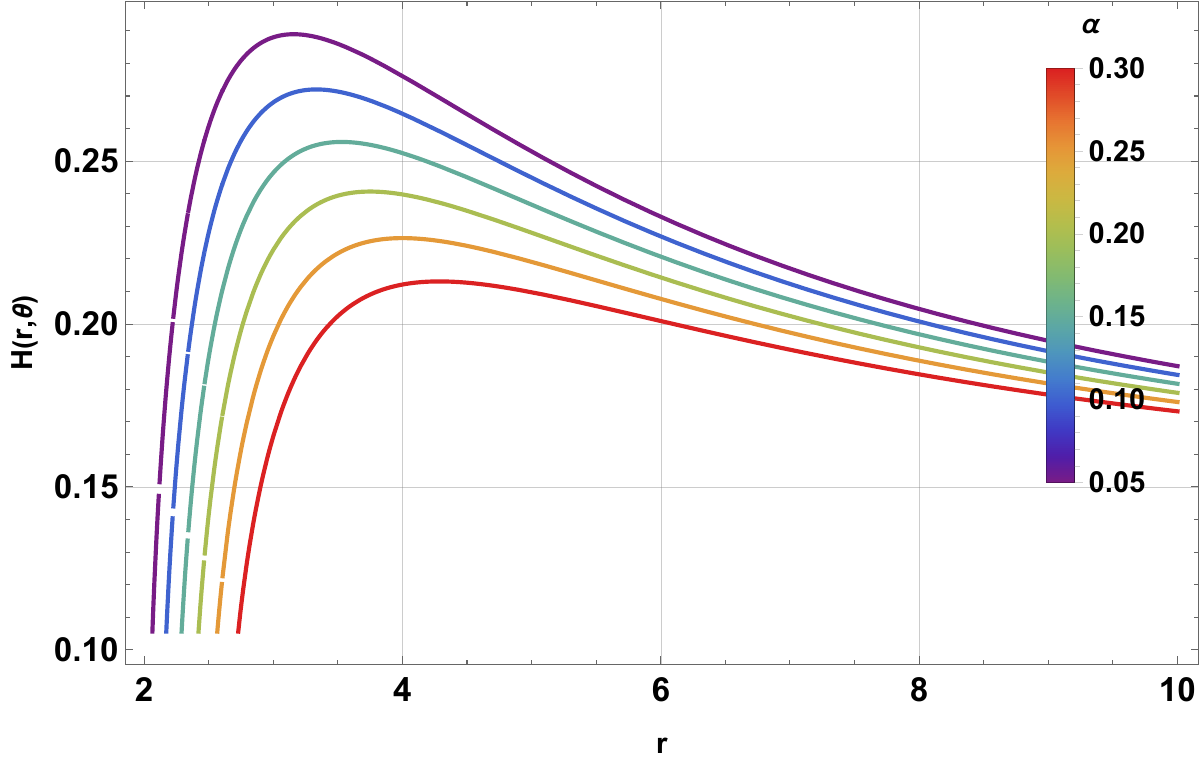}\\
(a) $\theta=\pi/6$ \hspace{6cm} (b) $\theta=\pi/4$\\
\includegraphics[width=0.45\linewidth]{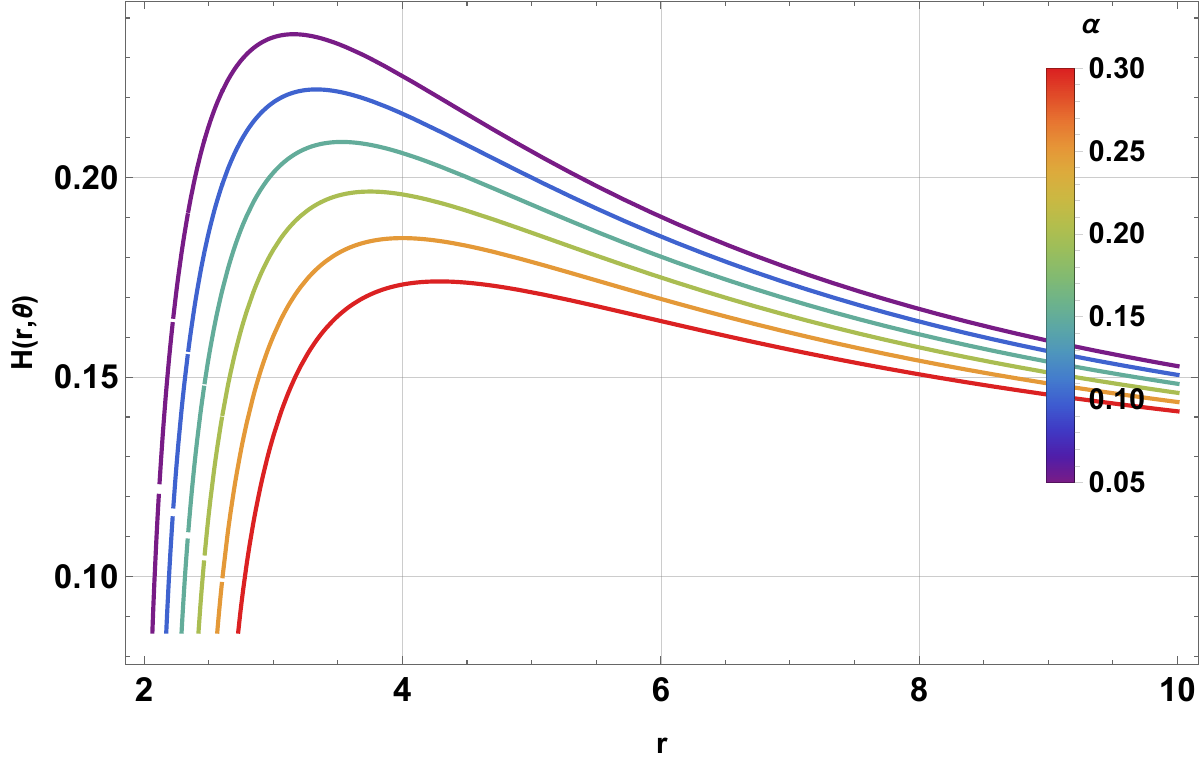}\qquad
\includegraphics[width=0.45\linewidth]{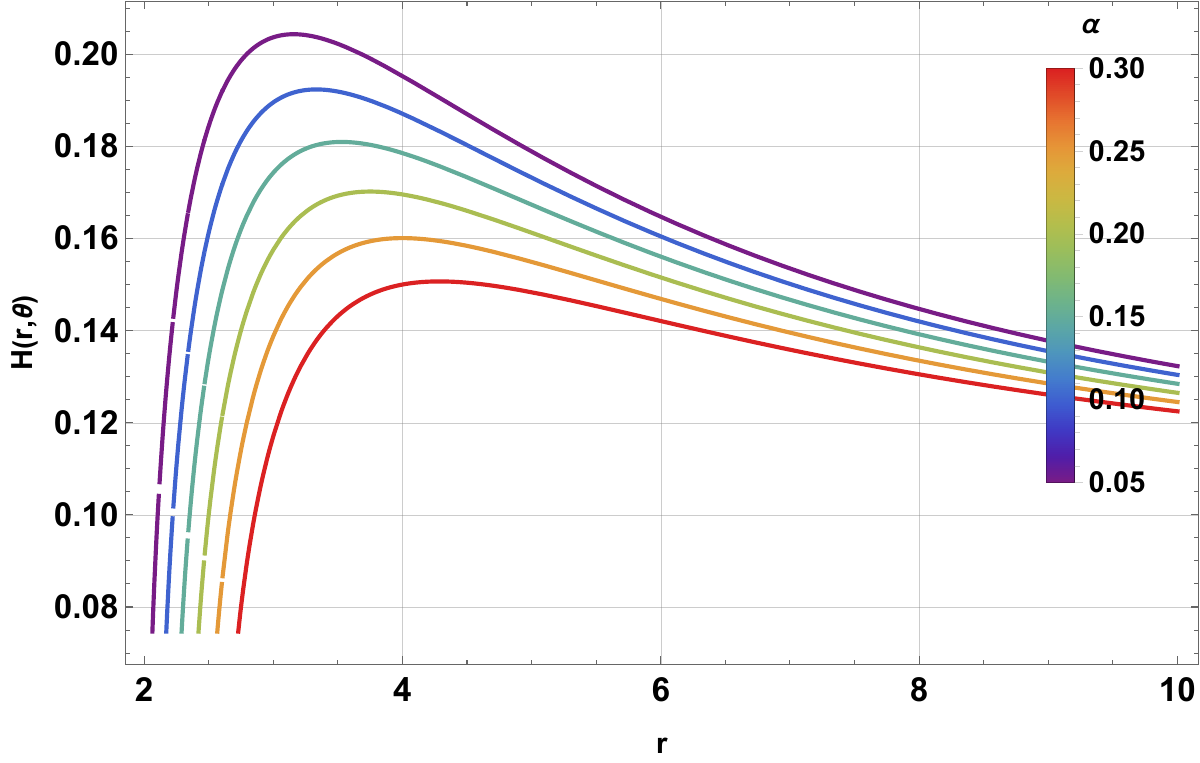}\\
(c) $\theta=\pi/3$ \hspace{6cm} (d) $\theta=\pi/2$
\caption{\footnotesize Behavior of the potential function $H(r,\theta)$ as a function of $r$ for different values of $\theta$ by varying the CoS parameter $\alpha$. Here, $M=1,\,\ell_p=10$.}
\label{fig:potential-function-1}
\end{figure*}

\begin{figure*}[ht!]
\centering
\includegraphics[width=0.45\linewidth]{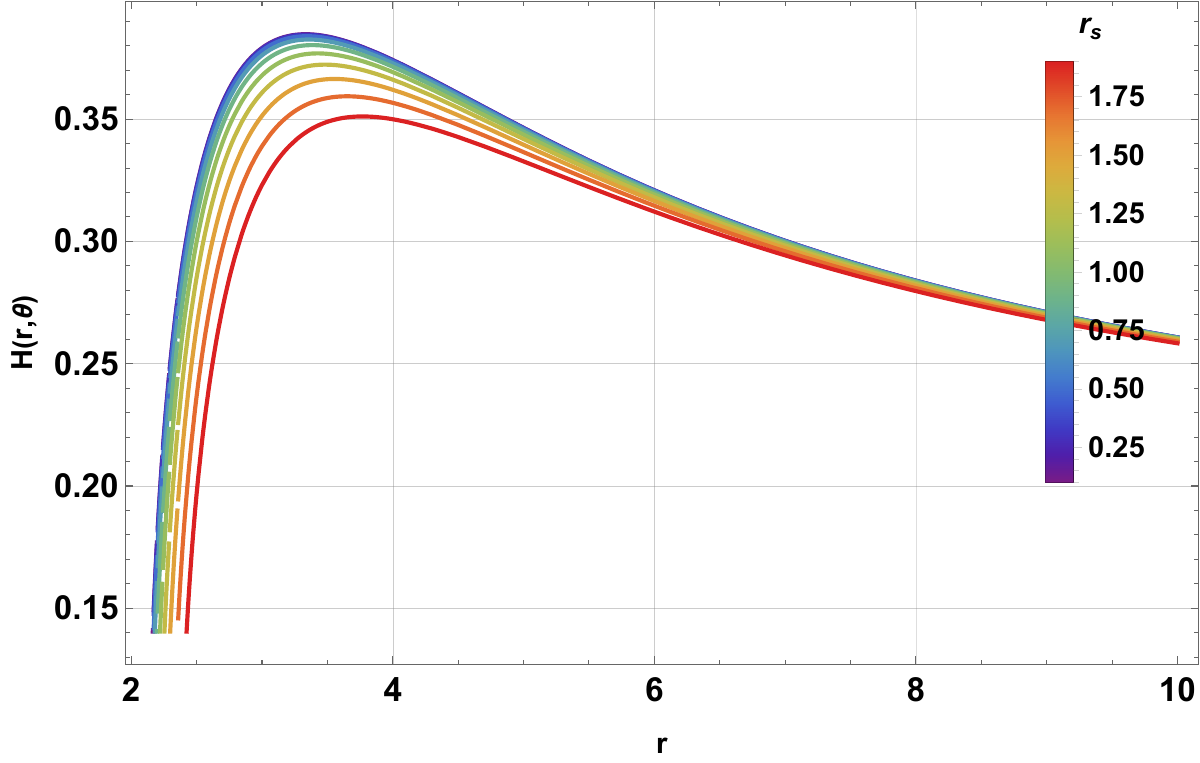}\qquad
\includegraphics[width=0.45\linewidth]{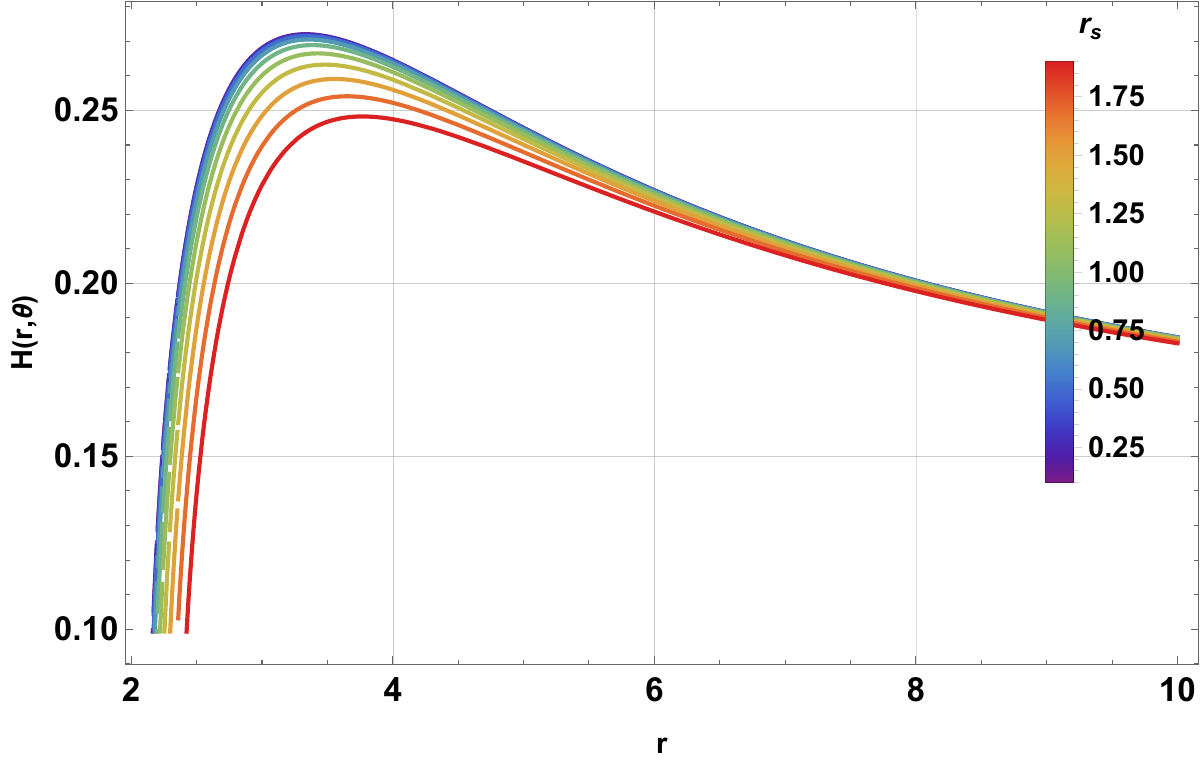}\\
(a) $\theta=\pi/6$ \hspace{6cm} (b) $\theta=\pi/4$\\
\includegraphics[width=0.45\linewidth]{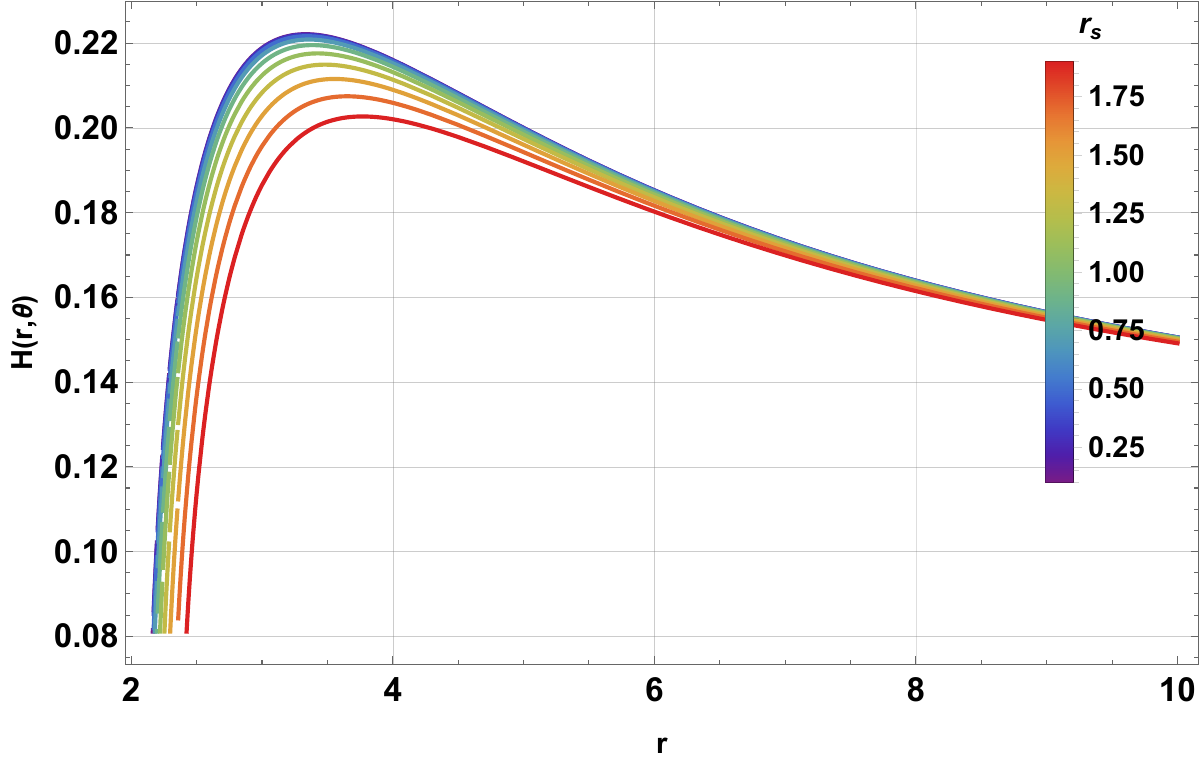}\qquad
\includegraphics[width=0.45\linewidth]{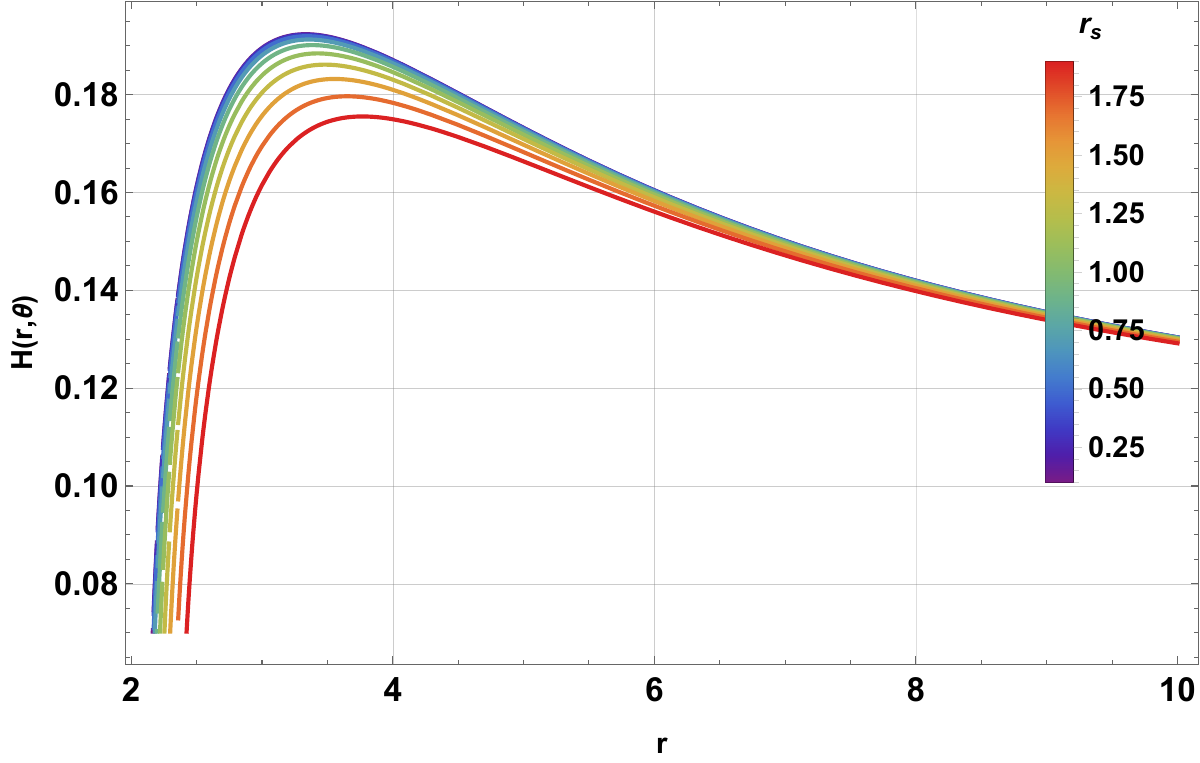}\\
(c) $\theta=\pi/3$ \hspace{6cm} (d) $\theta=\pi/2$
\caption{\footnotesize Behavior of the potential function $H(r,\theta)$ as a function of $r$ for different values of $\theta$ by varying the halo radius $r_s$. Here, $M=1,\,\ell_p=10$.}
\label{fig:potential-function-2}
\end{figure*}

\section{Topological Properties of Light Rings}\label{Sec:IV}

Topologically, photon rings are unstable closed null orbits, and this instability is universal: any small perturbation pushes photons either toward the horizon or out to infinity. This unstable character underlies the formation of the BH shadow, since the critical impact parameter associated with photon rings defines the shadow boundary \cite{perlick}. Moreover, recent topological analyses have shown that the number and stability type of photon rings are constrained by index theorems, ensuring that BHs generally possess at least one unstable photon ring \cite{liu2024}. These topological properties connect fundamental aspects of spacetime geometry with observational signatures such as shadows and lensing patterns.

To study the topological property of the light rings, one can introduce a potential function as \cite{BB2,BB3}
\begin{align}
    H(r,\theta)&=\sqrt{-\frac{g_{tt}}{g_{\theta\theta}}}\notag\\&=\frac{\sqrt{1-\alpha-\frac{2M}{r}-\rho_s\,r_s^2\,\mbox{ln} {\left(1+\frac{r_s}{r}\right)}+\frac{r^2}{\ell^2_p}}}{r\,\sin \theta},\label{cc1}
\end{align}
where the function $H(r, \theta)$ is regular for $r >r_h$, the horizon radius. One can show that the photon sphere radius can occur by the condition $\partial_r H(r, \theta)=0$.

From the above expression (\ref{cc1}), we observe that geometric and physical parameters, such as the BH mass $M$, the string cloud parameter $\alpha$, the curvature radius $\ell_p$, and the DM profile characterized by $(r_s,\rho_s)$, modify this potential function.

In Fig. \ref{fig:potential-function-1}, we illustrate the behavior of the potential function $H(r, \theta)$ for various values of $\theta$ by varying the CoS parameter $\alpha$, while keeping all other parameters fixed. The results show that the potential function decreases as $\alpha$ increases. As a result, the normalized vector is also influenced by changes in $\alpha$.

Similarly, Fig. \ref{fig:potential-function-2} presents the variation of $H(r, \theta)$ for different $\theta$ values by changing the halo radius $r_s$, with other parameters held constant. The potential function exhibits a similar decreasing trend with increasing $r_s$, consistent with the behavior observed in the previous figure.

\begin{figure*}[ht!]
\centering
\includegraphics[width=0.45\linewidth]{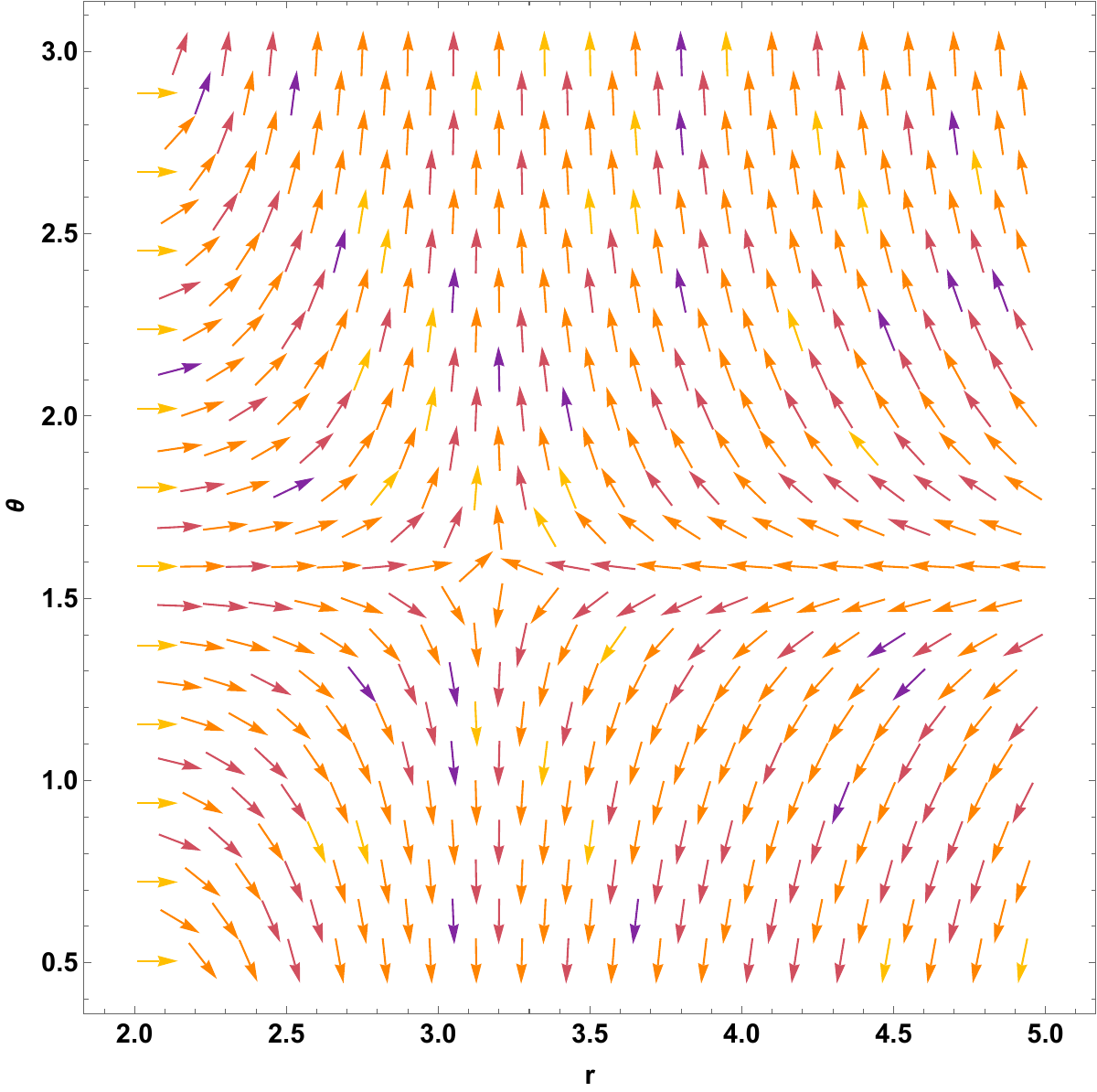}\qquad
\includegraphics[width=0.45\linewidth]{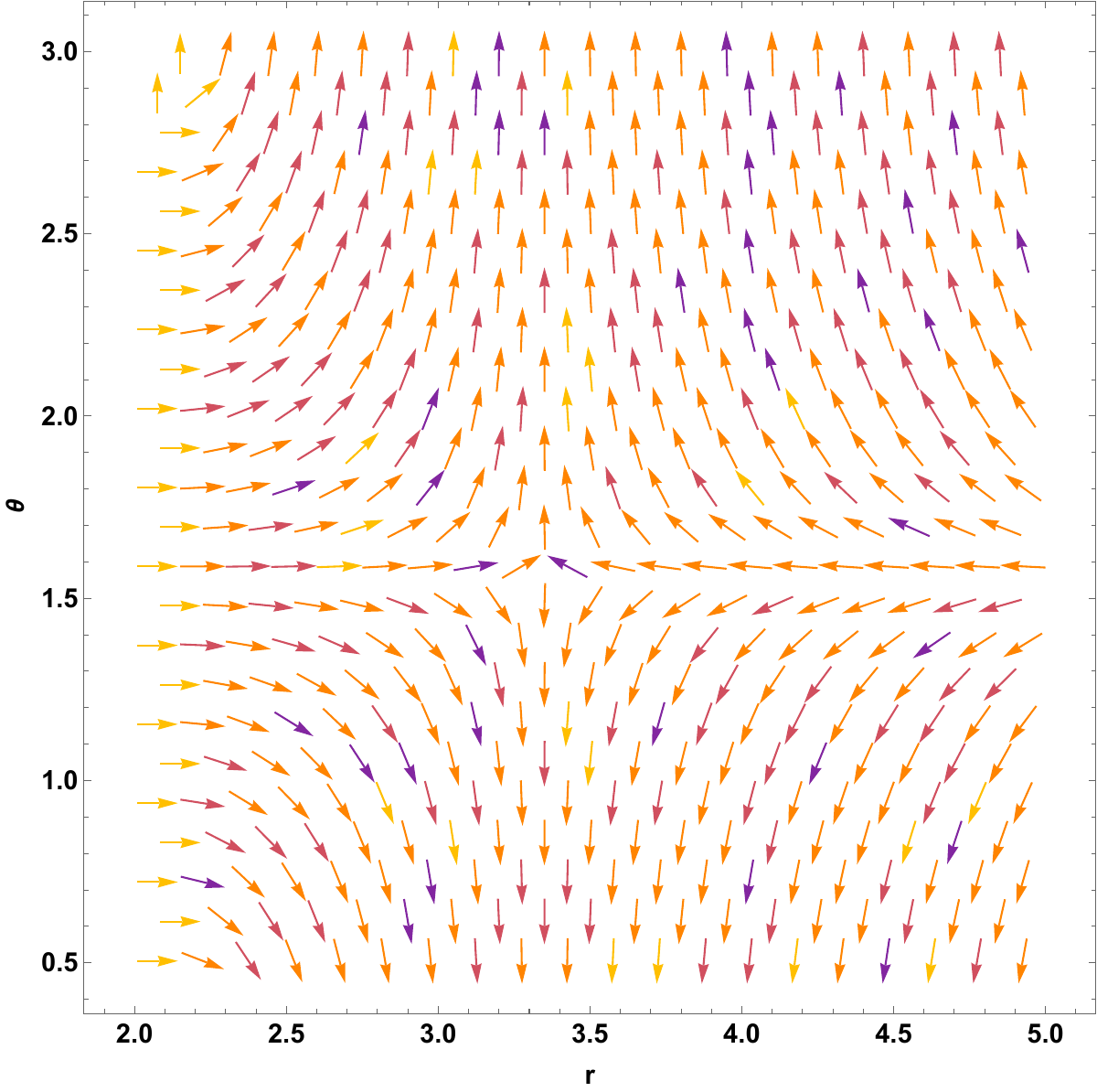}\\
(a) $\alpha=0.05$ \hspace{6cm} (b) $\alpha=0.1$\\
\includegraphics[width=0.45\linewidth]{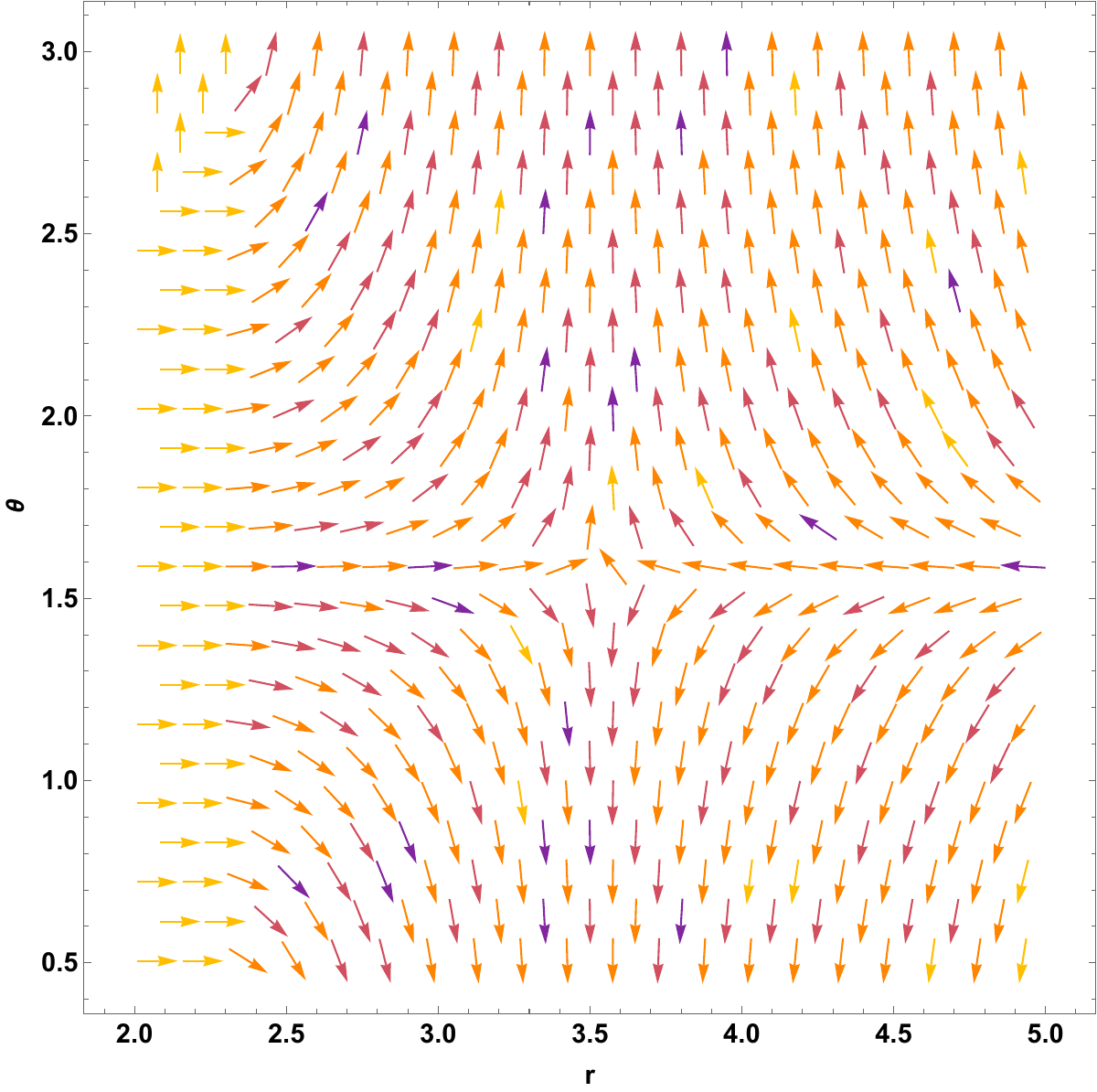}\qquad
\includegraphics[width=0.45\linewidth]{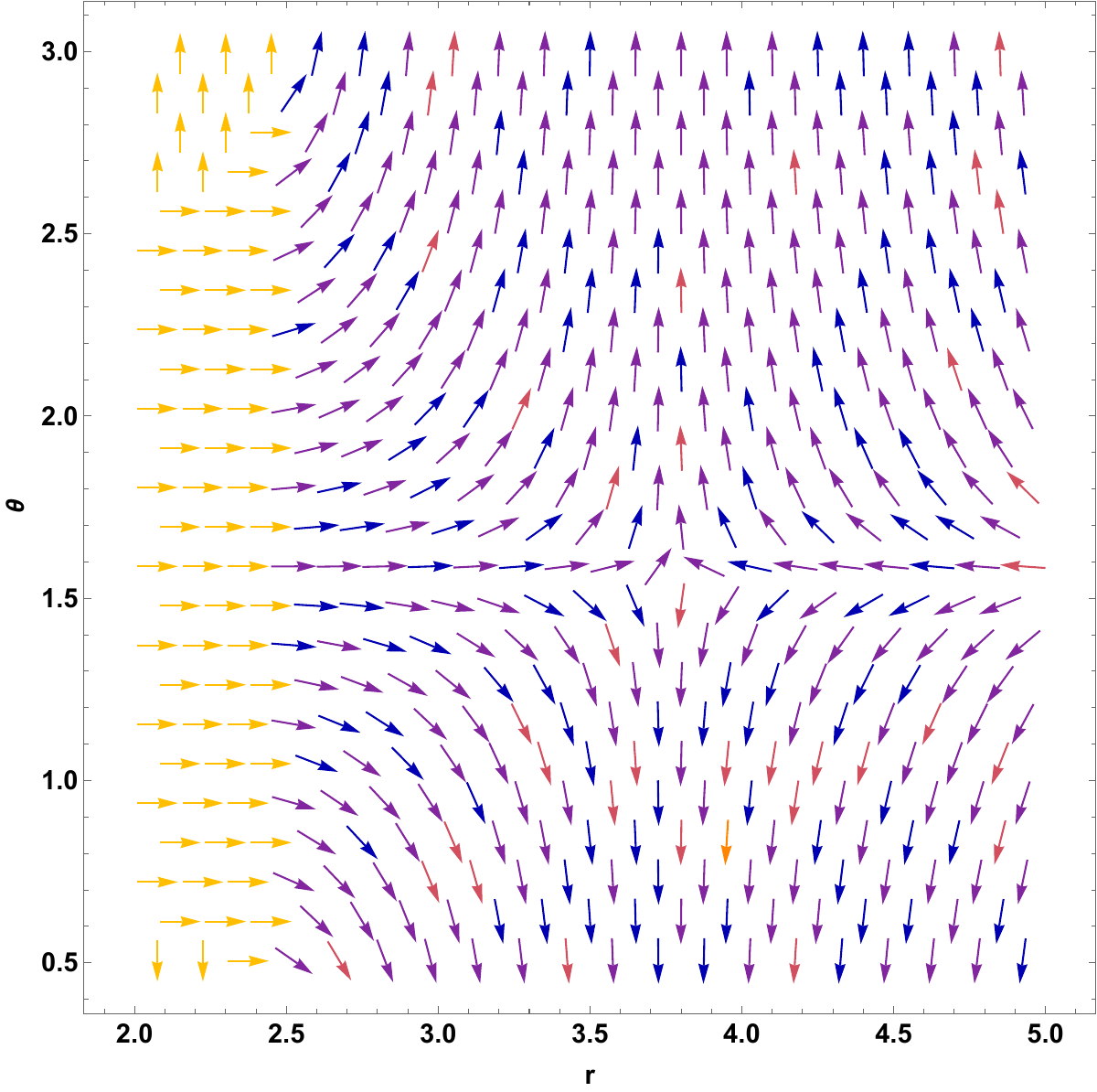}\\
(c) $\alpha=0.15$ \hspace{6cm} (d) $\alpha=0.20$
\caption{\footnotesize The arrows represent the unit vector field ${\bf n}$ on a portion of the $r-\theta$ plane for BH by varying the CoS parameter $\alpha$. Here $M=1,\,r_s=0.5,\,\rho_s=0.05$ and $\ell_p=25$.}
\label{fig:field-1}
\end{figure*}

\begin{figure*}[ht!]
\centering
\includegraphics[width=0.45\linewidth]{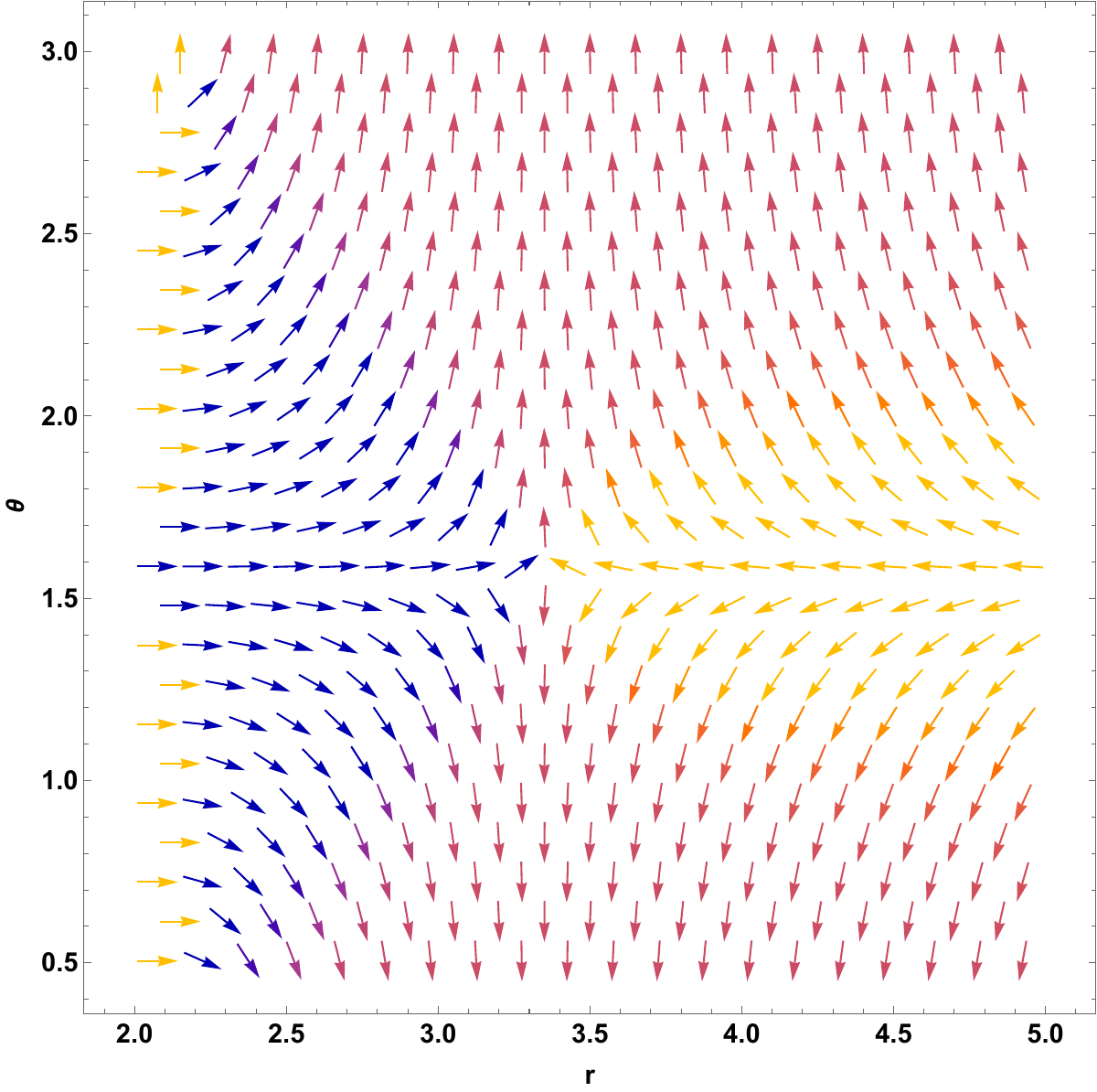}\qquad
\includegraphics[width=0.45\linewidth]{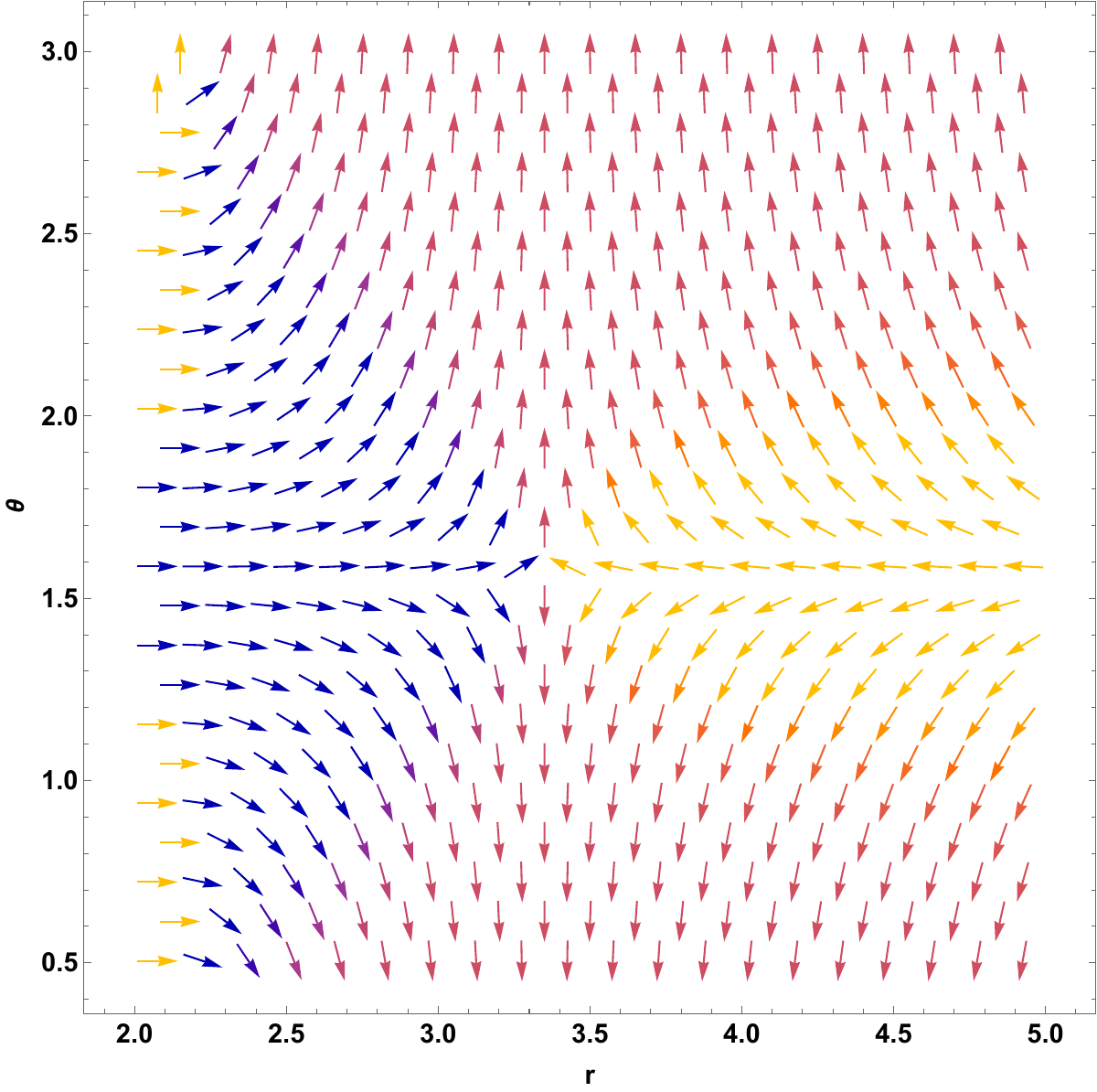}\\
(a) $r_s=0.3$ \hspace{6cm} (b) $r_s=0.6$\\
\includegraphics[width=0.45\linewidth]{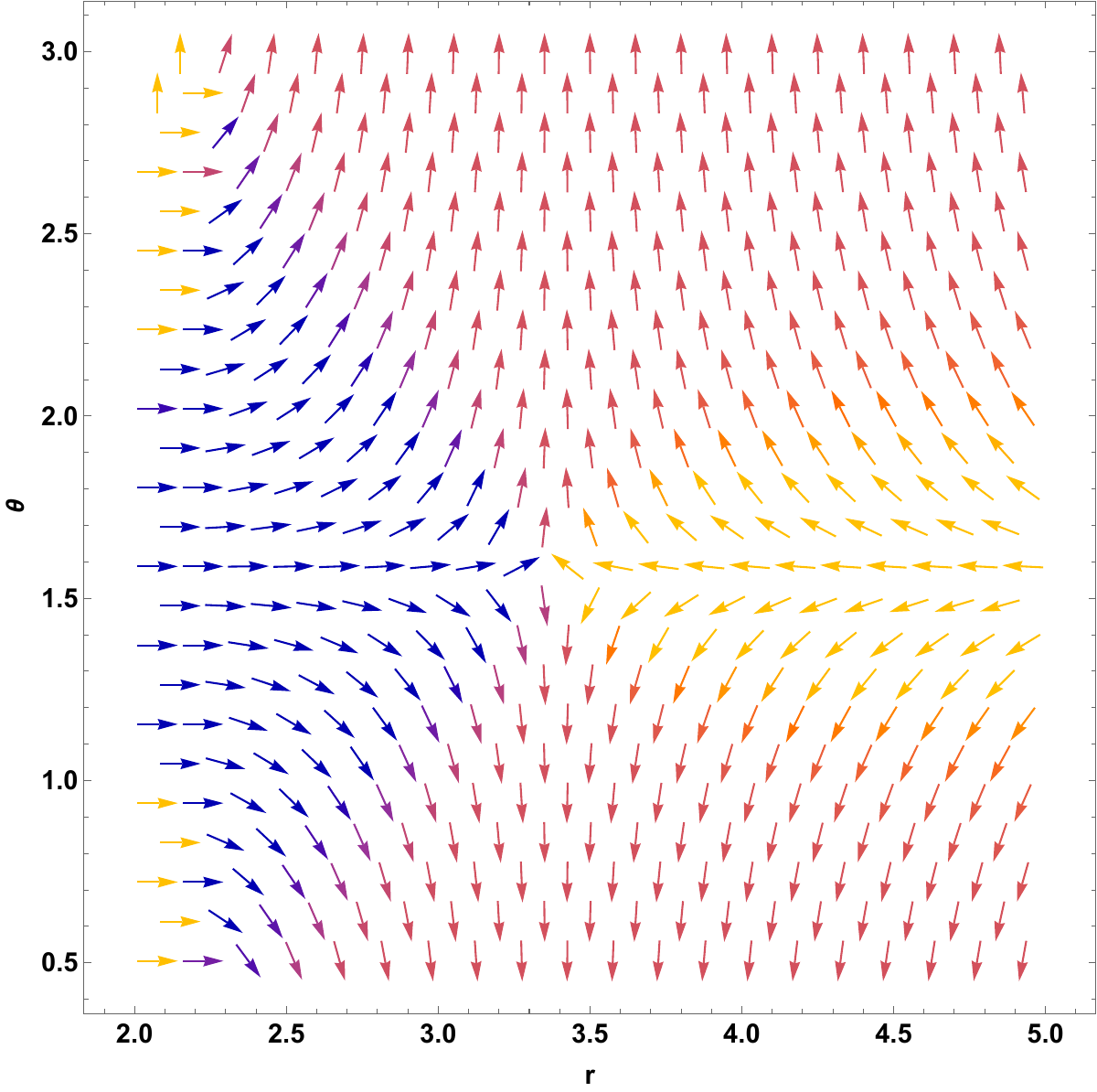}\qquad
\includegraphics[width=0.45\linewidth]{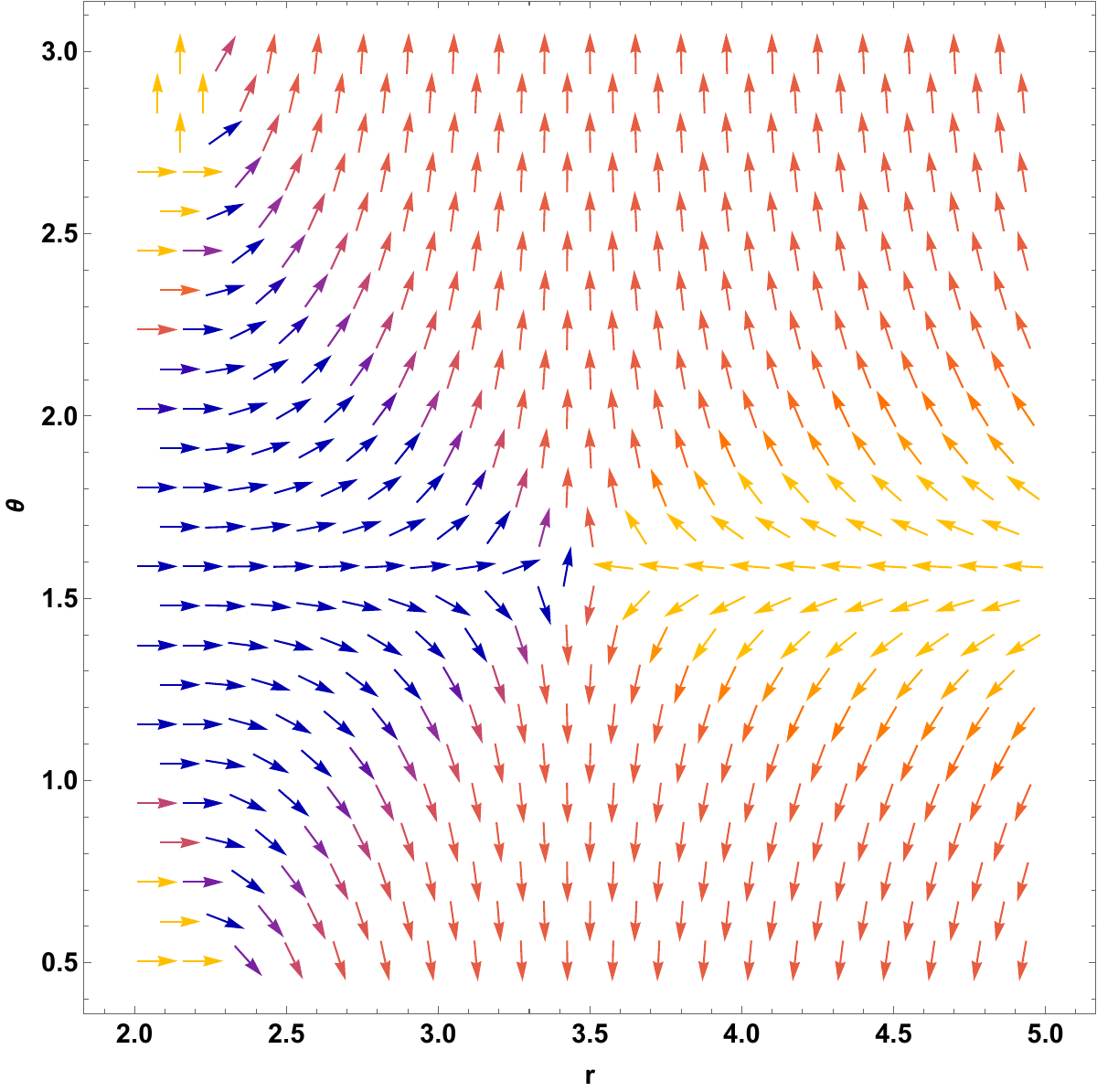}\\
(c) $r_s=0.9$ \hspace{6cm} (d) $r_s=1.2$
\caption{\footnotesize The arrows represent the unit vector field ${\bf n}$ on a portion of the $r-\theta$ plane for BH by varying the halo radius $r_s$. Here $M=1,\,\alpha=0.1,\,\rho_s=0.05$ and $\ell_p=25$.}
\label{fig:field-2}
\end{figure*}

The key vector field ${\bf v}=(v_r\,,\,v_{\theta})$ using the definition in \cite{AA2,AA3} is given as
\begin{align}
v_r&=-\frac{1}{r^2\,\sin\theta}\notag\\&\times \left(1-\alpha-\frac{3M}{r}-\rho_s\,r_s^2\,\mbox{ln} {\left(1+\frac{r_s}{r}\right)}-\frac{1}{2}\,\frac{\rho_s\,r^3_s}{r+r_s}\right),\label{cc2}\\
v_{\theta}&=-\frac{\sqrt{1-\alpha-\frac{2M}{r}-\rho_s\,r_s^2\,\mbox{ln} {\left(1+\frac{r_s}{r}\right)}+\frac{r^2}{\ell^2_p}}}{r^2}\,\frac{\cot \theta}{\sin \theta}.\label{cc3}
\end{align}

Consequently, the normalized field components read
\begin{widetext}
\begin{align}
n_r=-\frac{1-\alpha-\frac{3M}{r}-\rho_s r_s^2\ln\!\Big(1+\frac{r_s}{r}\Big)-\frac{1}{2}\frac{\rho_s r_s^3}{r+r_s}}
{\sqrt{\Big(1-\alpha-\frac{3M}{r}-\rho_s r_s^2\ln\!\Big(1+\frac{r_s}{r}\Big)-\frac{1}{2}\frac{\rho_s r_s^3}{r+r_s}\Big)^2+\Big(1-\alpha-\frac{2M}{r}-\rho_s r_s^2\ln\!\Big(1+\frac{r_s}{r}\Big)+\frac{r^2}{\ell_p^2}\Big)\cot^2\theta}},
\\
n_{\theta}=-\frac{\sqrt{\,1-\alpha-\frac{2M}{r}-\rho_s r_s^2\ln\!\Big(1+\frac{r_s}{r}\Big)+\frac{r^2}{\ell_p^2}}\,\cot \theta}
{\sqrt{\Big(1-\alpha-\frac{3M}{r}-\rho_s r_s^2\ln\!\Big(1+\frac{r_s}{r}\Big)-\frac{1}{2}\frac{\rho_s r_s^3}{r+r_s}\Big)^2+\Big(1-\alpha-\frac{2M}{r}-\rho_s r_s^2\ln\!\Big(1+\frac{r_s}{r}\Big)+\frac{r^2}{\ell_p^2}\Big)\cot^2\theta}}.   
\end{align}
\end{widetext}
At $(r,\theta)=(r_{\rm ph},\pi/2)$ one recovers the zero of the unit field ${\bf n}$ as expected.

One can see that BH mass $M$, the curvature radius $\ell_p$, the string cloud parameter $\alpha$, and the DM halo profile characterized by $(r_s,\rho_s)$ modify the normalized unit vector field.

Figure \ref{fig:field-1} illustrates the behavior of the normalized vector field ${\bf n}(r, \theta)$ on a portion of the $r$-$\theta$ plane for different values of the CoS parameter $\alpha$, with all other parameters kept fixed. 

Similarly, Fig. \ref{fig:field-2} shows the behavior of ${\bf n}(r, \theta)$ for different values of the halo radius $r_s$, holding the remaining parameters constant. In both figures, the arrows indicate the direction of the normalized vector field on the $r$-$\theta$ plane.

\section{Scalar Perturbations and QNMs}\label{Sec:V}

Scalar field perturbations in BH spacetimes are an important tool in understanding the dynamical stability and response of BHs to external disturbances. A minimally coupled scalar field $\Phi$ evolving on a fixed BH background typically satisfies the Klein-Gordon equation, which, under separation of variables, reduces to a Schrödinger-like wave equation with an effective potential that encodes the spacetime geometry and parameters like the field's mass and charge. The structure of this potential determines the behavior of the scalar field, including whether perturbations decay (stable) or grow (unstable) with time. One central application of scalar perturbations is the study of quasinormal modes (QNMs)-complex-frequency solutions representing damped oscillations that dominate the late-time behavior of perturbations and are key signatures in gravitational wave astronomy \cite{konoplya2011,berti2019}.

For a minimally coupled scalar $\Phi$ with mass $\mu$, the Klein-Gordon equation is given by the following form:
\begin{equation}
    \Box\Phi-\mu^2\Phi=0,\label{SP1}
\end{equation}
where
\begin{equation}
    \Box=\frac{1}{\sqrt{-g}}\,\partial_{\mu}\,\left(\sqrt{-g}\,g^{\mu\nu}\,\partial_{\nu}\right).\label{SP2}
\end{equation}
In this context, $ g_{\mu\nu} $ represents the metric tensor, $ g^{\mu\nu} $ its inverse (contravariant form), and $ g $ the determinant of the metric.

\begin{figure*}[ht!]
\centering
\includegraphics[width=0.45\linewidth]{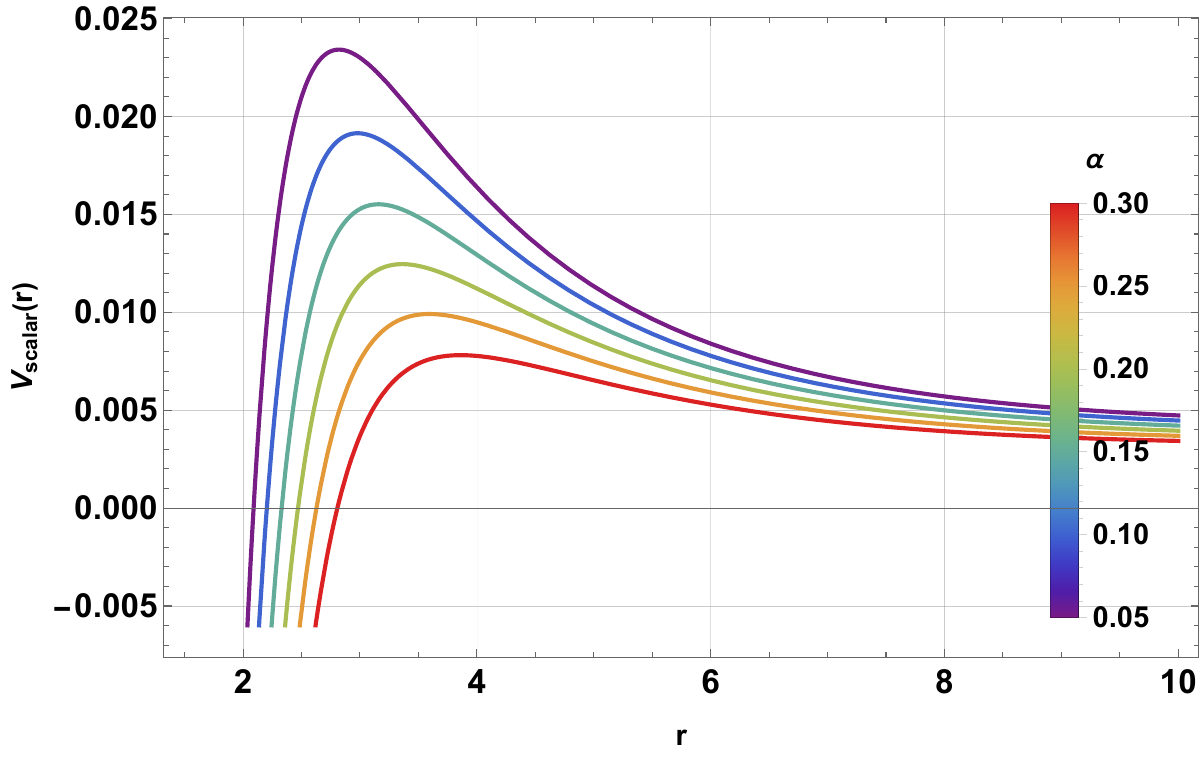}\qquad
\includegraphics[width=0.45\linewidth]{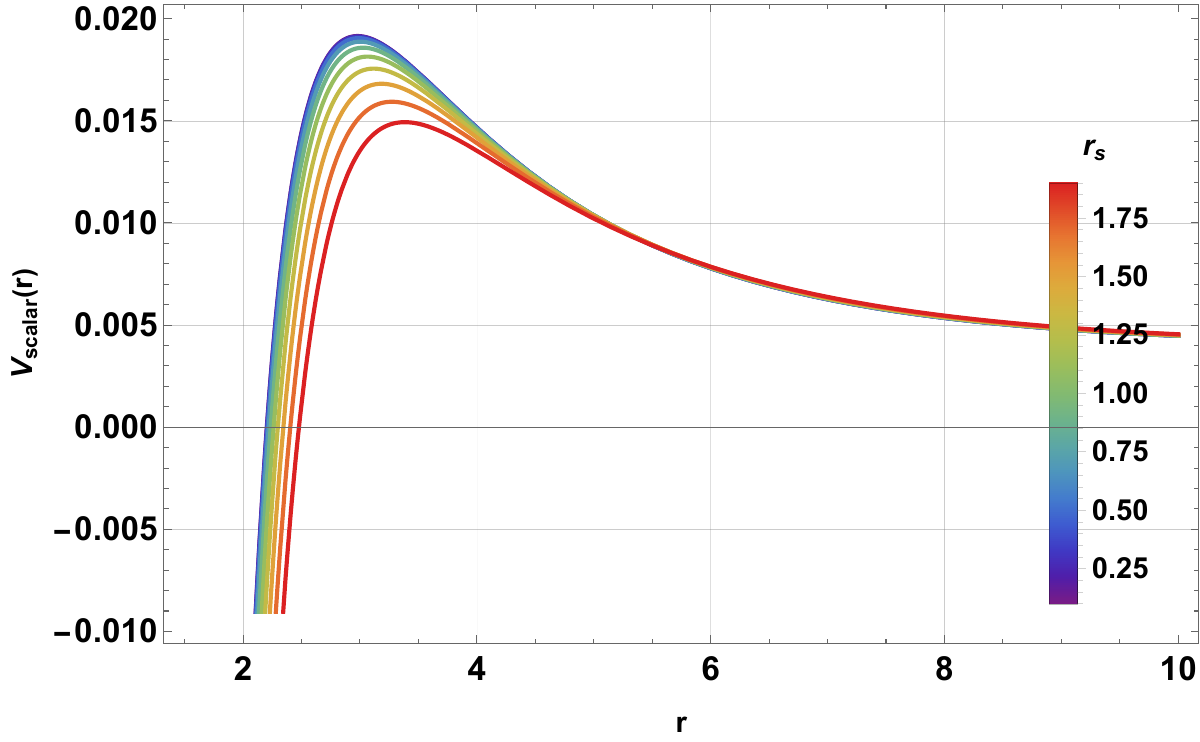}\\
(a) $r_s=0.2,\,\rho_s=0.02$ \hspace{5cm} (b) $\alpha=0.1,\,\rho_s=0.05$
\caption{\footnotesize Behavior of the scalar perturbative potential $V_\text{scalar}$ as a function of $r$ for dominant modes $\ell=0$ by varying the CoS parameter $\alpha$ and halo radius $r_s$. Here, $M=1,\,\ell_p=25$.}
\label{fig:scalar-potential}
\end{figure*}

Since the given space-time is static and spherically symmetric, without loss of generality, we choose the scalar field ansatz as follows:
\begin{equation}
\Phi=e^{-i\,\omega\, t}\,Y_{\ell m}(\theta,\phi)\,\psi(r)/r,\label{SP3}
\end{equation}
where $Y(\theta, \phi)$ is the spherical harmonics and $\omega$ is the QNMs frequency.

Introducing the tortoise coordinate via $dr_\ast=dr/f(r)$, we obtain the Schrödinger-like radial equation
\begin{equation}
    \frac{d^2\psi}{dr_\ast^2}+\left[\omega^2-V_s(r)\right]\psi=0,\label{SP4}
\end{equation}
where the scalar perturbative potential is given by
\begin{equation}
    V_s(r)=f(r)\left(\frac{\ell(\ell+1)}{r^2}+\frac{f'(r)}{r}+\mu^2\right),\quad \ell \geq 0\label{SP5}
\end{equation}
This $V_s(r)$ is the starting point for QNM computations (e.g., WKB or continued fractions).

For a massless scalar field, the scalar perturbative potential reads
\begin{align}
V_s(r)&=\frac{1}{r^2}\,\left(\ell(\ell+1)+\dfrac{2M}{r}+\dfrac{\rho_s r_s^3}{r+r_s}+\dfrac{2 r^2}{\ell_p^2}\right)\notag\\&\times \left(1-\alpha-\frac{2M}{r}-\rho_s\,r_s^2\,\mbox{ln} {\left(1+\frac{r_s}{r}\right)}+\frac{r^2}{\ell^2_p}\right).\label{SP6}
\end{align}

We observe that the scalar perturbative potential is affected by the BH mass $M$, the curvature radius $\ell_p$, the string cloud parameter $\alpha$, and the DM halo profile characterized by $(r_s,\rho_s)$. Moreover, the quantum number $\ell$ alters this perturbative potential.

In Fig. \ref{fig:scalar-potential}, we illustrate the behavior of this perturbative potential as a function of $r$ by varying CoS parameter $\alpha$ and the halo radius $r_s$, while keeping other parameters fixed. In both panels, we observe that this potential reduces with increasing values of both $\alpha$ and $r_s$, indicating the effects of these factors on the propagation of a massless scalar field in the gravitational field of a BH.

Expressing the above potential in terms of dimensionless variables via $x=r/M$, $y=r_s/M$, $k=M/\ell_p$ and writing $\lambda=M^2\,\rho_s$, we find
\begin{align}
M^2\,V_s&=\frac{1}{x^2}\,\left(\ell(\ell+1)+\dfrac{2}{x}+\dfrac{\lambda\,y^3}{x+y}+2\,k^2\,x^2\right)\notag\\ & \times \left(1-\alpha-\frac{2}{x}-\lambda\,y^2\,\mbox{ln} {\left(1+\frac{y}{x}\right)}+k^2\,x^2\right).\label{SP7}
\end{align}

\begin{figure*}[ht!]
\centering
\includegraphics[width=0.45\linewidth]{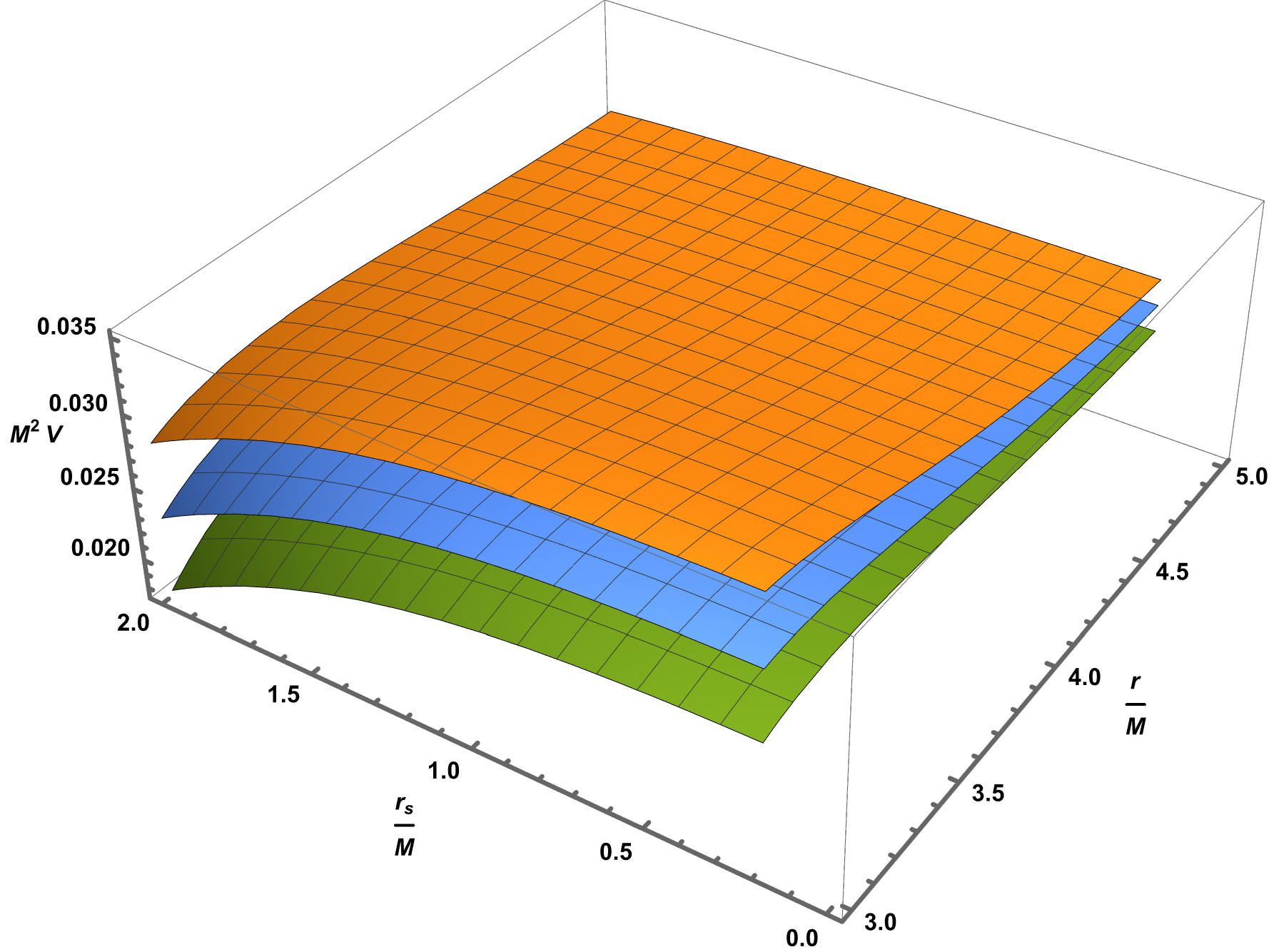}
\caption{\footnotesize Qualitative feature of $M^2\,V_s$ for the dominant mode $\ell=0$: three-dimensional plot of the quantity $M^2\,V_s$ as a function of $r/M$ and $r_s/M$. Here $\lambda=0.05,\,k=0.1$. Green: $\alpha=0.05$, Blue:$\alpha=0.10$, Orange: $\alpha=0.15$.}
\label{fig:3d-plot-potential}
\end{figure*}

\begin{figure*}[ht!]
\centering
\includegraphics[width=0.32\linewidth]{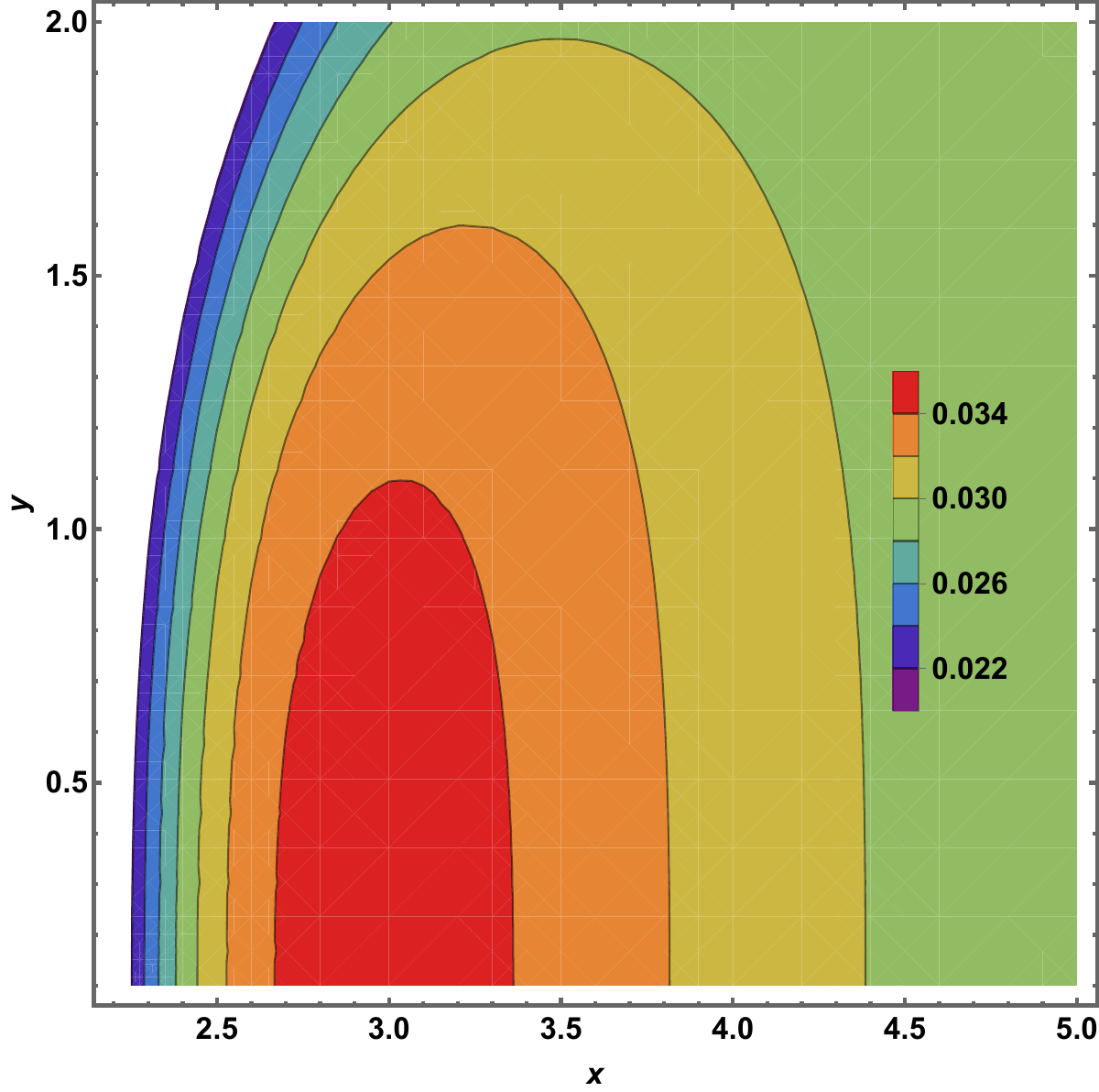}\quad
\includegraphics[width=0.32\linewidth]{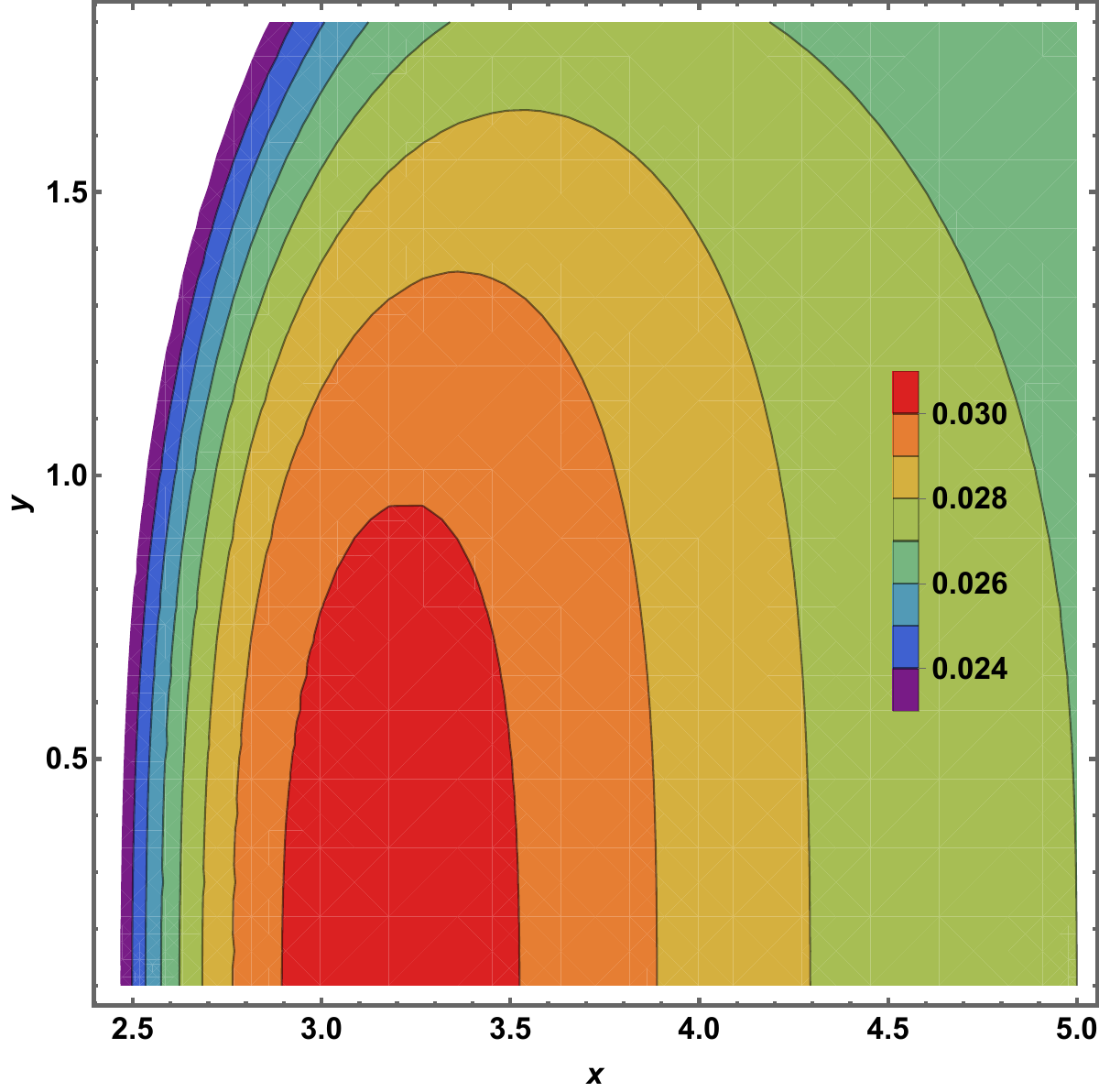}\quad
\includegraphics[width=0.32\linewidth]{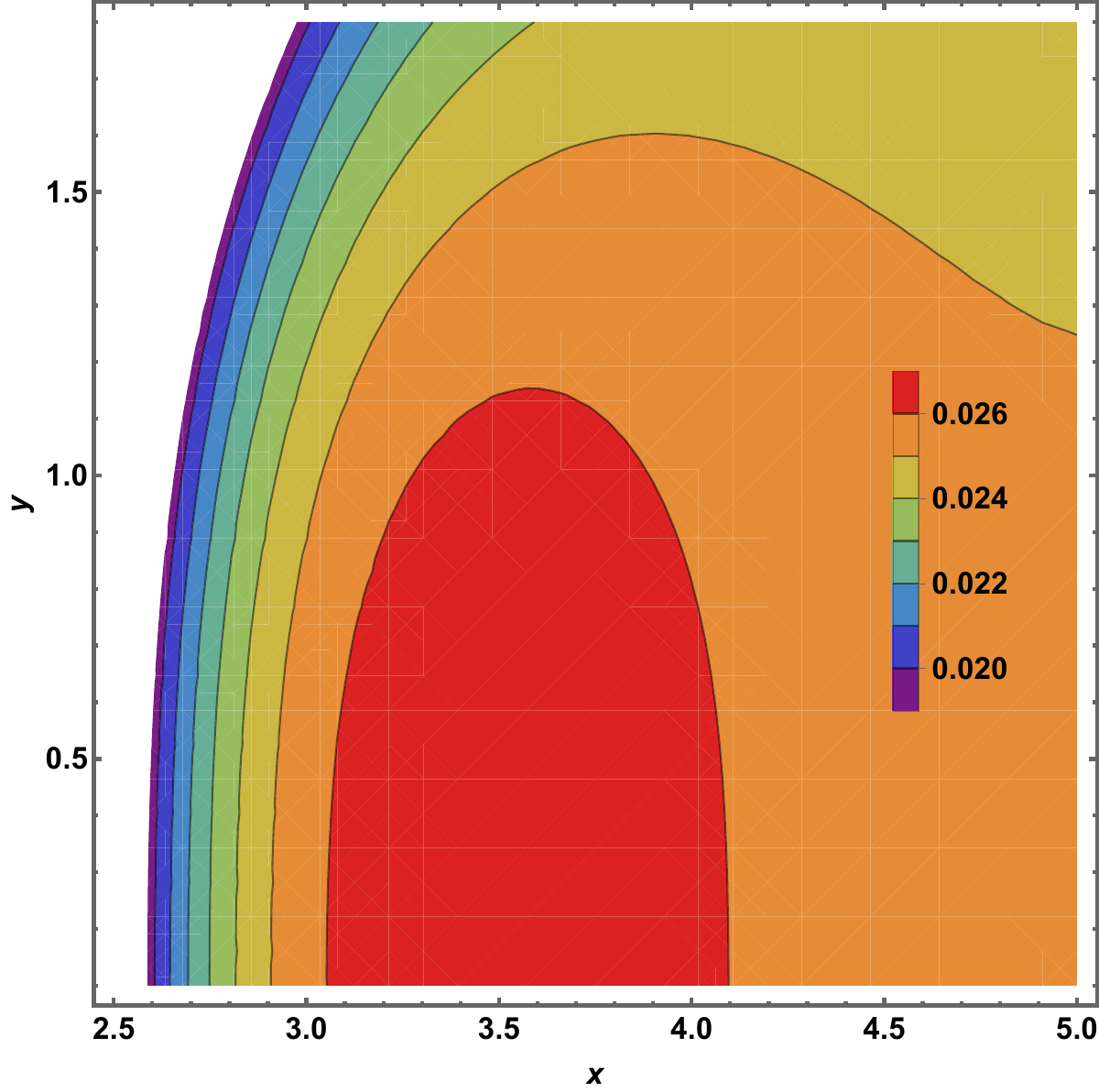}\\
(A) $\alpha=0.05$ \hspace{3cm} (b) $\alpha=0.10$ \hspace{3cm} (c) $\alpha=0.15$
\caption{\footnotesize Contour plot of the scalar perturbative potential for dominant mode $\ell=0$. Here $k=0.1,\,\lambda=0.05$.}
\label{fig:contour-plot}
\end{figure*}

In Fig. \ref{fig:3d-plot-potential}, we depict the qualitative feature of $M^2V$ for the dominant mode $\ell=0$ as a function of $r/M$ and $r_s/M$ for different values of CoS parameter $\alpha$. We observe that as the value of $\alpha$ increases, the quantity $M^2V$ also rises.

In Fig. \ref{fig:contour-plot}, we draw the contour plot of $M^2V$ for the dominant mode $\ell=0$, which corresponds to $s$-state wave by varying CoS parameter $\alpha$.

\subsection{Quasinormal modes (QNMs) spectra}

Quasinormal modes (QNMs) are characteristic damped oscillations that describe how BHs and other compact objects return to equilibrium after being perturbed. These modes are defined by specific boundary conditions: purely ingoing waves at the BH horizon and purely outgoing waves at spatial infinity (or at the AdS boundary in asymptotically AdS space-times). Unlike normal modes in closed systems, QNMs have complex frequencies, where the real part represents the oscillation frequency and the imaginary part encodes the damping rate due to energy loss \cite{konoplya2011,konoplya2}.

QNMs are central to BH spectroscopy and play a crucial role in gravitational wave astronomy, especially in the ringdown phase observed in events like those detected by LIGO and Virgo. The spectrum of QNMs depends only on the parameters of the BH (mass, charge, spin) and the underlying theory of gravity, making them powerful probes for testing general relativity and its alternatives \cite{berti2019,kanti,cardoso}.

In modified gravity theories and extra-dimensional models, QNMs can deviate significantly from general relativity predictions, providing potential observational signatures of new physics \cite{kanti, konoplya3}. Analytical and numerical methods such as the WKB approximation, continued fraction method, and time-domain integration are commonly used to compute QNM spectra \cite{konoplya2, konoplya3}.

\begin{table*}[ht!]
\centering
\begin{tabular}{|c|c|c|c|}
\hline
$r_s$ & $\alpha = 0.05$ & $\alpha = 0.10$ & $\alpha = 0.15$ \\
\hline
0.1 & $0.2057 - 0.3140i$ & $0.1840 - 0.2960i$ & $0.1633 - 0.2781i$ \\
\hline
0.2 & $0.2057 - 0.3139i$ & $0.1840 - 0.2959i$ & $0.1633 - 0.2781i$ \\
\hline
0.3 & $0.2056 - 0.3139i$ & $0.1840 - 0.2959i$ & $0.1633 - 0.2780i$ \\
\hline
0.4 & $0.2056 - 0.3138i$ & $0.1839 - 0.2958i$ & $0.1632 - 0.2779i$ \\
\hline
0.5 & $0.2055 - 0.3136i$ & $0.1838 - 0.2956i$ & $0.1632 - 0.2778i$ \\
\hline
0.6 & $0.2053 - 0.3133i$ & $0.1837 - 0.2954i$ & $0.1630 - 0.2775i$ \\
\hline
0.7 & $0.2051 - 0.3129i$ & $0.1835 - 0.2950i$ & $0.1628 - 0.2772i$ \\
\hline
0.8 & $0.2048 - 0.3125i$ & $0.1832 - 0.2945i$ & $0.1626 - 0.2767i$ \\
\hline
0.9 & $0.2044 - 0.3118i$ & $0.1829 - 0.2940i$ & $0.1623 - 0.2762i$ \\
\hline
1.0 & $0.2040 - 0.3111i$ & $0.1825 - 0.2932i$ & $0.1620 - 0.2755i$ \\
\hline
1.1 & $0.2034 - 0.3102i$ & $0.1820 - 0.2924i$ & $0.1615 - 0.2747i$ \\
\hline
1.2 & $0.2028 - 0.3091i$ & $0.1814 - 0.2914i$ & $0.1610 - 0.2738i$ \\
\hline
1.3 & $0.2021 - 0.3079i$ & $0.1808 - 0.2902i$ & $0.1605 - 0.2726i$ \\
\hline
1.4 & $0.2012 - 0.3065i$ & $0.1800 - 0.2889i$ & $0.1598 - 0.2714i$ \\
\hline
1.5 & $0.2003 - 0.3050i$ & $0.1792 - 0.2874i$ & $0.1590 - 0.2700i$ \\
\hline
1.6 & $0.1992 - 0.3032i$ & $0.1782 - 0.2857i$ & $0.1582 - 0.2684i$ \\
\hline
1.7 & $0.1981 - 0.3013i$ & $0.1772 - 0.2839i$ & $0.1572 - 0.2666i$ \\
\hline
1.8 & $0.1968 - 0.2991i$ & $0.1760 - 0.2818i$ & $0.1562 - 0.2646i$ \\
\hline
1.9 & $0.1954 - 0.2968i$ & $0.1747 - 0.2796i$ & $0.1551 - 0.2625i$ \\
\hline
2.0 & $0.1938 - 0.2943i$ & $0.1734 - 0.2772i$ & $0.1538 - 0.2602i$ \\
\hline
\end{tabular}
\caption{QNM frequencies $\omega$ for various values of the halo $r_s$ under three different $\alpha$ values with $\ell=1$. Here $M=1,\,\rho_s=0.02,\,\ell_p=25$.}
\label{tab:4}
\end{table*}

\begin{table*}[ht!]
\centering
\begin{tabular}{|c|c|c|c|}
\hline
$r_s$ & $\alpha = 0.05$ & $\alpha = 0.10$ & $\alpha = 0.15$ \\
\hline
0.1 & $0.4207 - 0.2834i$ & $0.3858 - 0.2670i$ & $0.3521 - 0.2506i$ \\
\hline
0.2 & $0.4207 - 0.2834i$ & $0.3858 - 0.2670i$ & $0.3520 - 0.2506i$ \\
\hline
0.3 & $0.4206 - 0.2833i$ & $0.3857 - 0.2670i$ & $0.3520 - 0.2506i$ \\
\hline
0.4 & $0.4205 - 0.2832i$ & $0.3856 - 0.2669i$ & $0.3519 - 0.2505i$ \\
\hline
0.5 & $0.4203 - 0.2831i$ & $0.3854 - 0.2667i$ & $0.3517 - 0.2503i$ \\
\hline
0.6 & $0.4199 - 0.2828i$ & $0.3851 - 0.2665i$ & $0.3514 - 0.2501i$ \\
\hline
0.7 & $0.4195 - 0.2825i$ & $0.3846 - 0.2661i$ & $0.3510 - 0.2498i$ \\
\hline
0.8 & $0.4189 - 0.2821i$ & $0.3841 - 0.2657i$ & $0.3505 - 0.2494i$ \\
\hline
0.9 & $0.4182 - 0.2815i$ & $0.3834 - 0.2652i$ & $0.3499 - 0.2489i$ \\
\hline
1.0 & $0.4173 - 0.2808i$ & $0.3826 - 0.2646i$ & $0.3492 - 0.2483i$ \\
\hline
1.1 & $0.4162 - 0.2800i$ & $0.3816 - 0.2638i$ & $0.3483 - 0.2475i$ \\
\hline
1.2 & $0.4149 - 0.2791i$ & $0.3804 - 0.2629i$ & $0.3472 - 0.2467i$ \\
\hline
1.3 & $0.4134 - 0.2780i$ & $0.3791 - 0.2619i$ & $0.3459 - 0.2457i$ \\
\hline
1.4 & $0.4117 - 0.2768i$ & $0.3775 - 0.2607i$ & $0.3445 - 0.2446i$ \\
\hline
1.5 & $0.4098 - 0.2754i$ & $0.3757 - 0.2593i$ & $0.3429 - 0.2433i$ \\
\hline
1.6 & $0.4077 - 0.2738i$ & $0.3738 - 0.2578i$ & $0.3411 - 0.2418i$ \\
\hline
1.7 & $0.4053 - 0.2721i$ & $0.3716 - 0.2562i$ & $0.3391 - 0.2402i$ \\
\hline
1.8 & $0.4027 - 0.2702i$ & $0.3692 - 0.2544i$ & $0.3370 - 0.2385i$ \\
\hline
1.9 & $0.3999 - 0.2681i$ & $0.3666 - 0.2524i$ & $0.3346 - 0.2366i$ \\
\hline
2.0 & $0.3968 - 0.2659i$ & $0.3638 - 0.2502i$ & $0.3320 - 0.2345i$ \\
\hline
\end{tabular}
\caption{QNM frequencies $\omega$ for various values of the halo $r_s$ under three different $\alpha$ values with $\ell=2$. Here $M=1,\,\rho_s=0.02,\,\ell_p=25$.}
\label{tab:5}
\end{table*}

Tables \ref{tab:4}-\ref{tab:5} present the numerical values of QNM frequencies calculated using the third-order WKB approximation for the modes $\ell=1$ and $\ell=2$. These results are obtained by varying the CoS parameter $\alpha$ and the halo radius $r_s$, while keeping all other parameters fixed. From the data, it is evident that both $\alpha$ and $r_s$ significantly affect the QNM spectra, with the frequencies decreasing as either parameter increases.

\section{Thermodynamics}\label{Sec:VI}

Black-hole thermodynamics provides a powerful bridge between geometry, quantum field theory, and statistical mechanics. The area law and Hawking radiation established BHs as genuine thermodynamic systems with entropy and temperature~\cite{Hawking1975}. In asymptotically AdS spacetimes, the existence of a well-defined canonical ensemble yields rich phase structure, including the Hawking-Page transition between thermal AdS and large AdS BHs~\cite{HawkingPage1983}. In the modern extended framework, the cosmological constant is promoted to a thermodynamic pressure and the ADM mass becomes enthalpy, completing a consistent first law and Smarr relation with a geometric thermodynamic volume~\cite{Dolan2011,CveticEtAl2011}. These ingredients underlie Van der Waals-like criticality in AdS BHs and associated swallow-tail phenomenology in Gibbs free energy~\cite{KubiznakMannTeo2017}.

In our geometry, the cloud-of-strings (CoS) parameter $\alpha $\cite{PSL} and the dark-matter (DM) halo parameters $(\rho_s,r_s)$ deform the lapse function, thereby shifting temperature extrema, heat-capacity divergences, and possible Hawking-Page/first-order transitions. We therefore present closed-form expressions for $T$, $S$, $M$ (enthalpy), $V$, and response functions, and we discuss their qualitative consequences, emphasizing how $\alpha$ and $(\rho_s,r_s)$ renormalize the effective ``attractive'' piece in the equation of state. We also align our analysis with recent developments on noncommutative/extended thermodynamics, e.g., criticality and Joule-Thomson expansion, highlighting which conclusions remain universal and which are model-dependent \cite{ElHadriJemri2025,OkcuAydiner2017,Johnson2014,WeiLiuPRL2015,WeiLiuPRD2019,Cong2021}.

\subsection{Horizon data and primary thermodynamic quantities}\label{Sec:thermo-1}

The event horizon $r_h$ is the largest real root in the lapse function and can be determined using the following condition:
\begin{equation}
f(r_h)=1-\alpha-\frac{2M}{r_h}-\rho_s r_s^2\ln\!\Big(1+\frac{r_s}{r_h}\Big)+\frac{r_h^2}{\ell_p^2}=0.
\label{SS1}
\end{equation}
This allows us to express the mass parameter $M$ as a function of $r_h$,
\begin{equation}
M=\frac{r_h}{2}\left[\,1-\alpha-\rho_s r_s^2\ln\!\Big(1+\frac{r_s}{r_h}\Big)+\frac{r_h^2}{\ell_p^2}\right].
\label{SS2}
\end{equation}

In geometrized units, the surface gravity $\kappa=\tfrac12 f'(r_h)$ yields the Hawking temperature \cite{Hawking1975}:
\begin{align}
T
&=\frac{f'(r_h)}{4\pi}
=\frac{1}{4\pi}\left[\frac{2M}{r_h^2}+\frac{\rho_s r_s^3}{r_h(r_h+r_s)}+\frac{2r_h}{\ell_p^2}\right]\nonumber\\
&=\frac{1}{4\pi r_h}\left[1-\alpha
-\rho_s r_s^2 \ln\!\Big(1+\frac{r_s}{r_h}\Big)
+\frac{\rho_s r_s^3}{r_h+r_s}
+\frac{3\,r_h^2}{\ell_p^2}\right],
\label{SS3}
\end{align}
where in the second line we eliminated $M$ using \eqref{SS2}. The Bekenstein-Hawking entropy and the (horizon) area read \cite{Bekenstein1973}
\begin{equation}
S=\frac{\mathcal A_h}{4}=\pi r_h^2,
\qquad \mathcal A_h=4\pi r_h^2.
\label{SS4}
\end{equation}

We observe that the Hawking temperature depends on the curvature radius $\ell_p$, the string cloud parameter $\alpha$, and the DM halo profile characterized by $(\rho_s,r_s)$.

\begin{figure*}[ht!]
\centering
\includegraphics[width=0.45\linewidth]{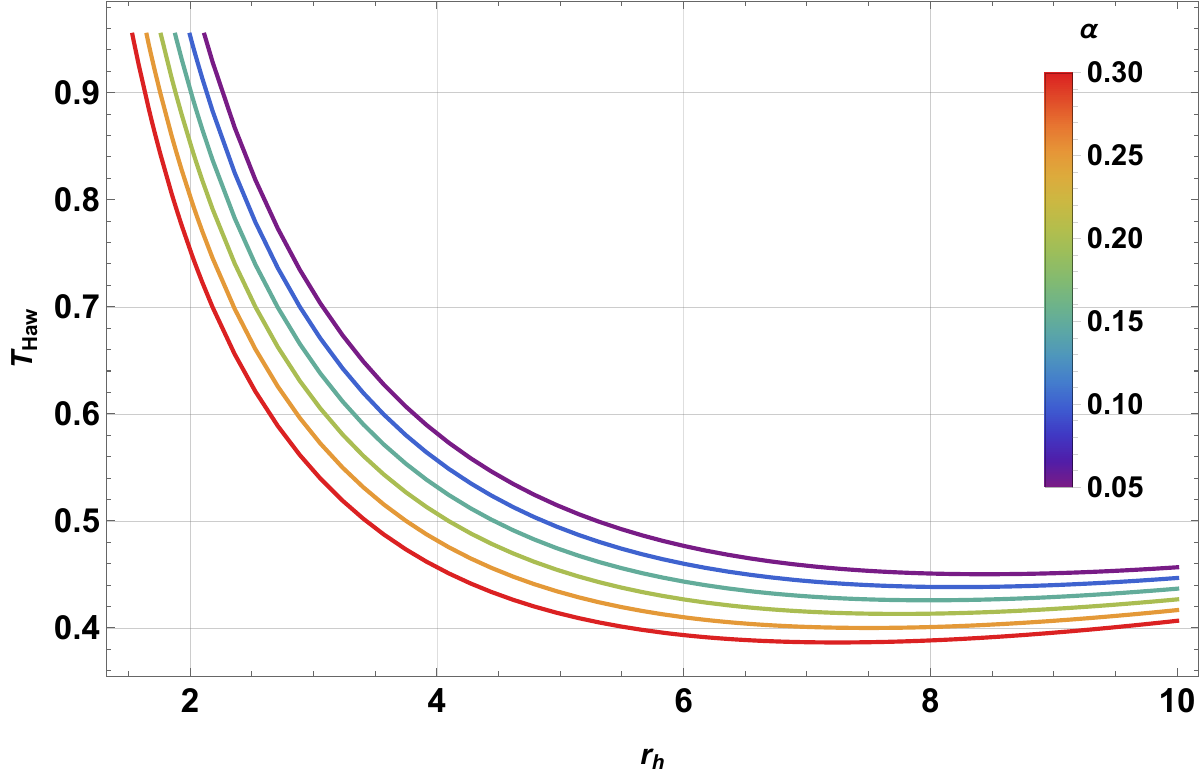}\qquad
\includegraphics[width=0.45\linewidth]{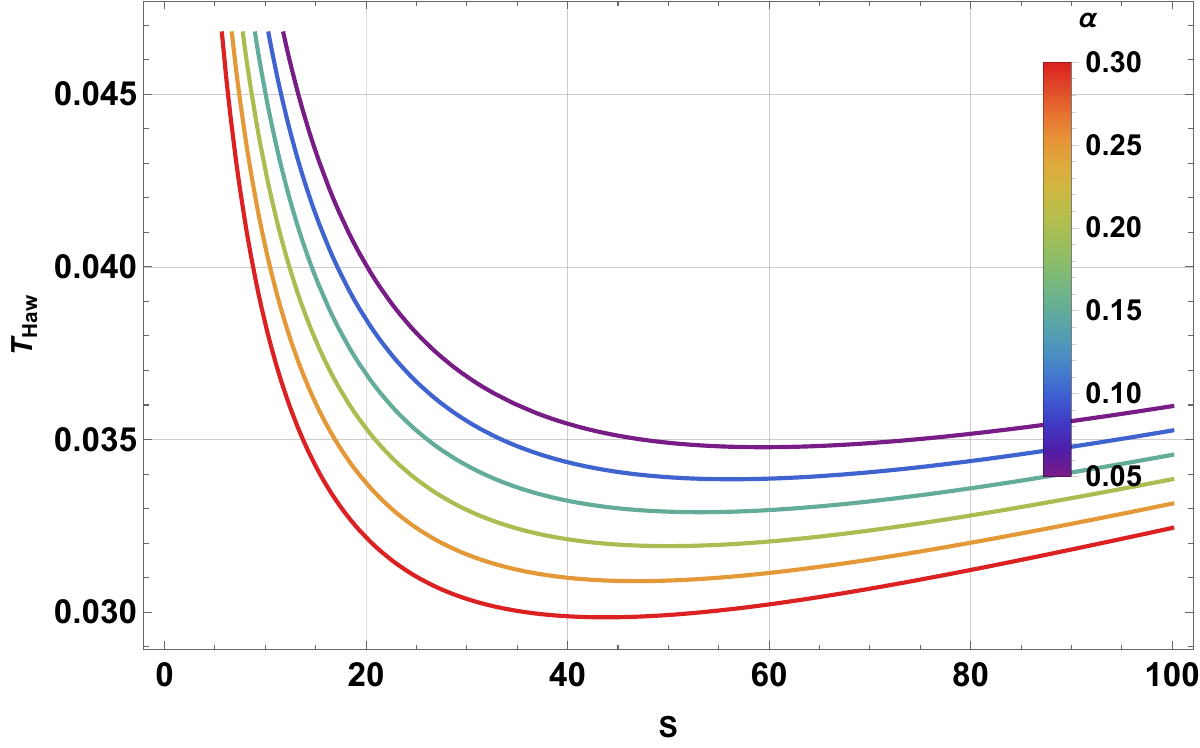}\\
(a) $\ell_p=15$ \hspace{6cm} (b) $P=0.002$
\caption{\footnotesize Behavior of the Hawking temperature $T_\text{Haw}$ as a function of horizon $r_h$  (left panel) and entropy $S$ (right panel)  by varying the CoS parameter $\alpha$. Here, $M=1,,r_s=0.2,,\rho_s=0.02$.}
\label{fig:temperature-1}
\includegraphics[width=0.45\linewidth]{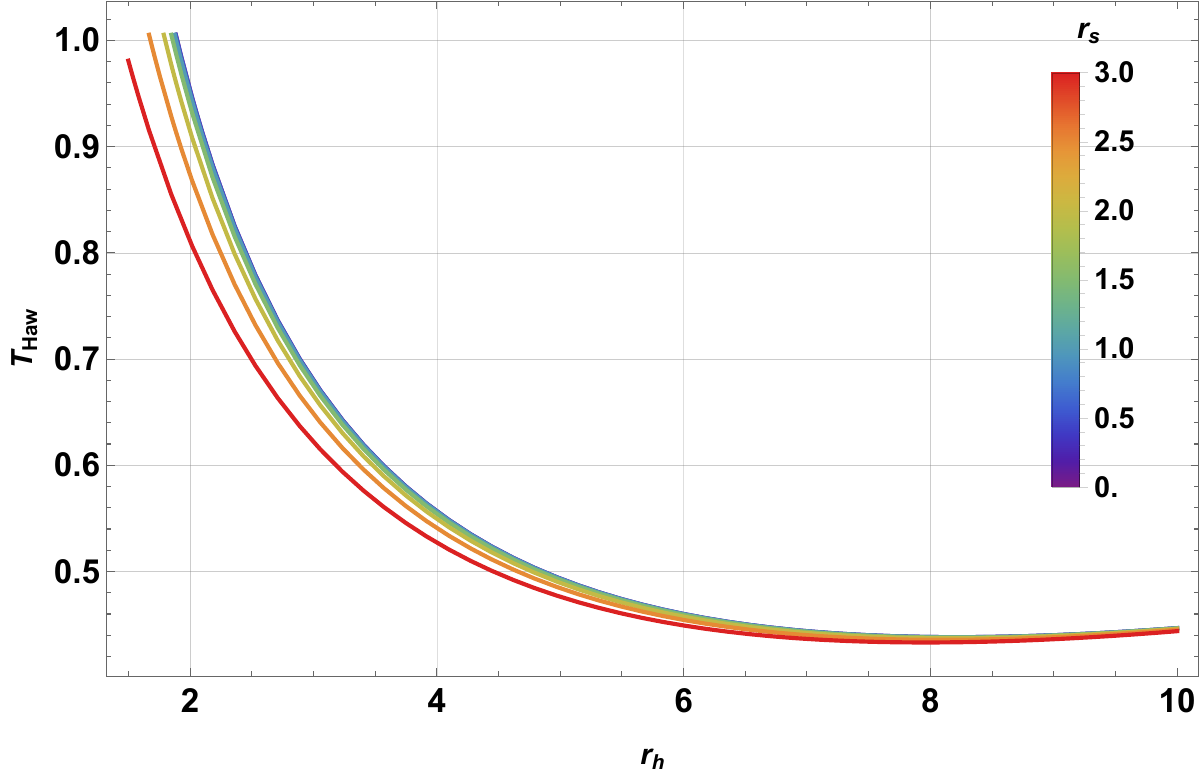}\qquad
\includegraphics[width=0.45\linewidth]{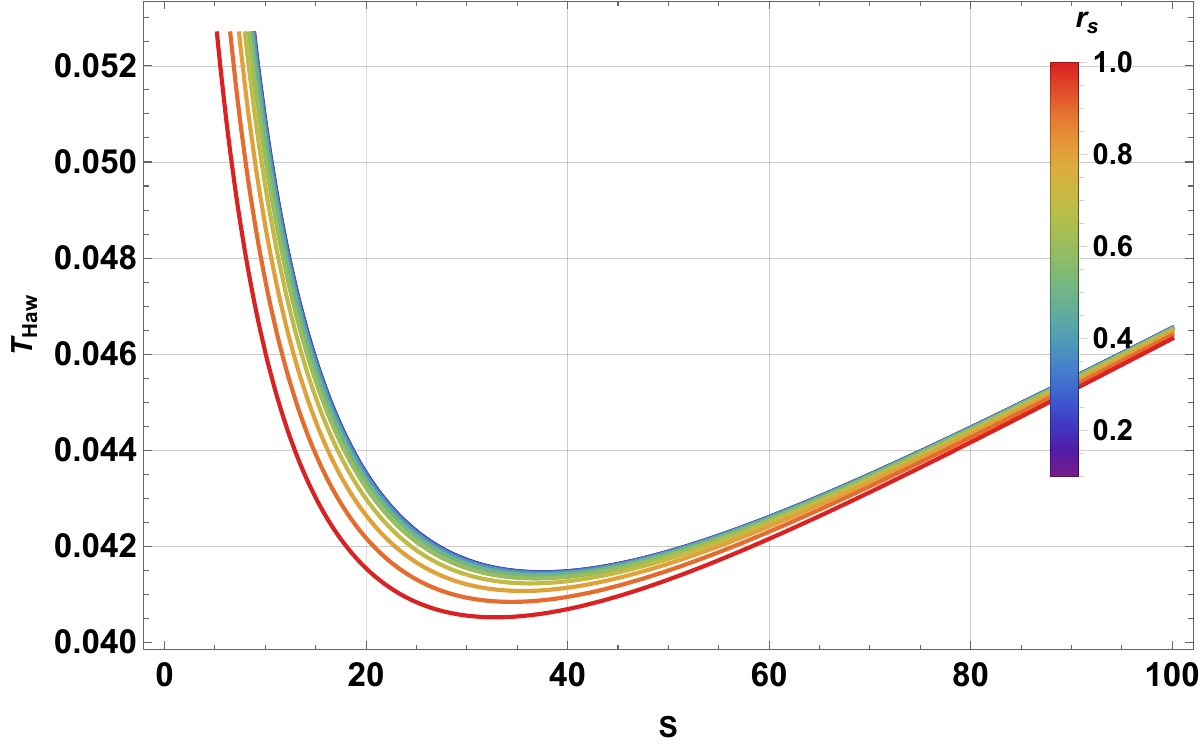}\\
(a) $\ell_p=15$ \hspace{6cm} (b) $P=0.003$
\caption{\footnotesize Behavior of the Hawking temperature $T_\text{Haw}$ as a function of horizon $r_h$ (left panel) and entropy $S$ (right panel) by varying the halo radius $r_s$. Here, $M=1,,\alpha=0.1,,\rho_s=0.05$.}
\label{fig:temperature-2}
\end{figure*}

In Fig. \ref{fig:temperature-1} (a), at fixed $(r_s,\rho_s,\ell_p)$, increasing $\alpha$ uniformly lowers $T(r_h)$ across all $r_h$, reflecting the effective deficit-angle contribution of the string cloud to the lapse; the zero of $T$ (if present) shifts to larger $r_h$ and the minimum of $T$ (where $dT/dr_h=0$) moves accordingly. In Figure \ref{fig:temperature-1} (b) (fixed $P$), the same trend persists, but the $+3r_h/\ell_p^2$ term enhances the growth at large $r_h$, so curves become nearly linear for $r_h\gg r_s$. 

In Fig. \ref{fig:temperature-2}, plotting $T$ versus $S=\pi r_h^2$ makes the small/large-$r_h$ behavior transparent: near small $S$, the halo term $-\rho_s r_s^2\ln(1+r_s/r_h)$ dominates and depresses $T$, while at large $S$ the linear-in-$S^{1/2}$ AdS contribution drives $T$ upward. Increasing $r_s$ mimics a stronger DM backreaction and pushes the unstable small-$r_h$ branch to larger $r_h$. These qualitative deformations are analogous to how extra matter sectors (charge, rotation, noncommutative smearing) shift temperature extrema in other AdS BHs~\cite{KubiznakMann2012,ElHadriJemri2025}.

\subsection{Extended thermodynamics and first law}

Extended thermodynamics is a framework in BH physics that incorporates the cosmological constant ($\Lambda=-3/\ell^2_p$) as a dynamical thermodynamic variable. In this context, $\Lambda$ is associated with pressure via the relation \cite{KastorRayTraschen2009,Dolan2011},
\begin{equation}
P\equiv -\Lambda=\frac{3}{\ell_p^2},
\label{SS5}
\end{equation}
which implies that a negative cosmological constant (as in AdS spacetimes) corresponds to a positive pressure. This approach extends the traditional first law of BH thermodynamics to include a pressure-volume term, making it analogous to the thermodynamics of ordinary systems. The BH mass is then interpreted as enthalpy ($\mathcal{H}$) rather than internal energy. Extended thermodynamics provides deeper insights into BH phase transitions, critical phenomena, and holographic dualities.

Thereby, the mass $M$ from Eq.~\eqref{SS2} in terms of pressure can be expressed (with $8\pi=1$) as
\begin{align}
&M(r_h,P,\alpha,\rho_s,r_s)=\frac{r_h}{2}\notag\\& \times \left[\,1-\alpha-\rho_s r_s^2\ln\!\Big(1+\frac{r_s}{r_h}\Big)+\frac{P}{3}\,r_h^2\right].
\label{SS6}
\end{align}
Moreover, in terms of entropy $S$, the BH mass $M$ can be written as
\begin{align}
&M(S,P,\alpha,\rho_s,r_s)=\frac{1}{2}\sqrt{\frac{S}{\pi}} \notag\\ & \times\left[\,1-\alpha-\rho_s r_s^2\ln\!\Big(1+r_s\sqrt{\frac{\pi}{S}}\Big)+\frac{P}{3}\,S\right].
\label{SS7}
\end{align}

\begin{figure*}[ht!]
\centering
\includegraphics[width=0.45\linewidth]{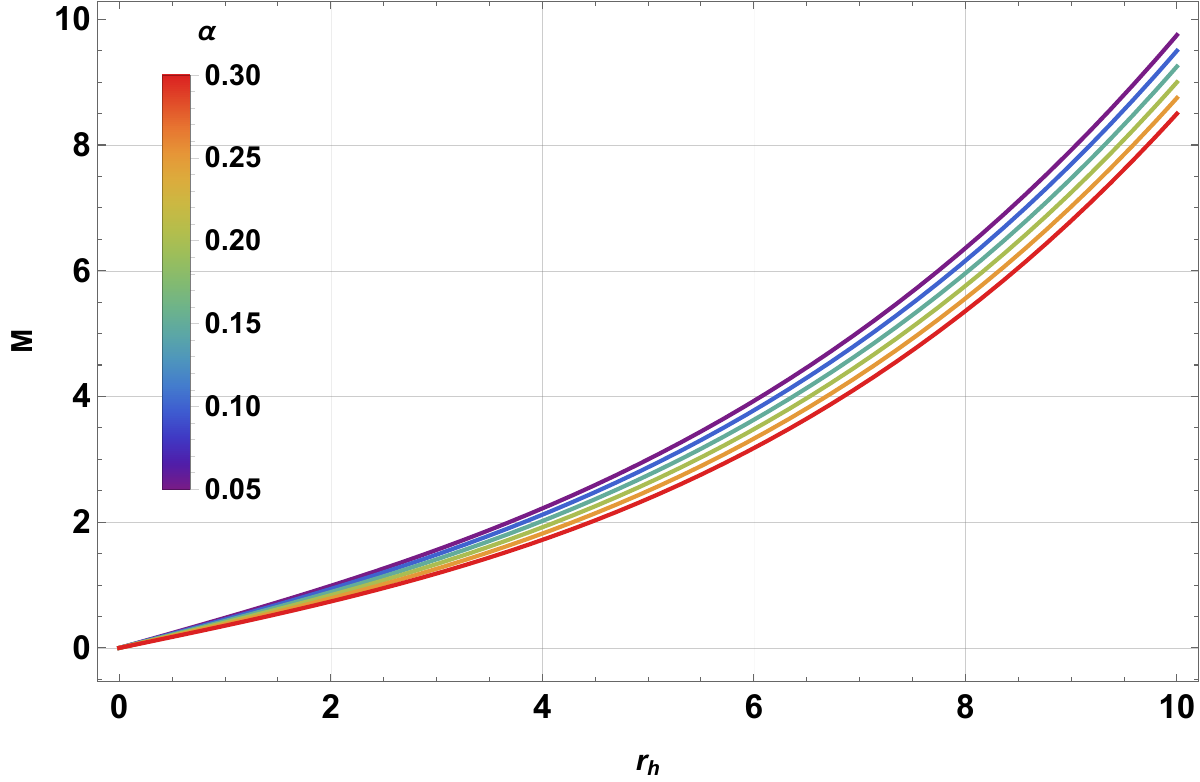}\qquad
\includegraphics[width=0.45\linewidth]{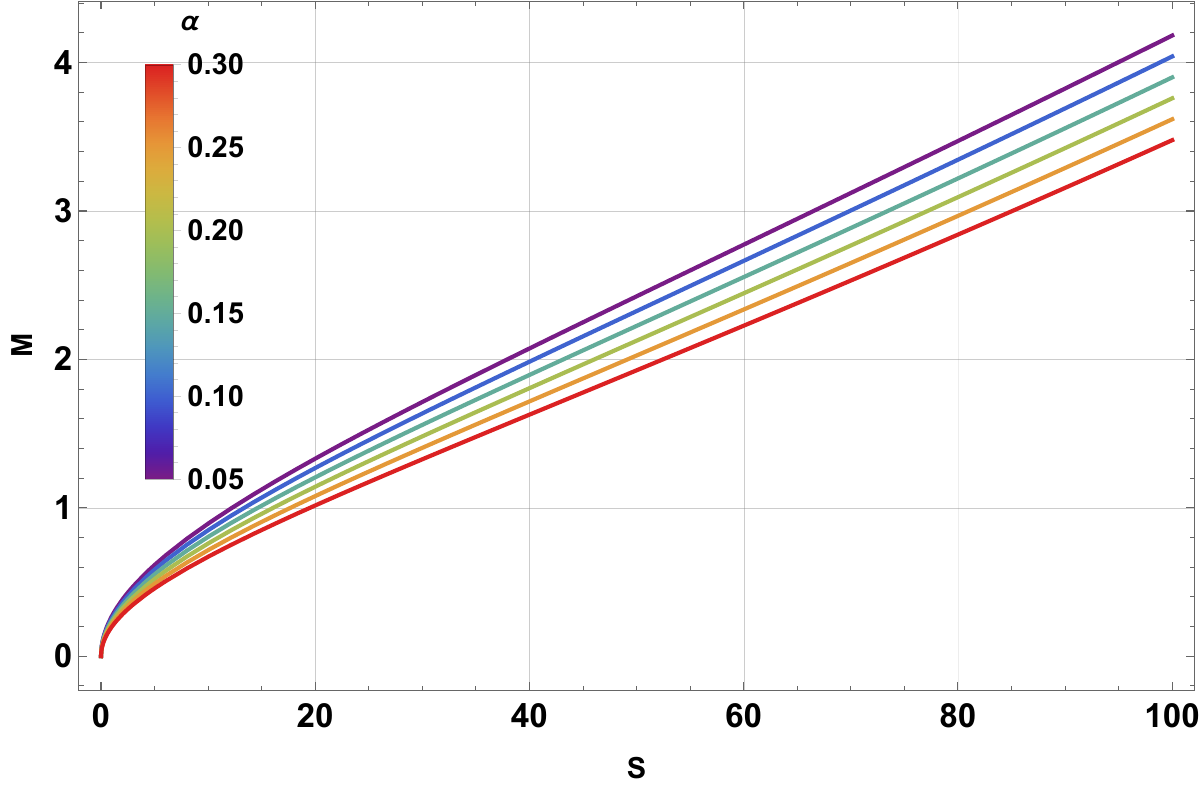}\\
(a) $\ell_p=10$ \hspace{6cm} (b) $P=0.002$
\caption{\footnotesize Behavior of the BH mass $M$ as a function of horizon $r_h$ (left panel) and entropy $S$ (right panel) by varying the CoS parameter $\alpha$. Here, $M=1,,r_s=0.2,,\rho_s=0.02$.}
\label{fig:mass-1}
\includegraphics[width=0.45\linewidth]{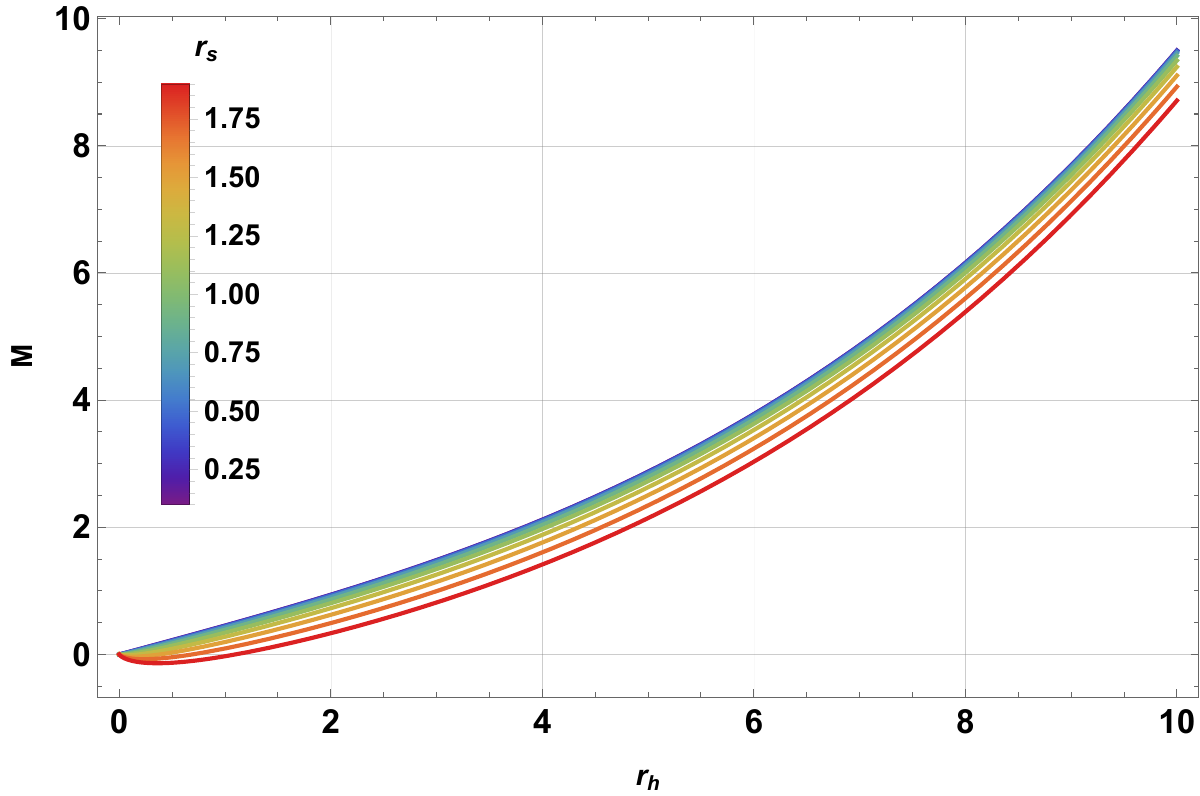}\qquad
\includegraphics[width=0.45\linewidth]{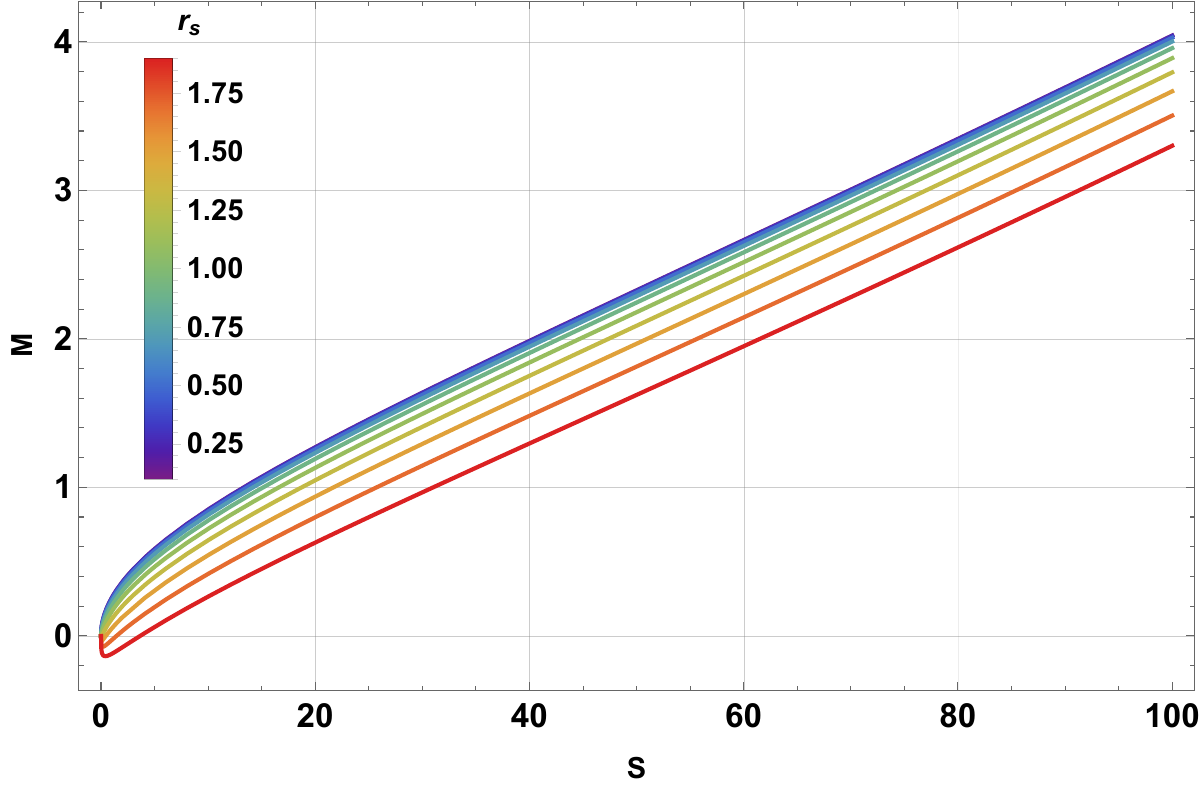}\\
(a) $\ell_p=10$ \hspace{6cm} (b) $P=0.002$
\caption{\footnotesize Behavior of the BH mass $M$ as a function of horizon $r_h$ (left panel) and entropy $S$ (right panel) by varying the halo radius $r_s$. Here, $M=1,,\alpha=0.1,,\rho_s=0.01$.}
\label{fig:mass-2}
\end{figure*}

The enthalpy $M$ inherits a linear growth $\sim r_h^3/\ell_p^2$ at large radius due to the $P,V$ contribution, while the CoS/DM sector produces a negative offset through the logarithmic halo term. In Fig. \ref{fig:mass-1} (a), larger $\alpha$ lowers $M(r_h)$ for fixed $(P,\rho_s,r_s)$, consistent with $\Theta_\alpha<0$ below. In Fig.~\ref{fig:mass-1} (b), plotting $M$ against $S$ highlights the enthalpic nature of AdS BHs: the volume term $\propto P S^{3/2}$ dominates at large $S$, whereas the matter-sector work terms control the small-$S$ behavior.

In Fig. \ref{fig:mass-2}, we generate graphs showing the behavior of BH mass $M$ as a function of horizon $r_h$ and entropy $S$ by varying the halo radius $r_s$. Here also, we observe a similar trend in the BH mass as in Fig. \ref{fig:mass-1}.

In the BH mass expression given by Eq.~(\ref{SS7}), considering the cloud of strings parameter $\alpha$ and DM halo profile parameters ($r_s, \rho_s$) as extensive thermodynamic parameters, we can write the first law of BH thermodynamics as \cite{JMT1,JMT2,JMT3}
\begin{equation}
dM=T\,dS+V\,dP+\Theta_{\alpha}\,d\alpha+\Theta_{\rho}\,d\rho_s+\Theta_{r_s}\,dr_s,
\label{SS8}
\end{equation}
where the intensive thermodynamic variables conjugate to the parameters $\alpha$, $r_s$ and $\rho_s$, respectively, are given by
\begin{widetext}
\begin{align}
\Theta_{\alpha}&\equiv\left(\frac{\partial M}{\partial \alpha}\right)_{S,P,\rho_s,r_s}=-\frac{1}{2}\sqrt{\frac{S}{\pi}}=-\frac{r_h}{2},\label{SS9}\\[2pt]
\Theta_{\rho_s}&\equiv\left(\frac{\partial M}{\partial \rho_s}\right)_{S,P,\alpha,r_s}=-4\pi r_s^2 \sqrt{\frac{S}{\pi}}\, \ln\!\left(1 + r_s \sqrt{\frac{\pi}{S}} \right)
=-4\pi r_h r_s^2\,\ln\!\Big(1+\frac{r_s}{r_h}\Big),\label{SS10}\\[2pt]
\Theta_{r_s}&\equiv\left(\frac{\partial M}{\partial r_s}\right)_{S,P,\alpha,\rho_s}=-4\pi \rho_s \sqrt{\frac{S}{\pi}} \left[ 2r_s \ln\!\left(1 + r_s \sqrt{\frac{\pi}{S}} \right) + \frac{r_s^2}{\sqrt{\frac{S}{\pi}} + r_s} \right]
=-4\pi r_h \rho_s\!\left[2\,r_s\ln\!\Big(1+\frac{r_s}{r_h}\Big)+\frac{r_s^2}{r_h+r_s}\right],\label{SS11}\\[2pt]
T &\equiv \left(\frac{\partial M}{\partial S}\right)_{P,\alpha,r_s,\rho_s}=\frac{1}{4\sqrt{\pi S}} \left[ 1 - \alpha - \rho_s r_s^2 \ln\!\left(1 + r_s \sqrt{\frac{\pi}{S}} \right) \right] + \frac{2\pi \rho_s r_s^3}{S \left(1 + r_s \sqrt{\frac{\pi}{S}} \right)} + \frac{2P \sqrt{S}}{\sqrt{\pi}}.\label{SS12}
\end{align}
\end{widetext}
Moreover, the thermodynamic volume is given by
\begin{equation}
V \equiv \left(\frac{\partial M}{\partial P}\right)_{S,\alpha,r_s,\rho_s} =\frac{4 S^{3/2}}{3 \sqrt{\pi}}=\frac{4\pi}{3}\,r_h^3,
\label{SS13}
\end{equation}
which coincides with the geometric volume for static, spherically symmetric BHs \cite{KastorRayTraschen2009}.

Now, using \eqref{SS3}, entropy $S=\pi r_h^2$, thermodynamic volume $V=\tfrac{4\pi}{3}r_h^3$ and thermodynamic pressure $P=\tfrac{3}{\ell_p^2}$, we find
\begin{equation}
2TS-2PV=\frac{r_h}{2}\,\Big[1-\alpha-\rho_s r_s^2\ln\!\Big(1+\frac{r_s}{r_h}\Big)+\frac{\rho_s r_s^3}{r_h+r_s}\Big]+\frac{r_h^3}{2\ell_p^2}.
\label{SS18}
\end{equation}
Comparing with BH mass $M=\dfrac{r_h}{2}\Big[1-\alpha-\rho_s r_s^2\ln\!\Big(1+\frac{r_s}{r_h}\Big)+\dfrac{r_h^2}{\ell_p^2}\Big]$, we obtain
\begin{equation}
M=2TS-2PV-\frac{r_h}{2}\,\frac{\rho_s r_s^3}{\,r_h+r_s\,}.
\label{SS19}
\end{equation}
Thus, DM halo supplies a finite additional (work-like) contribution compared to the standard form. In the limit at $r_s \to 0$, one recovers the standard Smarr formula, $M=2TS-2PV$ \cite{KastorRayTraschen2009}.

The signs $\Theta_\alpha<0$ and $\Theta_{\rho_s}<0$ indicate that increasing the CoS density or the halo density at fixed $(S,P)$ reduces the enthalpy $M$, i.e., the matter sector \emph{does work on} the spacetime. The mixed term $\Theta_{r_s}$ shows that enlarging the halo size at fixed $\rho_s$ also lowers $M$, with a rational correction $\propto (r_h+r_s)^{-1}$ tied to the finite halo mass. These features mirror the role of $Q$ in RN-AdS and deformation parameters in noncommutative BHs, where additional couplings enter as natural work terms in the first law~\cite{ElHadriJemri2025}.

\subsection{Equation of state and criticality}

Combining \eqref{SS5} with \eqref{SS3} yields an equation of state $P=P(T,r_h)$,
\begin{equation}
P(T,r_h)=\frac{T}{2r_h}-\frac{1}{r_h^2}
\Bigg[(1-\alpha)-\rho_s r_s^2\ln\!\Big(1+\frac{r_s}{r_h}\Big)
+\frac{\rho_s r_s^3}{r_h+r_s}\Bigg].
\label{SS14}
\end{equation}
Introducing the specific volume $v\equiv 2r_h$, one can analyze isotherms $P(v)$ and search for mean-field criticality via $(\partial P/\partial r_h)_T=(\partial^2 P/\partial r_h^2)_T=0$. In the $\alpha,\rho_s\to 0$ limit, the Van der Waals picture of charged AdS BHs \cite{KubiznakMann2012} is recovered; here, the CoS/DM sector shifts the attractive $1/r_h^2$-like contributions through the square brackets, thereby deforming the location (and possibly the existence) of critical points.

\paragraph*{Criticality diagnostics and coexistence.} Define
\begin{equation}
A(r_h)\equiv (1-\alpha)\;-\;\rho_s r_s^2\,\ln\!\Big(1+\frac{r_s}{r_h}\Big)\;+\;\frac{\rho_s r_s^3}{r_h+r_s},
\end{equation}
so that
\begin{equation}
P(T,r_h)=\frac{T}{2r_h}-\frac{A(r_h)}{r_h^{2}}.
\end{equation}
Inflection-point conditions at fixed temperature,
\begin{equation}
\left(\frac{\partial P}{\partial r_h}\right)_{T}=0,
\qquad
\left(\frac{\partial^{2} P}{\partial r_h^{2}}\right)_{T}=0,
\end{equation}
determine $(T_c,r_c)$ and reduce to the RN-AdS values when $A(r_h)\to 1$~\cite{KubiznakMann2012}. Because $A'(r_h)<0$ for small $r_h$ (halo-enhanced attraction) and $A'(r_h)\to 0$ for $r_h\gg r_s$, the coexistence region widens and can shift to larger radii as $(\rho_s,r_s)$ increases. Along the small/large-BH coexistence curve, the Clausius-Clapeyron relation
\begin{equation}
\frac{dP}{dT}=\frac{\Delta S}{\Delta V}
\end{equation}
gives, using $S=\pi r_h^2$ and $V=\tfrac{4\pi}{3}r_h^3$,
\begin{equation}
\frac{dP}{dT}
=\frac{\pi\big(r_{\mathrm L}^{2}-r_{\mathrm S}^{2}\big)}{\tfrac{4\pi}{3}\big(r_{\mathrm L}^{3}-r_{\mathrm S}^{3}\big)}
=\frac{3}{4}\,\frac{r_{\mathrm L}+r_{\mathrm S}}{\,r_{\mathrm L}^{2}+r_{\mathrm L}r_{\mathrm S}+r_{\mathrm S}^{2}\,},
\end{equation}
a universal geometric expression independent of $A(r_h)$ (the latter only selects which pair $(r_{\mathrm S},r_{\mathrm L})$ coexist)~\cite{KubiznakMann2012,WeiLiuPRL2015}. Maxwell equal-area construction on the $P$-$V$ plane is likewise applicable.

\subsection{Heat capacity and local stability}

The heat capacity governs local thermal stability in the canonical ensemble at fixed $P$
   
\begin{widetext}
\begin{equation}
C_{P}=T\left(\frac{\partial S}{\partial T}\right)_{P}
= -\frac{2\pi\left[1-\alpha
-\rho_s r_s^2 \ln\!\Big(1+\frac{r_s}{r_h}\Big)
+\frac{\rho_s r_s^3}{r_h+r_s}
+ P\,r_h^2\right]}{\left[1-\alpha
-\rho_s r_s^2\ln\!\Big(1+\frac{r_s}{r_h}\Big)
+\frac{\rho_s r_s^3\,r_h}{(r_h+r_s)^2}
- P\,r_h^2\right]}.
\label{SS15}
\end{equation}
\end{widetext}
Divergences of $C_P$ at
\[
1-\alpha
-\rho_s r_s^2\ln\!\Big(1+\frac{r_s}{r_h}\Big)
+\frac{\rho_s r_s^3\,r_h}{(r_h+r_s)^2}
-\frac{3r_h^2}{\ell_p^2}=0
\]
mark continuous phase transitions between locally unstable ($C_P<0$) and stable ($C_P>0$) branches. In the pure SAdS limit, one recovers the standard pattern where small BHs are unstable and large ones are stable; here, $\alpha>0$ and the DM halo shifts the transition radius.

In terms of entropy $S$, the specific heat capacity becomes   
\begin{widetext}
\begin{equation}
C_{P}=-\frac{2\pi \left[1 - \alpha - \rho_s r_s^2 \ln\!\left(1 + r_s \sqrt{\frac{\pi}{S}}\right) + \frac{\rho_s r_s^3}{\sqrt{\frac{S}{\pi}} + r_s} +  P S\right]}{\left[1 - \alpha - \rho_s r_s^2 \ln\!\left(1 + r_s \sqrt{\frac{\pi}{S}}\right) + \frac{\rho_s r_s^3 \sqrt{\frac{S}{\pi}}}{\left(\sqrt{\frac{S}{\pi}} + r_s\right)^2} -  P S\right]}.
\label{SS15a}
\end{equation}
\end{widetext}
\begin{figure*}[ht!]
\centering
\includegraphics[width=0.45\linewidth]{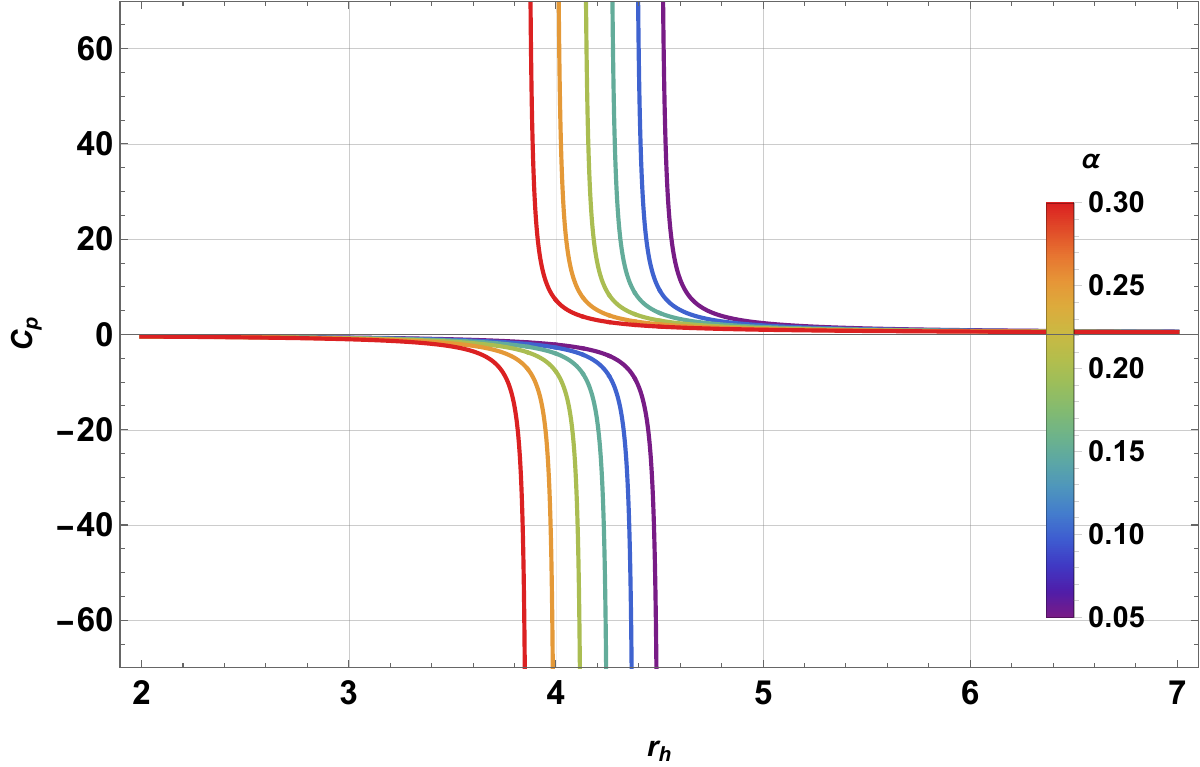}\qquad
\includegraphics[width=0.46\linewidth]{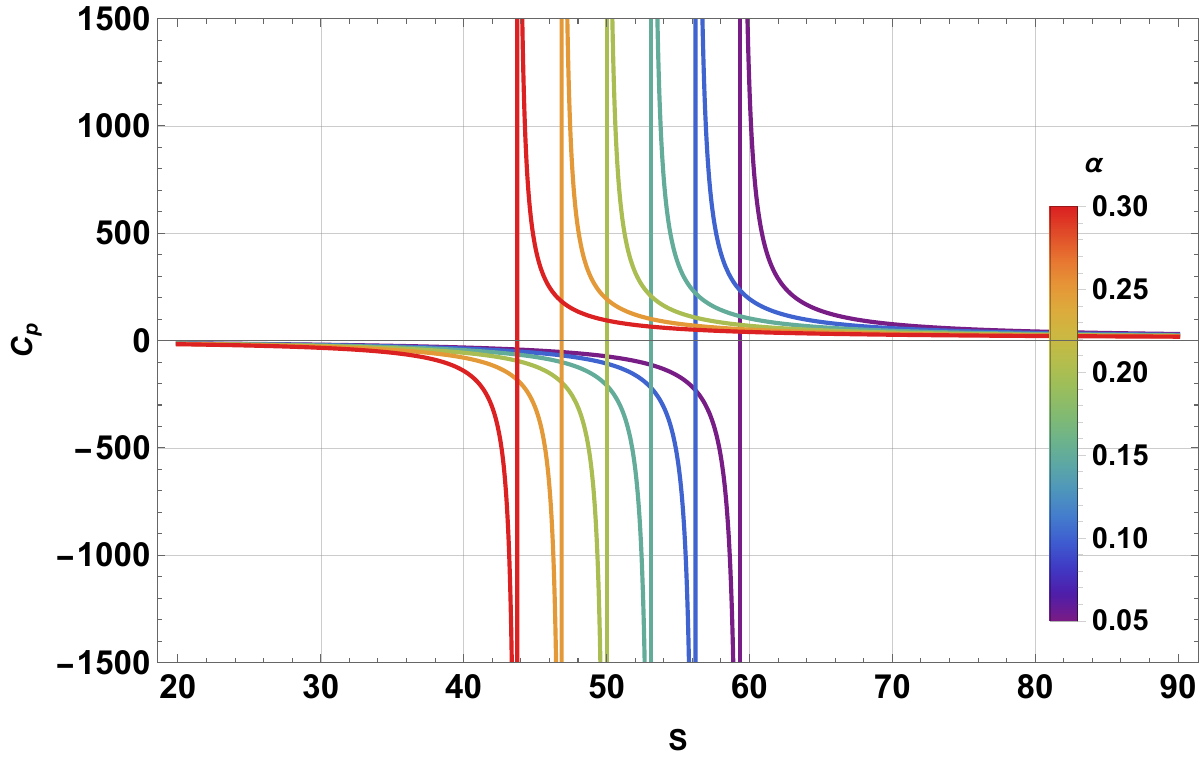}\
(a) $\ell_p=8$ \hspace{6cm} (b) $P=0.002$
\caption{\footnotesize Behavior of the specific heat capacity $C_p$ as a function of horizon $r_h$ (left panel) and entropy $S$ (right panel) by varying CoS parameter $alpha$. Here, $M=1,,r_s=0.5,,\rho_s=0.05$.}
\label{fig:capacity}
\end{figure*}

In Fig. \ref{fig:capacity}, the sign of $C_P$ in panel (a) tracks local stability: a negative dip between two positive branches signals the usual small/large-BH structure separated by a divergence at the spinodal radius. Raising $\alpha$ (or $r_s$) shifts the divergence rightwards and reduces the stable-small-BH window, in line with the temperature trends. Panel (b) shows the same physics in the $(C_P,S)$ plane; the denominator zero marks the onset of instability. These patterns mirror those seen in RN-AdS and in noncommutative AdS families \cite{KubiznakMann2012,ElHadriJemri2025}.

\begin{figure*}[ht!]
\centering
\includegraphics[width=0.45\linewidth]{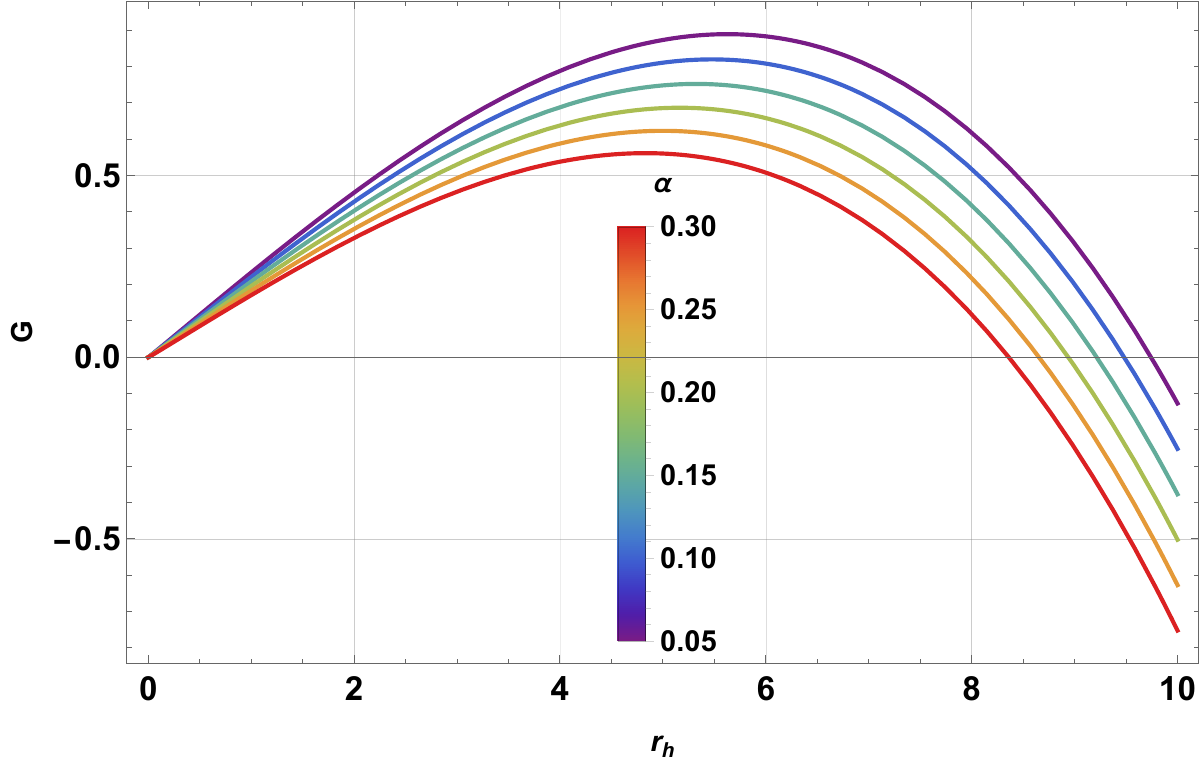}\qquad
\includegraphics[width=0.45\linewidth]{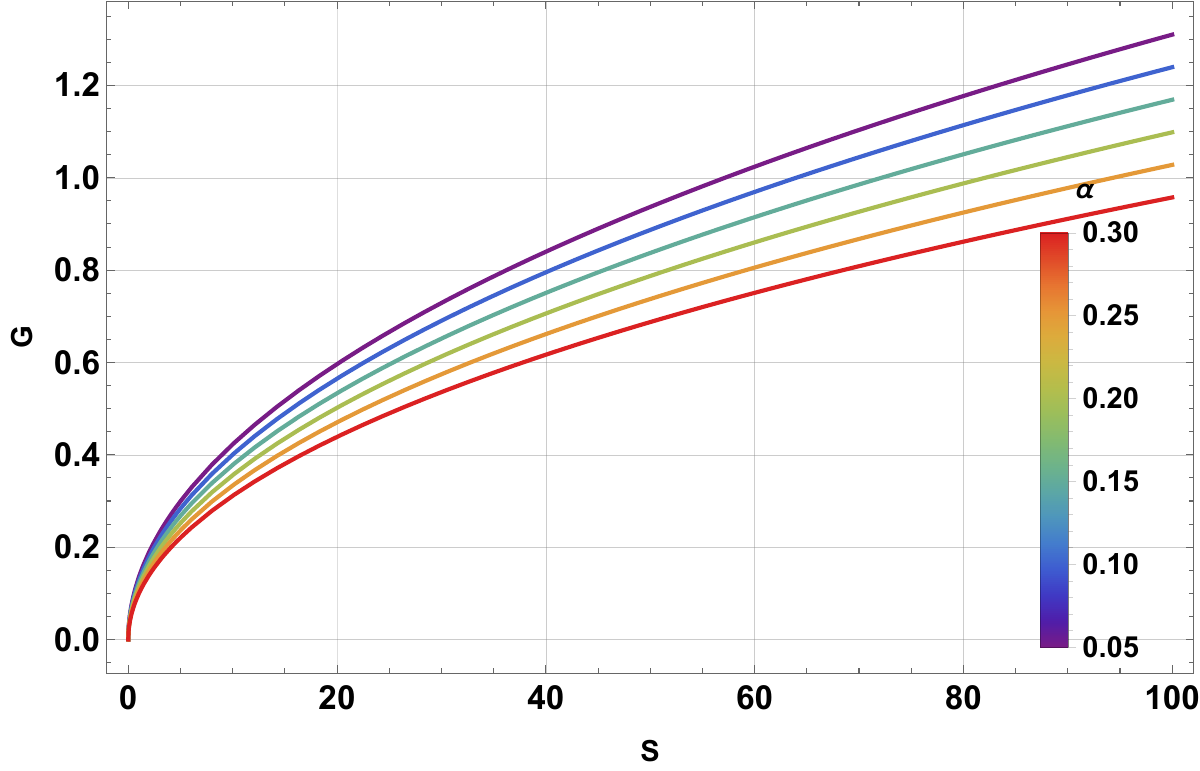}\\
(a) $\ell_p=15$ \hspace{6cm} (b) $P=0.002$
\caption{\footnotesize Behavior of the Gibb's free energy $G$ as a function of horizon $r_h$  (left panel) and entropy $S$ (right panel)  by varying the CoS parameter $\alpha$. Here, $M=1,,r_s=0.2,,\rho_s=0.02$.}
\label{fig:gibbs-1}
\includegraphics[width=0.45\linewidth]{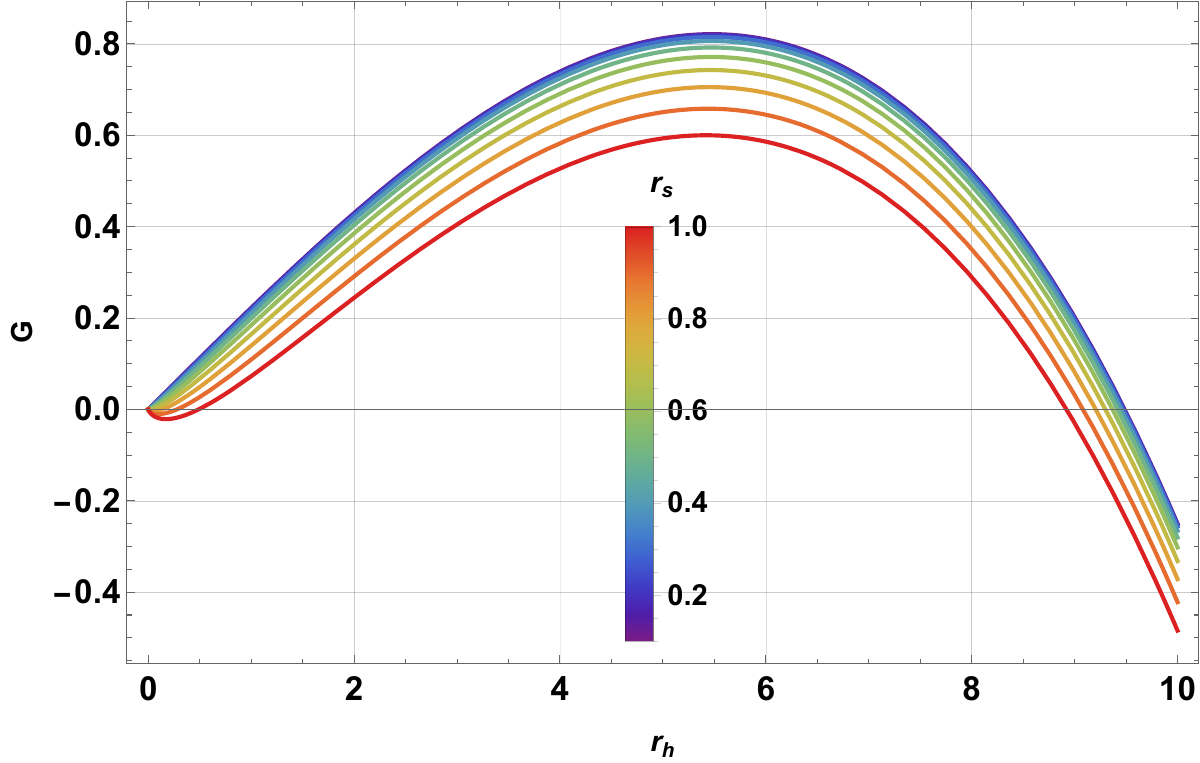}\qquad
\includegraphics[width=0.45\linewidth]{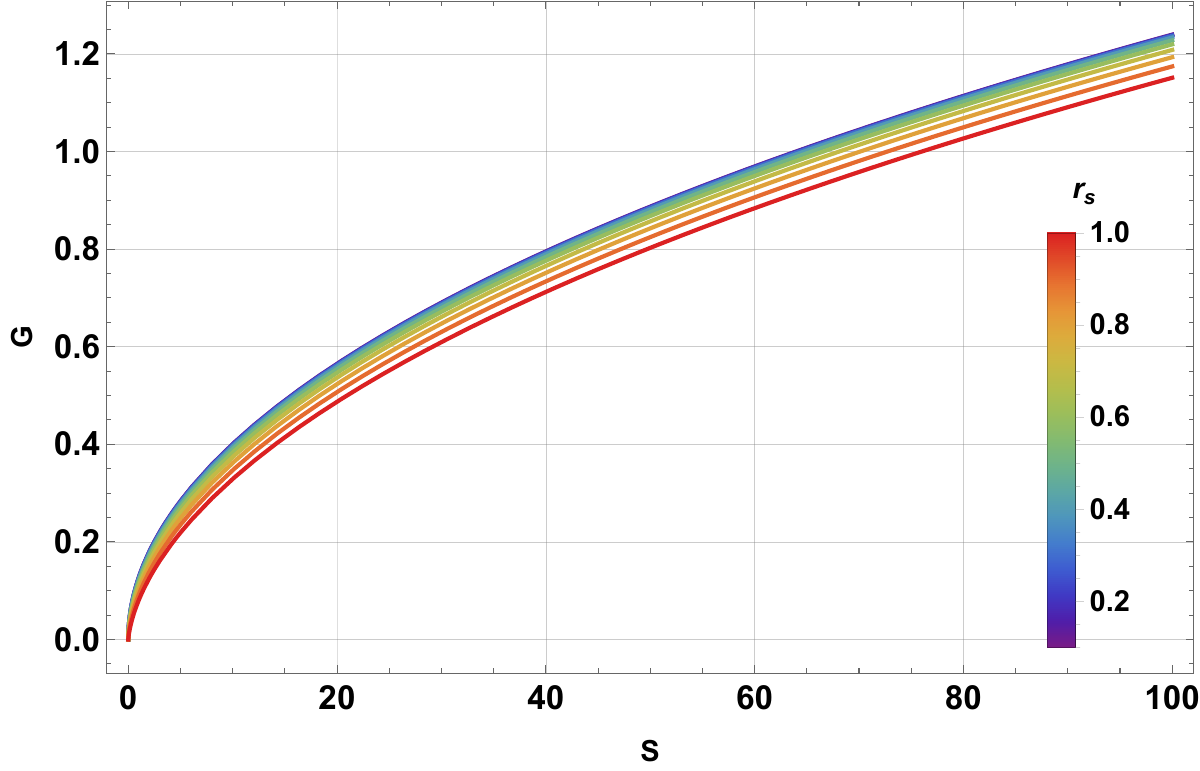}\\
(a) $\ell_p=15$ \hspace{6cm} (b) $P=0.003$
\caption{\footnotesize Behavior of the Gibb's free energy $G$ as a function of horizon $r_h$ (left panel) and entropy $S$ (right panel) by varying the halo radius $r_s$. Here, $M=1,,\alpha=0.1,,\rho_s=0.05$.}
\label{fig:gibbs-2}
\end{figure*}

\subsection{Gibbs free energy and Hawking-Page transition}

At fixed pressure $P$, the Gibbs free energy is
\begin{equation}
G(T,P)=M-TS,
\label{SS16}
\end{equation}
which, as a function of $r_h$, admits a compact form after inserting \eqref{SS3} and \eqref{SS6}:
\begin{align}
&G(r_h;P,\alpha,\rho_s,r_s)=\frac{r_h}{4}\notag\\ & \times \left[
1-\alpha
-\rho_s r_s^2\ln\!\Big(1+\frac{r_s}{r_h}\Big)
-\frac{\rho_s r_s^3}{r_h+r_s}
-\frac{P}{3}\,r_h^2
\right].
\label{SS17}
\end{align}
In terms of entropy, we have
\begin{align}
&G(S;P,\alpha,\rho_s,r_s)=\frac{1}{4}\sqrt{\frac{S}{\pi}}\notag\\ & \times \left[
1-\alpha
-\rho_s r_s^2\ln\!\Big(1+r_s\sqrt{\frac{\pi}{S}}\Big)
-\frac{\rho_s r_s^3}{\sqrt{S/\pi}+r_s}
-\frac{P}{3}\,S
\right].
\label{SS17a}
\end{align}

In Fig. \ref{fig:gibbs-1}, we depict the Gibbs free energy $G$ as a function of horizon and entropy by varying the CoS parameter $\alpha$. We observe that the Gibbs free energy decreases with increasing values of $\alpha$. 

Figure \ref{fig:gibbs-2} depicts the Gibbs free energy $G$ as a function of horizon and entropy by varying the halo radius $r_s$. Here also, we observe that the Gibbs free energy decreases with increasing values of $r_s$, showing similar behavior as does that of Fig. \ref{fig:gibbs-1}. 

\paragraph*{Global phases.} For $\alpha=\rho_s=0$, $G=\tfrac{r_h}{4}\big(1-r_h^2/\ell_p^2\big)$ and the Hawking-Page (HP) temperature occurs where $G$ changes sign, $r_h=\ell_p$ \cite{HawkingPage1983}. The CoS/DM sector lowers $G$ by a finite amount at fixed $r_h$, hence: (i) the HP point shifts to slightly smaller $T$ (large BH becomes globally preferred earlier); (ii) any swallow-tail structure (first-order small/large transition at fixed $P$) is displaced in the $(T, P)$ plane. This is the same qualitative effect seen when extra charges or noncommutative smearing are turned on Ref.     \cite{ElHadriJemri2025}.    

\subsection{Limiting regimes and qualitative trends}

In the Schwarzschild-AdS limit, obtained by setting $\alpha=\rho_s=0$, the standard relations are recovered,
\begin{align}
T&=\frac{1}{4\pi}\!\left(\frac{1}{r_h}+\frac{3r_h}{\ell_p^2}\right),\notag\\ 
V&=\frac{4\pi}{3}r_h^3, \notag\\
M&=\frac{r_h}{2}\!\left(1+\frac{r_h^2}{\ell_p^2}\right),  \notag   
\end{align}
with the Hawking--Page transition at $r_h=\ell_p$~\cite{HawkingPage1983}. For large BHs ($r_h\gg r_s$), the dark-matter corrections are suppressed as $\ln(1+r_s/r_h)\sim r_s/r_h$ and $r_s^3/(r_h+r_s)\sim r_s^3/r_h$, so the temperature approaches
\begin{equation}
 T\simeq \frac{1}{4\pi}\!\left(\frac{1-\alpha}{r_h}+\frac{3r_h}{\ell_p^2}\right)   
\end{equation}and the thermodynamics tends to the cloud-of-strings-deformed SAdS case~\cite{PSL}. In the small-black-hole regime ($r_h\ll r_s$), the enhancement $\ln(1+r_s/r_h)\sim \ln(r_s/r_h)$ magnifies the halo contribution, reduces $T$, and shifts the divergence of $C_P$; this reshapes the small/unstable branch and the onset of global dominance in $G$, altering the phase structure relative to the Van der Waals picture~\cite{KubiznakMann2012}.

Beyond the compact summary above, a few analytic expansions help clarify how each limit reorganizes the thermodynamics. It is convenient to isolate the combination as
\begin{equation}
A(r_h)\equiv (1-\alpha)-\rho_s r_s^2\ln\!\Big(1+\frac{r_s}{r_h}\Big)+\frac{\rho_s r_s^3}{r_h+r_s},   
\end{equation}
so that the temperature reads 
\begin{equation}
T=\frac{1}{4\pi r_h}\big[A(r_h)+3r_h^2/\ell_p^2\big],    
\end{equation}
 the equation of state is
 \begin{equation}
 P(T,r_h)=\frac{T}{2r_h}-\frac{A(r_h)}{r_h^2},     
 \end{equation}
 and the Gibbs free energy (at fixed $P$) is 
 \begin{equation}
 G=\frac{r_h}{4}\big[A(r_h)-\frac{P}{3}\,r_h^2\big]    
 \end{equation}
 up to the same $A(r_h)$ -controlled halo/CoS corrections used above.

\textit{Large-black-hole regime} $\mathbf{(r_h\gg r_s)}$. A systematic expansion gives
\begin{align}
\ln\!\Big(1+\frac{r_s}{r_h}\Big)&=\frac{r_s}{r_h}-\frac{r_s^2}{2r_h^2}+\frac{r_s^3}{3r_h^3}+\cdots,\\ 
\frac{1}{r_h+r_s}&=\frac{1}{r_h}\Big(1-\frac{r_s}{r_h}+\frac{r_s^2}{r_h^2}-\cdots\Big),   
\end{align}

which yields the cancellation of all $O(r_h^{-1})$ terms and the leading correction
\[
A(r_h)=(1-\alpha)-\frac{\rho_s r_s^4}{2\,r_h^2}+\frac{2\rho_s r_s^5}{3\,r_h^3}+O\!\left(\frac{r_s^6}{r_h^4}\right).
\]
Hence, the DM/halo sector decouples quadratically, and
\begin{align}
T&=\frac{1}{4\pi}\!\left(\frac{1-\alpha}{r_h}+\frac{3r_h}{\ell_p^2}\right)
-\frac{\rho_s r_s^4}{8\pi\,r_h^3}+O\!\left(\frac{r_s^5}{r_h^4}\right),
\\
G&=\frac{r_h}{4}\!\left[(1-\alpha)-\frac{P}{3}\,r_h^2\right]
-\frac{\rho_s r_s^4}{8\,r_h}+ \cdots.    
\end{align}

Thermally, the system approaches the cloud-of-strings deformation of SAdS, with only subleading $r_h^{-2}$ (and higher) halo corrections. In particular, the Hawking-Page (HP) radius is shifted only mildly:
solving $G=0$ gives $r_{HP}^2\simeq \frac{3}{P}(1-\alpha)\big[1+O(\rho_s r_s^4/r_{HP}^4)\big]$, i.e. the CoS deficit $\alpha$ lowers the HP radius while DM effects are suppressed by $(r_s/r_{HP})^4$.

\textit{Small-black-hole regime} $\mathbf{(r_h\ll r_s)}$. Here,
\begin{align}
\ln\!\Big(1+\frac{r_s}{r_h}\Big)&=\ln\!\Big(\frac{r_s}{r_h}\Big)+O\!\Big(\frac{r_h}{r_s}\Big),\\ 
\frac{r_s^3}{r_h+r_s}&=r_s^2\Big(1-\frac{r_h}{r_s}+O\!\Big(\frac{r_h^2}{r_s^2}\Big)\Big),  
\end{align}

so that
\[
A(r_h)\simeq (1-\alpha)-\rho_s r_s^2\Big[\ln\!\Big(\frac{r_s}{r_h}\Big)-1\Big].
\]
The logarithmic enhancement dominates, making $A(r_h)$ decrease as $r_h$ shrinks. Consequently,
\[
T\simeq \frac{A(r_h)}{4\pi r_h}\quad(\text{since }3r_h^2/\ell_p^2\ll A),
\]
is strongly reduced relative to SAdS. If parameters are such that $A(r_h)$ crosses zero, the temperature exhibits a minimum and vanishes at a finite radius $r_{0}$ determined implicitly by $A(r_0)=0$, i.e.
\[
r_{0}\;\simeq\; r_s\,\exp\!\left[-1+\frac{1-\alpha}{\rho_s r_s^2}\right].
\]
Below (or near) this point, the small-$r_h$ branch ceases to exist or becomes thermodynamically pathological, and the divergence of $C_P$ is pushed to larger $r_h$ compared to SAdS. The Gibbs free energy is likewise suppressed at small radii,
$
G\simeq \tfrac{r_h}{4}A(r_h)<\tfrac{r_h}{4}(1-\alpha)
$,
delaying the onset of global dominance of the BH phase against thermal AdS.

\textit{Schwarzschild-AdS recovery}. Setting $\alpha=\rho_s=0$ gives back
\begin{align}
T&=\frac{1}{4\pi}\big(\frac{1}{r_h}+\frac{3r_h}{\ell_p^2}\big), \\
V&=\frac{4\pi}{3}r_h^3, \label{vsads}\\
M&=\frac{r_h}{2}\big(1+\frac{r_h^2}{\ell_p^2}\big), 
\end{align}
and the HP transition at $r_h=\ell_p$~\cite{HawkingPage1983}. Turning on $\alpha$ acts as a uniform deficit that lowers the small-$r_h$ contribution to $T$ and reduces $r_{HP}$, whereas the DM halo produces a scale-dependent correction: negligible for $r_h\gg r_s$ but logarithmically important for $r_h\ll r_s$. In combination, these trends quantitatively reshape the unstable small-BH branch, shift the $C_P$ divergence, and move the HP line to lower temperature, while leaving the large-BH sector close to SAdS with a simple $(1-\alpha)$ renormalization and tiny $O\!\big((r_s/r_h)^4\big)$ halo tails.

\subsection{Joule-Thomson expansion }
For completeness, we note that the Joule-Thomson (JT) coefficient at fixed enthalpy $M$ is
\[
\mu_{\rm JT}\equiv \left(\frac{\partial T}{\partial P}\right)_{M}
=\frac{1}{C_P}\left[T\left(\frac{\partial V}{\partial T}\right)_{P}-V\right],
\]
with inversion curve determined by $\mu_{\rm JT}=0\Rightarrow T_i=V\left(\frac{\partial T}{\partial V}\right)_{P}$. Using Eqs. \eqref{SS14} and (\ref{vsads}), one can obtain $T_i(r_h)$ implicitly. As in charged and noncommutative AdS families \cite{OkcuAydiner2017,ElHadriJemri2025}, we find a single-branch inversion curve with $T_i$ increasing monotonically with $P$; larger $(\rho_s,r_s)$ typically lower $T_i$ at fixed $P$ by enhancing the effective attraction encoded in $A(r_h)$ (cf. Eq. \eqref{SS14}). This qualitative behavior is universal across a wide class of AdS BHs \cite{OkcuAydiner2017,Cong2021}.

Altogether, Eqs. \eqref{SS2}-\eqref{SS17a} provide a complete thermodynamic description of the CoS+DM AdS BH. The extended first law \eqref{SS8} and the modified Smarr relation~\eqref{SS19} hold with natural work terms for the matter sector, while local/global diagnostics ($C_P$, $G$) neatly capture how $\alpha$ and $(\rho_s,r_s)$ reshape the phase diagram. The deformations we observe parallel those reported in RN-AdS, rotating AdS, and noncommutative AdS BHs~\cite{KubiznakMann2012,KubiznakMannTeo2017,ElHadriJemri2025}, reinforcing the robustness of the AdS ``BH chemistry'' paradigm~\cite{KubiznakMannTeo2017}.

\section{Conclusions}\label{Sec:VII}

We studied a static, spherically symmetric Schwarzschild-AdS black hole modified by a cloud of strings and a Dehnen-type dark-matter halo. We showed that the metric function (\ref{function})
solves Einstein’s equations for the composite source $T^\mu{}_\nu(\text{CoS})+T^\mu{}_\nu(\text{DM})-\Lambda g^\mu{}_\nu$, yielding a halo density that falls off as $r^{-4}$ at large radii and a finite asymptotic mass offset $M_{\rm halo}=\rho_s r_s^3/2$. The geodesic analysis showed that, for photons, increasing the string-cloud parameter $\alpha$ and the halo scale $r_s$ lowers the effective barrier and weakens the effective radial force while enlarging the photon-sphere radius $r_{\rm ph}$ and the shadow size $R_s=b_c$, thereby increasing the capture cross section. For massive particles, the specific energy and angular momentum of circular motion grow with $\alpha$ and with $r_s$, whereas the ISCO radius exhibits a clear competition: it moves outward as $\alpha$ increases and inward as $r_s$ grows, reflecting respectively the angular-deficit effect of the string cloud and the logarithmic halo term.

In the topological description of light rings, the potential $H(r,\theta)=\sqrt{-g_{tt}/g_{\theta\theta}}$ enabled us to track the critical points of the unit vector field on the $(r,\theta)$ plane, confirming the persistence of an unstable circular null orbit throughout the explored parameter space and its continuous deformation with $(\alpha,\rho_s,r_s)$. Scalar perturbations indicated that the effective potential $V_s(r)$ decreases as either $\alpha$ or $r_s$ increases, and the corresponding quasinormal modes shift to lower oscillation frequencies with weaker damping, leading to a longer ringdown phase. These trends point to observational imprints in both shadow measurements and gravitational-wave spectroscopy. In extended thermodynamics, we derived closed-form expressions for $M(r_h)$, $T(r_h)$, and $S=\pi r_h^2$ with $P=3/\ell_p^2$, and we established a consistent first law and Smarr relation with natural work terms conjugate to $(\alpha,\rho_s,r_s)$. The string cloud uniformly lowers the temperature and moves extrema of $T(r_h)$ and specific-heat divergences to larger radii, while the halo suppresses $T$ at small $r_h$ and the AdS term dominates at large scales. As a consequence, the Gibbs free energy and the Hawking-Page transition are deformed, reshaping the thermal phase structure relative to standard Schwarzschild-AdS.

From a phenomenological perspective, the growth of $R_s$ and $b_c$ with $(\alpha,r_s)$ suggests that EHT-like shadow measurements can constrain combinations of the angular deficit and halo scale; the opposite trends of $r_{\rm ISCO}$ with $\alpha$ and $r_s$ imply distinct signatures in accretion efficiency and X-ray spectra, while ringdown frequencies provide complementary bounds through gravitational waves. Natural continuations include a rotating extension (Kerr-AdS with cloud and halo) to probe superradiance, photon regions, and shadow deformations; the assessment of alternative halo profiles and possible CoS-DM couplings confronted with galactic-center data; higher-order and time-domain QNM calculations, including massive and charged fields and tests of the eikonal link to photon rings; and a full mapping of the $(P,T)$ diagram with criticality and Joule-Thomson expansion. Altogether, the string cloud and the Dehnen halo furnish a controlled and astrophysically motivated laboratory where geodesic, perturbative, and thermodynamic observables coherently probe the near-horizon environment of AdS black holes.

\section*{Acknowledgments}

F.A. acknowledges the Inter University Centre for Astronomy and Astrophysics (IUCAA), Pune, India for granting visiting associateship. E. O. Silva acknowledges the support from grants CNPq/306308/2022-3, FAPEMA/UNIVERSAL-06395/22, FAPEMA/APP-12256/22, and (CAPES) - Brazil (Code 001).

\section*{Data Availability}

No data were generated or created in this article.

\section*{Conflicts of interest statement}

The authors declare(s) no conflicts of interest.

\bibliographystyle{apsrev4-2}
\bibliography{Refs}

\end{document}